\documentclass[epsf,10pt]{article}
\usepackage{amssymb, amsmath}
\begin{document}

\makeatletter
\@addtoreset{equation}{section}
\def\theequation{\thesection.\arabic{equation}}
\def\@maketitle{\newpage
 \null
 {\normalsize \tt \begin{flushright} 
  \begin{tabular}[t]{l} \@date  
  \end{tabular}
 \end{flushright}}
 \begin{center} 
 \vskip 2em
 {\LARGE \@title \par} \vskip 1.5em {\large \lineskip .5em
 \begin{tabular}[t]{c}\@author 
 \end{tabular}\par} 
 \end{center}
 \par
 \vskip 1.5em} 
\makeatother
\topmargin=-1cm
\oddsidemargin=1.5cm
\evensidemargin=-.0cm
\textwidth=15.5cm
\textheight=22cm
\setlength{\baselineskip}{16pt}
\title{Second-Order Formalism for 3D Spin-3 Gravity}
\author{
Ippei~{\sc Fujisawa} and
Ryuichi~{\sc Nakayama}
       \\[1cm]
{\small
    Division of Physics, Graduate School of Science,} \\
{\small
           Hokkaido University, Sapporo 060-0810, Japan}
}
\date{
EPHOU-12-006  \\
September  2012 
}
%
%
\maketitle

\begin{abstract} 
A second-order formalism for the theory of 3D spin-3 gravity is considered. 
Such a formalism is obtained by solving the torsion-free condition for 
the spin connection $\omega^a_{\mu}$, and substituting the result 
into the action integral. In the first-order formalism of the spin-3 
gravity defined in terms of $SL(3,R) \times SL(3,R)$ Chern-Simons (CS) 
theory, however, the generalized torsion-free condition cannot be easily 
solved for the spin connection, because the vielbein $e^a_{\mu}$ itself is 
not invertible. To circumvent this problem, extra vielbein-like fields 
$e^a_{(\mu\nu)}$ are introduced {\em as a functional of} $e^a_{\mu}$. 
New set of 
affine-like connections $\Gamma_{\mu M}^N$ are defined in terms of the 
metric-like fields, and a generalization of the Riemann curvature 
tensor is also presented. In terms of this generalized Riemann tensor 
the action integral in the second-order formalism is expressed. 
The transformation rules of the metric and the 
spin-3 gauge field under the generalized diffeomorphims are obtained 
explicitly. As in Einstein gravity, the new affine-like connections 
are related to the spin connection by a certain  gauge 
transformation, and a gravitational CS term  
expressed in terms of the new connections is also presented.

\end{abstract}
\newpage
\setlength{\baselineskip}{18pt}

\newcommand {\beq}{\begin{equation}}

\newcommand {\eeq}{\end{equation}}
\newcommand {\beqa}{\begin{eqnarray}}
\newcommand {\eeqa} {\end{eqnarray}}
\newcommand{\bm}[1]{\mbox{\boldmath $#1$}}
\newcommand{\Sq}{D(X)}
\newcommand{\al}{2\pi \alpha'}
\newcommand{\RR}{{\mathsf R\hspace*{-0.9ex}%
  \rule{0.15ex}{1.5ex}\hspace*{0.9ex}}}
\section{Introduction}
\hspace{5mm}
Gravity theories coupled to massless higher spin fields have been studied 
extensively in recent years.\cite{HR}\cite{Campoleoni}\cite{GGS}\cite{GG}
\cite{GaHa}\cite{GK}\cite{Ahn}\cite{GGHR}\cite{CY}\cite{AGKP}\cite{KP}
\cite{C2}\cite{GGR}\cite{GHJ}\cite{HGJR}\cite{BCT}\cite{CHLM}\cite{PTT}
\cite{review}
  This is due to the conjectured holographic 
relation between these theories in AdS$_4$ and the O(N) vector model in 
3D.\cite{Gubser} 
These higher-spin theories contain an infinite number of fields.
\cite{Fronsdal}\cite{FV1}\cite{FV2}\cite{Vasiliev} 

Three dimensional higher-spin theories\cite{Blencowe} are simpler to work with 
than those 
in higher dimensions, because higher-spin fields can be truncated to only 
those with spin $s \leq N$ and the theory can be defined in terms of the 
Chern-Simons action.\cite{Campoleoni}
 By using the $SL(3,R) \times SL(3,R)$ invariant 
Chern-Simons (CS) theory, the black hole solution with spin-3 charge in 
spin-3 gravity was obtained and studied.\cite{GK}\cite{AGKP}\cite{KP}\cite{CHLM}

The CS theory for gravity is very efficient for obtaining solutions. The 
condition of flat connections is easy to implement. The asymptotic 
behavior of the solution was found to be not necessarily ${\cal O}(r^2)$.
One needs to perform spin-3 gauge transformations to transform the black 
hole solution into a form with a manifest event horizon.\cite{GK}\cite{AGKP} 
The geometry of 3D higher spin gravity, which must be a generalization of the 
Riemannian geometry, is not well understood. It is difficult to understand 
this within the CS approach. Therefore, it is necessary to understand the 
geometry of higher-spin gravity in more details by a different approach. 
Additionally, general integral formulae for the spin-3 charge and 
the entropy are not yet derived. The explicit action integrals for 
matter fields 
coupled to higher-spin gravity is also not known.  

In the spin-2 gravity theory there exist first-order and second-order 
formalisms. 
In the first-order formalism, a spin connection and a vielbein field are 
introduced, and the action integral contains only the first-order derivatives 
of the fields. In the second-order formalism, the spin connection is 
eliminated by solving the torsion-free condition, and the solution is 
substituted into the action integral. Then the action integral becomes
quadratic in the derivatives. Both formalisms are equivalent. 
The CS formulation of the spin-3 gravity is the first-order one. It is 
expected that a second-order formalism also exists for the spin-3 gravity. 
In order to tackle this problem it is necessary to rewrite the theory 
of higher-spin gravity as a geometrical theory 
in terms of the metric-like fields. For this purpose one needs to define  
affine-like connections, covariant derivatives and the curvature tensors 
in the spin-3 gravity theory by using the vielbein fields. 

Apparently, this problem is difficult to solve, because in $SL(N,R) 
\times SL(N,R)$ CS theory (with $N \geq 3$), the dimension of the algebra 
$sl(N,R)$ is larger than that of spacetime. The vielbein $e^a_{\mu}$ is not 
a square matrix and does not have an inverse.\footnote{In \cite{Campoleoni}
 the torsion-free condition in spin-s gravity is solved to first order of 
expansion around a spin-2 background. In this paper the torsion-free condition
 will be solved explicitly without using perturbative expansions.} For example, 
in the $N=3$ case which corresponds to spin-3 gravity, $a$ runs from 1 to 8 
and $\mu$ from 0 to 2. In this paper, to compensate this gap of the numbers of 
components, auxiliary vielbein fields $e^a_{(\mu\nu)}$  will be introduced. 
With the traceless and symmetry conditions $g^{\mu\nu} \, e_{(\mu\nu)}^a=0$,  
$e^a_{(\mu\nu)}=e^a_{(\nu\mu)}$, the entire vielbein field becomes an 8 $\times$ 8
 matrix. So if this generalized vielbein field is non-degenerate, an inverse 
vielbein exists and the torsion-free condition can be solved. 

Now, many concepts and geometrical quantites can be introduced into the 
spin-3 gravity in parallel with the Einstein gravity. The purpose of this 
paper is to pursue this possibility. For example, two connections, 
$\Gamma_{\mu\nu}^{\lambda}$ and $\Gamma_{\mu\nu}^{(\lambda\rho)}$, are 
obtained by generalization of the vielbein postulate,  
$\partial_{\mu} \, e^a_{\nu}+ {f^a}_{bc} \, \omega^b_{\mu} \, e^c=
\Gamma_{\mu\nu}^{\lambda} \, e^a_{\lambda}+\frac{1}{2} \, 
\Gamma_{\mu\nu}^{(\lambda\rho)} \, e^a_{(\lambda\rho)}$, instead of the 
Christoffel symbol $\hat{\Gamma}_{\mu\nu}^{\lambda}$ in the Einstein gravity. 
These connections can be expressed {\em purely in terms of metric-like 
fields}, although somewhat formally. These connections 
are expected to be used to describe the geometry of spin-3 gravity. 
By using these connections appropriate covariant derivatives $\nabla_{\mu}$ 
can be introduced in such a way that the full covariant derivatives $D_{\mu}$ 
of the vielbeins, $D_{\mu} \, e^a_{\nu}=\nabla_{\mu} \, e^a_{\nu}+{f^a}_{bc} \, 
\omega^b_{\mu} \, e^c_{\nu}$ and  $D_{\mu} \, e^a_{(\nu\lambda)}= \nabla_{\mu} \, 
e^a_{(\nu\lambda)}+{f^a}_{bc} \, \omega^b_{\mu} \, e^c_{(\nu\lambda)}$, vanish. In 
the definition of $\nabla_{\mu} \, e^a_{(\nu\lambda)}$, two more connections, 
$\Gamma_{\mu, (\nu\lambda)}^{\rho}$ and $\Gamma_{\mu, (\nu\lambda)}^{(\rho\sigma)}$, 
are also defined in terms of metric-like fields. 
It then turns out convenient to combine  $e^a_{\mu}$ and  $e^a_{(\mu\nu)}$ into
a single vielbein field $e^a_M$, where $M$ takes two kinds of indices, 
$M=\mu$ and $M=(\mu\nu)$.  Here $(\mu\nu)$ denotes a traceless, symmetric 
pair of base-space indices. Similarly, the above four affine-like connections 
can be combined as $\Gamma_{\mu M}^N$. The metric tensor can also be 
generalized: defining $G_{MN}=e^a_M \, e_{a N}$, which is generalization of 
the metric tensor $g_{\mu\nu}=e^a_{\mu} \, e_{a \nu}$, we will find that $G_{MN}$ 
is compatible with the covariant derivative $\nabla_{\mu}$. 

The metric-like fields in spin-3 gravity theory include the spin-3 gauge 
field $\phi_{\mu\nu\lambda} =(1/2) \, d_{abc} \, e^a_{\mu} \, e^b_{\nu} \, e^c_{\lambda}$
\cite{Campoleoni} in addition to the ordinary metric field $g_{\mu\nu}$. Here 
$d_{abc}$ is the completely symmetric invariant tensor for $sl(3,R)$. 
This spin-3 gauge field is related to the generalized metric 
$G_{\mu(\nu\lambda)}$ 
defined above. It will be pointed out in sec.C of this paper that one can 
construct more metric-like fields by using invariant tensors, $d_{abc}$ and 
$f_{abc}$. Naturally, one expects that there will be relations among these 
metric-like fileds. For spin-3 geometry near AdS$_3$ vacuum, it is expected 
that only $g_{\mu\nu}$ and $\phi_{\mu\nu\lambda}$ are independent degrees of 
freedom and
other metric-like fields can be expressed in terms of them. Indeed, by using 
perturbation expansions around AdS$_3$ vacuum, it is possible to convince 
oneself that this expectation is correct.  However, closed form expressions 
for such relations among metric-like fields are not easy to obtain. One needs 
to introduce these extra fields to do all the rewriting of them in terms of 
$g_{\mu\nu}$ and $\phi_{\mu\nu\lambda}$.  It is also 
unclear if for {\em any} spin-3 geometry, all the metric-like fields can 
always be expressed in terms of $g_{\mu\nu}$ and $\phi_{\mu\nu\lambda}$. These 
problems will not be solved in this paper. Therefore in our second-order 
formalism the results are not expressed just in terms of the metric-like 
fields, but vielbein fields and structure constants will be left out in the 
final expressions. 

The covariant-constancy conditions for $e^a_{\mu}$ and $e^a_{(\mu\nu)}$ will be 
solved to yield the spin connection $\omega^a_{\mu}(e)$ as a functional of 
the vielbeins. 
\begin{equation}
\omega^a_{\mu \ c}(e) \equiv {f^a}_{bc} \, \omega^b_{\mu}(e)=-E_c^{\nu} \, 
\nabla_{\mu} 
\, e^a_{\nu}-\frac{1}{2} \, E_c^{(\nu\lambda)} \, \nabla_{\mu} \, e^a_{(\nu\lambda)}
\end{equation}
Here $E^{\mu}_a$ and $E^{(\mu\nu)}_a$ are inverse vierbeins. By substituting 
this into the CS action, the second-order action will be obtained. 
Furthermore, from the connections, 
$\Gamma_{\mu M}^N$, generalized curvature tensors, ${R^M}_{N \lambda\rho}(\Gamma)$, 
can be defined and the action integral can be expressed in terms of these 
curvature tensors. 
\begin{equation}
S_{\text{ 2nd order}} = \frac{k}{12\pi} \ \int \, d^3x \, \left\{ -
\epsilon^{\mu\nu\lambda} \, 
{(f^a}_{bc} \,  e^c_{\mu} \, e^b_M \, E^N_a) \, {R^M}_{N\nu\lambda}(\Gamma)+4 \, 
\epsilon^{\mu\nu\lambda} \, {f^a}_{bc} \, e^a_{\mu} \, e^b_{\nu} \, e^c_{\lambda} 
\right\} 
\end{equation}  
As this eq shows, the vielbein fields and the structure constants $f_{abc}$ 
still remain, and the action integral is not expressed purely in terms of 
the metric-like fields. This is due to the reason presented in the previous 
paragraph. However, the spin connection $\omega^a_{\mu}$ is eliminated and 
the action integral is 
expressed as the second-order forms of the vielbein fields. 
In this sense the formulation we obtained is the second-order one. What 
remains to be done is to reexpress the action only in terms of the metric-like 
fields. This will not be attempted in this paper.

As in Einstein gravity, the connections $\Gamma_{\mu M}^N$ and the spin 
connection $\omega^a_{\mu}$ turn out to be related 
by a gauge transformation. By using this fact the gravitational CS term 
$S_{\text{GCS}}(\Gamma)$ can be explicitly expressed in terms of the 
connections ($\Gamma$'s) and the topologically massive spin-3 gravity 
theory is defined. 
\begin{eqnarray}
S_{\text{GCS}}^{\text{spin-3}}(\Gamma) &=&  \frac{k}{8\pi \mu} \, \int \, d^3x \, 
\epsilon^{\mu\nu\lambda} \, 
\left(\Gamma^{\rho}_{\mu\sigma} \partial_{\nu} \, \Gamma^{\sigma}_{\lambda \rho}
+\frac{1}{2} \, \Gamma^{(\rho\sigma)}_{\mu\kappa} \, \partial_{\nu} \, 
\Gamma^{\kappa}_{\lambda, (\rho\sigma)}+\frac{1}{2} \, 
\Gamma^{\kappa}_{(\rho\sigma)} \, \partial_{\nu} \, 
\Gamma_{\lambda\kappa}^{(\rho\sigma)}\right. \nonumber \\
&&+\frac{1}{4} \, \Gamma^{(\rho\sigma)}_{\mu, (\kappa\tau)} \, 
\partial_{\nu} \, \Gamma^{(\kappa\tau)}_{\lambda, (\rho\sigma)}
+\frac{2}{3} \, \Gamma_{\mu\sigma}^{\rho} \, 
\Gamma_{\nu \kappa}^{\sigma} \, 
\Gamma_{\lambda\rho}^{\kappa} + \, \Gamma_{\mu, \sigma}^{(\rho\tau)} 
\, \Gamma_{\nu \kappa}^{\sigma} \, 
\Gamma_{\lambda, (\rho\tau)}^{\kappa} \nonumber \\ && \left. +\frac{1}{2} \, 
\Gamma_{\mu, (\sigma\eta)}^{(\rho\tau)} \, \Gamma_{\nu \kappa}^{(\sigma\eta)} \, 
\Gamma_{\lambda, (\rho\tau)}^{\kappa}+\frac{1}{12} \, \Gamma_{\mu,
 (\sigma\eta)}^{(\rho\tau)} \, \Gamma_{\nu (\kappa\alpha)}^{(\sigma\eta)} \, 
\Gamma_{\lambda, (\rho\tau)}^{(\kappa\alpha)} \right).
\end{eqnarray}

From the gauge transformations of the CS theory the generalized 
diffeomorphism of the metric field $g_{\mu\nu}$ and the spin-3 gauge field
 $\phi_{\mu\nu\lambda}$ can be defined and computed.  
These generalized diffeomorphisms are the ordinary 3D diffeomorphism and the 
spin-3 gauge transformation.  It will be checked that these fields 
appropriately transform as spin-two and spin-three fields, respectively, 
under the ordinary diffeomorphism. New transformation rules of these fields 
under the spin-3 gauge transformation will also be obtained explicitly. The 
results are compactly written as
\begin{eqnarray}
\delta \, g_{\mu\nu}& =& \nabla_{\mu} \, \xi_{\nu}+\nabla_{\nu} \, \xi_{\mu},
\\
\delta \, \phi_{\mu\nu\lambda}  
&=& \nabla_{\mu} \, \left(\xi_{(\nu\lambda)}+ \frac{1}{3} \, \rho^a \, 
\Lambda_a 
\, g_{\nu\lambda}\right)    
+\nabla_{\nu} \, \left(\xi_{(\lambda\mu)}+ \frac{1}{3} \, \rho^a \, 
\Lambda_a \, g_{\lambda\mu}\right) \nonumber \\ &&
+\nabla_{\lambda} \, \left(\xi_{(\mu\nu)} + \frac{1}{3} 
\, \rho^a \, \Lambda_a \, g_{\mu\nu}\right), 
\end{eqnarray}
where $\xi_{\mu}$ and $\xi_{(\mu\nu)}$ are parameters of the generalized 
diffeomorphisms, and related to a gauge function $\Lambda^a$ (\ref{xi}), 
(\ref{zeta}). $\rho^a$ is defined by $\rho^a=\frac{1}{2} \, {d^a}_{bc} \, 
e^b_{\lambda} \, e^c_{\rho} \, g^{\lambda\rho}$. 
 
This paper is organized as follows. 
Sec.2 is a brief review of 3D spin-3 gravity as a CS theory. In sec.3 a 
method for an extension of the vielbein field is explained. Then two 
connections, $\Gamma_{\mu\nu}^{\lambda}$, $\Gamma_{\mu\nu}^{(\lambda\rho)}$, 
analogous to the Christoffel symbol are introduced 
and it will be shown that they can be expressed in terms of metric-like fields.
In sec.4 the spin connection is determined as a functional of the vielbein.  
The covariant derivatives of the vielbeins are defined, and two more 
connections, 
$\Gamma_{\mu, (\nu\lambda)}^{\rho}$, $\Gamma_{\mu, (\nu\lambda)}^{(\rho\sigma)}$, are 
introduced and related to the old ones, $\Gamma_{\mu\nu}^{\lambda}$, 
$\Gamma_{\mu\nu}^{(\lambda\rho)}$.  We obtain an invariant action integral 
for the spin-3 gravity in the second-order formalism. This action is 
surely the one in the second-order formalism, because the spin connection 
is eliminated, but this action still contains the vielbeins and the structure 
constants $f_{abc}$, and is not expressed purely in terms of metric-like fields. 
We point out that the indices of tensors are shown to have 
a novel pairing  structure, $\mu$ and $(\mu\nu)$.  
In sec.5 the transformation rules of the  metric field $g_{\mu\nu}$ and the 
spin-3 gauge field $\phi_{\mu\nu\lambda}$ under the generalized diffeomorphism 
will be computed. 
In sec.6 curvature tensors for the spin-3 geometry is defined.
In sec.7 the spin-3 gravity version of the gravitational CS term is derived. 
Finally, sec.8 is reserved for summary and discussions. 
There are appendices A-E. Appendix A summarizes the $sl(3,R)$ formulae.
In appendix B it will be shown that for the AdS$_3$ background the vielbein 
system $e^a_{\mu}$, $e^a_{(\mu\nu)}$ is actually non-degenerate. The inverse 
vielbeins and other quantities are obtained. Killing vectors $\xi_{\mu}$ 
and tensors $\xi_{(\mu\nu)}$ for AdS$_3$ are also presented. In appendix C it 
is pointed out that in addition to the metric $g_{\mu\nu}$ and the spin-3 
gauge field $\phi_{\mu\nu\lambda}$, extra gauge fields such as 
$g_{(\mu\nu)(\mu\nu)}$, $g_{(\mu\nu)(\lambda\rho)\sigma}$, 
$g_{(\mu\nu)(\lambda\rho)(\sigma\kappa)}$ with the number of indices up to 6 
must be introduced. It is argued, by perturbation expansion to the first 
non-trivial order, that for spin-3 geometry near AdS$_3$ vacuum
all metric-like fields would be expressed in terms of $g_{\mu\nu}$ and 
$\phi_{\mu\nu\lambda}$. 
In appendix D the complicated part $S_{\mu\nu,\lambda\rho}$ 
of the connection $\Gamma_{\mu\nu}^{(\lambda\rho)}$ is obtained explicitly,  
if somewhat formally.  In appendix E a metric tensor $G_{MN}$ for 8D 
space, which corresponds to the two types of indices $M=\mu, (\mu\nu)$ 
and composed of the metric, the spin-3 gauge field and another  
gauge field $g_{(\mu\nu)(\lambda\rho)}$, is introduced and some formulae for this metric tensor are derived.

\vspace*{.5cm}

{\it Note Added:} While this work was being completed, we found that a 
work\cite{CFPT} appeared in the arXiv,  which attempts to formulate 
3D spin-3 gauge theory in terms of the metric-like fields, by means of the 
perturbation in powers of the spin-3 gauge field $\phi_{\mu\nu\lambda}$ up to
 ${\cal O}(\phi_{\mu\nu\lambda})^2$. The action integral and various 
transformation rules of metric-like fields obtained in the present paper 
are not based on perturbations in $\phi_{\mu\nu\lambda}$. Further, we also 
generalize the geometrical notions of Einstein gravity, such as the 
connections and the curvature tensors, to the spin-3 geometry.

\section{3D spin-3 gravity as Chern-Simons theory}
\hspace{5mm}
Let us start by briefly reviewing the 3D spin-3 gravity defined in terms 
of the Chern-Simons theory. 

The 3D spin-3 gravity with a negative cosmological constant 
is defined by the $SL(3,R) \times SL(3,R)$ Chern-Simons (CS) 
actions.\cite{AT}\cite{WittenCS} 
\begin{eqnarray}
S_{CS} = \frac{k}{4\pi} \, \int \mbox{tr} (A \wedge dA+\frac{2}{3} A \wedge 
A \wedge A)- \frac{k}{4\pi} \, \int \mbox{tr} (\bar{A} \wedge d\bar{A}+
\frac{2}{3} \bar{A} \wedge \bar{A} \wedge \bar{A})
\label{CS1}
\end{eqnarray}
Here $A$ and $\bar{A}$ are gauge-field one-forms and live in the fundamental 
representation of $sl(3,R)$. The constant $k$ is the level of the CS actions 
and is related to the three dimensional Newton constant $G_N$. The AdS 
length $\ell$ is given by $k= \ell/4 G_N$. The above action is invariant under 
$SL(3,R) \times SL(3,R)$ gauge transformations up to boundary terms.
\begin{eqnarray}
\delta A &=& d\Lambda + [A, \Lambda], \nonumber \\
\delta \bar{A} &=& d \bar{\Lambda} +[\bar{A}, \bar{\Lambda}]
\end{eqnarray}
Here $\Lambda$ is an $sl(3, R)$ matrix. These gauge fields $A$, $\bar{A}$ are 
related to the vielbein one-form $e=e_{\mu} \, dx^{\mu}$ and the spin connection 
one-form $\omega=\omega_{\mu} \, dx^{\mu}$ by the relations. 
\begin{eqnarray}
A=\omega+\frac{1}{\ell} \, e, \qquad \bar{A}= \omega- \frac{1}{\ell} \, e
\end{eqnarray}
Here $x^{\mu} \ (\mu=0,1,2)$ is the coordinate of 3D 
spacetime. The Greek indices $\mu, \nu, ..$ will be used for spacetime indices 
and the Roman letters $a, b, ..$ will be used for internal space (local frame) 
indices. 

In what follows $\ell$ will be set to 1. The above 
action is then written in terms of $e$ and $\omega$.
\begin{eqnarray}
S_{CS} = \frac{k}{\pi} \ \int \mbox{tr} \ e \wedge \left(d\omega+\omega 
\wedge \omega+\frac{1}{3} \  e \wedge e\right)
\label{CS}
\end{eqnarray}
The vielbein and the spin connection are expanded in terms of the $sl(3, R)$ 
generators $t_a$. 
(See appendix A) \footnote{In \cite{Campoleoni} a notation $e^a_{\mu}$ 
($a=1,2,3$) is used for the vielbein and the extra field $e^{ab}_{\mu}$ 
($a,b=1,2,3$) is called spin-3 gauge field. We will not adopt this notation.}
\begin{eqnarray}
e = e^a \, t_a=  e^a_{\mu} \, t_a \, 
dx^{\mu}, \qquad \omega =\omega^a \, t_a= \omega^a_{\mu} \, t_a \, dx^{\mu}
\end{eqnarray}

The gauge transformations on $e$ and $\omega$ fall into two groups.

\begin{description}
\item [a. Local frame rotation (or, extended Lorentz transformation)] 
\begin{eqnarray}
\delta e &=&  [e, \, \Lambda_+], \label{elocal}\\
\delta\omega &=& d \, \Lambda_++
 [\omega,\, \Lambda_+], \qquad \Lambda_+ \equiv \frac{1}{2} \, 
(\Lambda+\bar{\Lambda}) \label{local}
\end{eqnarray}
These transformations are Lorentz transformations and extended rotations 
in local frame.  

\item [b. Generalized diffeomorphism (or, extended local translation)] 
\begin{eqnarray}
\delta e &=&  d \, \Lambda_-+  [\omega, \, \Lambda_-], \label{e} \\
\delta \omega &=&  [e, \, \Lambda_-], \qquad \Lambda_- \equiv \frac{1}{2} \, 
(\Lambda-\bar{\Lambda})
\label{diffeo}
\end{eqnarray} 
These transformations are the ordinary spacetime diffeomorphism and the 
spin-3 gauge transformations. 
\end{description}

As usual, the metric tensor $g_{\mu\nu}$ of the spacetime is defined in 
terms of 
the vielbein $e_{\mu}= e^a_{\mu} \, t_a$.
\begin{eqnarray}
g_{\mu\nu} = \frac{1}{2} \, \mbox{tr} \ e_{\mu} \, e_{\nu}= h_{ab} \ 
e^a_{\mu} \ e^b_{\nu}= e^a_{\mu} \ e_{a \nu}
\label{metric}
\end{eqnarray} 
For the definition of the Killing metric $h_{ab}$ for the $sl(3,R)$ algebra see 
appendix A. Throughout this paper $g_{\mu\nu}$ is assumed to be non-degenerate. 
It is also assumed that its signature is $(-,+,+)$. 
Its inverse is denoted as $g^{\mu\nu}$. Also in the 
literature\cite{Campoleoni},  the spin-3 gauge field 
$\phi_{\mu\nu\lambda}$ 
is defined.
\begin{eqnarray}
\phi_{\mu\nu\lambda} = \frac{1}{4} \  \mbox{tr} \ \{e_{\mu}, e_{\nu} \} \ 
e_{\lambda}
\label{spin3}
\end{eqnarray}
Here $\{ \ , \  \}$ is an anti-commutator. This tensor and $g_{\mu\nu}$ are 
supposed to be independent fields. The number of components of $e^a_{\mu}$ 
is 24 and there are 8 independent local frame transformations. So there 
must be 16 independent degrees of freedom for the metric-like fields. The 
metric $g_{\mu\nu}$ and the spin-3 gauge field $\phi_{\mu\nu\lambda}$ have 6 
and 10 independent components, respectively. Thus it is expected that 
$g_{\mu\nu}$ and $\phi_{\mu\nu\lambda}$ are sufficient.  However, in appendix C 
it will be shown that extra metric-like tensor fields must be introduced for 
describing spin-3 gravity. It is expected that these extra metric-like 
fields are not independent 
of the above fields, although explicit formulae relating them are nested and 
complicated.\footnote{A method for obtaining some of these relations is 
suggested at the end of appendix E.}  

One can resort to perturbation expansions around the AdS vacuum to the first 
non-trivial order, and argue that all these extra metric-like fields can be 
expressed in terms of $g_{\mu\nu}$ and $\phi_{\mu\nu\lambda}$. See appendix C for 
discussion. Full-order closed expressions are, however, not available. It is 
also unclear if these extra metric-like fields can be described in terms 
of $g_{\mu\nu}$ and $\phi_{\mu\nu\lambda}$, even when the vielbeins are near general
 backgrounds with non-vanishing spin-3 gauge field. 

In this paper, it will be assumed that all the metric-like fields can be 
expressed in terms of $g_{\mu\nu}$ and $\phi_{\mu\nu\lambda}$. However, because 
actually expressing them in terms of the metric fields and the spin-3 gauge 
fields seems very complicated, for technical reasons, we will leave the extra 
fields as they are and not try to represent them in terms of $g_{\mu\nu}$ and 
$\phi_{\mu\nu\lambda}$.

\section{Extension of the vielbein}
\hspace{5mm}
The eqs of motion for the CS theory (\ref{CS}) are given by the 
conditions of flat connections,
\begin{eqnarray}
F \equiv dA+ A \wedge A=0, \qquad  \bar{F} \equiv d\bar{A}+\bar{A} 
\wedge \bar{A}=0,
\end{eqnarray}
and in terms of $e$ and $\omega$ these eqs are rewitten as a 
torsion-free condition 
\begin{equation}
{\cal T} \equiv \frac{1}{2} \, (F-\bar{F}) = de + e \wedge \omega+\omega 
\wedge e=0 
\label{torsionless}
\end{equation}
and an Einstein-like eq with negative cosmological constant.
\begin{equation}
{\cal R} \equiv \frac{1}{2} \, (F-\bar{F}) =d\omega+ \omega \wedge \omega+ e 
\wedge e=0 \label{Einstein-like}
\end{equation}
This eq describes more degrees of freedom than those of gravity. 

In the ordinary (spin-2) 3D gravity described by $SL(2,R) \times SL(2,R)$ 
CS action, the condition (\ref{torsionless}) can be solved for $\omega$, if 
the dreibein $e^a_{\mu}$ is invertible. Then this solution is substituted into 
(\ref{Einstein-like}) and the Einstein eq for Anti-de-Sitter (AdS) space
is obtained. Similarly, to formulate a second-order theory of spin-3 
gravity in terms 
of the metric-like fields, $g_{\mu\nu}$, $\phi_{\mu \nu \lambda}$, .., it is 
necessary to first solve the torsion-free condition and express the spin 
connection $\omega$ in terms of $e$ and then, 
substitute the result into the action (\ref{CS}). The vielbein $e^a_{\mu}$ 
$(a=1, \ldots, 8; \mu=0,1,2)$, however, has a form of a rectangular matrix and 
is non-invertible, even if $g_{\mu\nu}$ is non-degenerate.

\subsection{New vielbein $e^a_{(\mu\nu)}$  }
\hspace{5mm}
To resolve this difficulty, 5 more basis vectors in the local frame must be 
introduced. Let us define the $SL(3,R)$ matrices
\begin{eqnarray}
\hat{e}_{\mu\nu} = \frac{1}{2} \ \{e_{\mu}, e_{\nu} \}-\frac{2 }{3} \
 g_{\mu\nu} \ \bm{I}.
\end{eqnarray} 
The second term on the right proportional to the identity matrix $\bm{I}$ 
is added to ensure tracelessness of $\hat{e}$ as a matrix of $sl(3,R)$. 
This is symmetric in the indices $\mu$, $\nu$ and there are 6 independent 
components. We are now going to regard the set ($e^a_{\mu}$, 
$\hat{e}^a_{(\mu\nu)}$) as an 8 $\times$ 8 matrix and define its inverse matrix. 
For this purpose we need to reduce the number of components by one, and we 
choose to subtract 
the trace of $\hat{e}_{\mu\nu}$ with respect to the indices $\mu$, $\nu$. 
\begin{eqnarray}
e_{(\mu\nu)} &\equiv& \hat{e}_{\mu\nu}- \frac{1}{3} \ g_{\mu\nu} \ \rho=\frac{1}{2} \ 
\{e_{\mu}, e_{\nu} \}- \frac{1}{3} \ g_{\mu\nu} \ \rho-\frac{2 }{3} \ 
g_{\mu\nu} \ \bm{I} \label{emunu}, \\
\rho & \equiv & g^{\lambda \rho} \ \hat{e}_{\lambda \rho} \label{rho}
\end{eqnarray}
The matrix $\rho$ is the trace part of $\hat{e}_{\mu\nu}$, and $e_{(\mu\nu)}$
satisfies $g^{\mu\nu} \, e_{(\mu\nu)}=0$.\footnote{
It is easy to find  background 
vielbeins for which actually the matrix $\rho \neq 0$.}
Although these additional matrices $e_{(\mu\nu)}$ functionally depend on 
$e_{\mu}$, 8D vectors $e^a_{(\mu\nu)}$ ($a=1, \ldots, 8$) defined by 
$e_{(\mu\nu)}=e^a_{(\mu\nu)} \, t_a$ are assumed to span the 8D space together with 
$e^a_{\mu}$. In appendix B the vielbeins for AdS$_3$ vacuum are examined and it 
is shown that this is the case. It can also be shown that extented vielbeins
($e^a_{\mu}$, $e^a_{(\mu\nu)}$) for the BTZ black hole embedded in the spin-3 
gravity\cite{Campoleoni}  are also non-degenerate. 

However, for general $e^a_{\mu}$, the extended vielbeins ($e^a_{\mu}$, 
$e^a_{(\mu\nu)}$) may be degenerate at some points. Even if this is the case, 
we may employ the usual method of the fibre bundles to avoid 
the singularity. Let us consider the case where the CS gauge field one-forms 
take the forms, $A=b^{-1} \, a \, b+b^{-1} \, db$, $\bar{A}=b \, \bar{a} \, 
b^{-1}+b \, db^{-1}$ with $b =\exp (r \, L_0)$\cite{Campoleoni}. $r$ is the 
radial coordinate, and $a$ and $\bar{a}$ are one-forms independent of $r$. 
Let the extended vielbeins be degenerate only at $r=r_0$. We cover the base 
manifold by two open coordinate neighborhoods, $U_1=\{(r,t,\phi)| \,  r > 
\alpha \}$, $U_2 =\{(r,t,\phi)| \, r < \beta\}$ with $r_0 < \alpha < \beta$.
The gauge fields on $U_i$ will be denoted as $A^i$ and $\bar{A}^i$, and 
we set $A^1(r) \equiv A(r)$, $\bar{A}^1(r) \equiv \bar{A}(r)$. 
We choose as a transition function (gauge transformation) on the overlap $U_1 
\cap U_2=\{(r,t,\phi)| \, \alpha < r < \beta\}$, an extended local translation 
$V_-=\exp \Lambda_-=\exp (-y L_0)$, where $L_0$ is one of the $sl(3, R)$ 
generators (\ref{generators}) and $y$ is a constant satisfying $y > 
\beta-r_0 > \beta-\alpha$. The transition function $V_-$ has an effect of a 
translation $r \rightarrow r-y$ on the gauge fields $A^{1}(r)$ and 
$\bar{A}^{1}(r)$ in $U_1$, and the gauge fields on $U_2$ are given by 
$A^2(r)=A^1(r-y)$ and 
$\bar{A}^2(r)=\bar{A}^1(r-y)$. Then, we have $r-y<\beta-y<r_0$ in  
$U_2$ and the extended vielbeins ($e^a_{\mu}$, $e^a_{(\mu\nu)}$) computed from 
$A^2(r)$ and $\bar{A}^2(r)$ are non-degenerate in $U_2$. If there are
more degenerate points, the extended vielbeins may also be made  
non-degenerate by the same procedure.

Now if $e^a_{\mu}$ and $e^a_{(\mu\nu)}$ are combined, these can be regarded as
 an $8 \times 8$ matrix and it has an inverse matrix. Let us define inverse 
vielbeins $E^{\mu}_a$ and $E^{(\mu\nu)}_a$ by
\begin{eqnarray}
E^{\mu}_a \ e^a_{\nu} &=& \delta^{\mu}_{\nu}, \qquad 
E^{\mu}_a \ e^a_{(\nu\lambda)} = 0,  \nonumber \\
E^{(\mu\nu)}_a \ e^a_{\lambda} &=&0, \qquad 
E^{(\mu\nu)}_a \ e^a_{(\lambda \rho)} = \delta^{\mu}_{\lambda} \, 
\delta^{\nu}_{\rho}+
\delta^{\nu}_{\lambda} \, \delta^{\mu}_{\rho}
-\frac{2}{3} \ g_{\lambda\rho} \ g^{\mu\nu}.
\label{Ee}
\end{eqnarray}
The right-hand side of the last eq ensures the tracelessness of 
$E^{(\mu\nu)}_a$ (and of $e^a_{(\mu\nu)}$): $g_{\mu\nu} \, E^{(\mu\nu)}_a=0$.
They also satisfy the relation 
\begin{eqnarray}
e^a_{\mu} \, E^{\mu}_b+\frac{1}{2} \, e^a_{(\mu\nu)} \, E^{(\mu\nu)}_b={\delta^a}_b.
\label{completeness}
\end{eqnarray}
In front of the second term on the left-hand side, a factor  $1/2$ appears 
because of the symmetry $\mu \leftrightarrow \nu$. 

\subsection{New connections }
\hspace{5mm}

Let us turn to the torsion-free condition (\ref{torsionless}) now expressed in 
terms of components. 
\begin{equation}
\partial_{\mu} \, e_{\nu}-\partial_{\nu} \, e_{\mu}+[ \omega_{\mu}, 
e_{\nu}]-[\omega_{\nu} , \, e_{\mu}]=0
\label{torsionfree}
\end{equation}
This relation implies that $\partial_{\mu} \, e_{\nu}
+[ \omega_{\mu}, e_{\nu}]$ is 
symmetric for interchange of $\mu$ and $\nu$. Because this is a traceless 
matrix, it can be expanded in terms of $e_{\lambda}$ and $e_{(\lambda\rho)}$. 
\begin{equation}
\partial_{\mu} \, e_{\nu}+[ \omega_{\mu}, e_{\nu}]= \Gamma_{\mu\nu}^{\rho} 
\ e_{\rho}
+\frac{1}{2} \ \Gamma_{\mu\nu}^{(\sigma\rho)} \ e_{(\sigma\rho)}
\label{GammaGamma}
\end{equation} 
Here $\Gamma_{\mu\nu}^{\lambda}$ and $\Gamma_{\mu\nu}^{(\lambda\rho)}$ are two 
connections to be determined later, {\em as functions of the metric-like fields 
and their derivatives}. These are symmetric in the lower indices
$\mu$ and $\nu$. In (\ref{GammaGamma}) one can choose 
$\Gamma_{\mu\nu}^{(\lambda\rho)}$ such that it satisfies 
$g_{\lambda\rho} \,  \Gamma_{\mu\nu}^{(\lambda\rho)}=0$, since it is multiplied by 
$e_{(\lambda\rho)}$. Note that the existence of $\Gamma_{\mu\nu}^{\lambda}$ and 
$\Gamma_{\mu\nu}^{(\lambda\rho)}$ are ensured by the assumed linear independence 
of $e_{\mu}=t_a \, e_{\mu}^a$ and $ e_{(\mu\nu)}=t_a \, e^a_{(\mu\nu)}$. One may 
represent these connections in terms of $\omega^a_{\mu}$ and the vielbeins. 
If this is the case, the introduction of the new connections will not be of much use.  Certainly, this is not our 
purpose. {\em We will express them only in terms of the vielbeins.} 

In order to derive such expressions, let us multiply (\ref{GammaGamma}) by 
$e_{\lambda}$ from the right. 
\begin{eqnarray}
\partial_{\mu} \, e_{\nu} \ e_{\lambda}+[ \omega_{\mu}, e_{\nu}]\ e_{\lambda}
= \Gamma_{\mu\nu}^{\rho} \ e_{\rho}\ e_{\lambda}
+\frac{1}{2} \ \Gamma_{\mu\nu}^{(\sigma\rho)} \ e_{(\sigma\rho)}\ e_{\lambda}
\label{eee}
\end{eqnarray}
Then by replacing $\nu$ by $\lambda$ in (\ref{GammaGamma}), multiplying it
by $e_{\nu}$ from the left and adding the result to (\ref{eee}) the following 
relation is obtained.
\begin{eqnarray}
\partial_{\mu} \ (e_{\nu} \, e_{\lambda}) +[ \omega_{\mu}, \,  e_{\nu} \, 
e_{\lambda}]
&=&\Gamma_{\mu\nu}^{\sigma} \ e_{\sigma} \, e_{\lambda}
+\Gamma_{\mu\lambda}^{\sigma} \ 
e_{\nu} \, e_{\sigma} \nonumber \\
&&+ \frac{1}{2} \, \Gamma_{\mu\nu}^{(\sigma\kappa)} \, e_{(\sigma\kappa)} \, 
e_{\lambda}
+\frac{1}{2} \, \Gamma_{\mu\lambda}^{(\sigma\kappa)}\, e_{\nu} \, 
e_{(\sigma\kappa)} 
\label{delee}
\end{eqnarray}
By taking the trace of this eq and using (\ref{metric}) the following 
relation is obtained.
\begin{equation}
\partial_{\mu} \, g_{\nu\lambda} = \Gamma_{\mu\nu}^{\sigma} \, 
g_{\sigma\lambda}+
\Gamma_{\mu\lambda}^{\sigma} \, g_{\sigma\nu} 
+\frac{1}{2} \, \Gamma_{\mu\nu}^{(\sigma\kappa)} \ g_{(\sigma\kappa)\lambda}+
\frac{1}{2} \, \Gamma_{\mu\lambda}^{(\sigma\kappa)} \ g_{(\sigma\kappa)\nu}
\label{delg}
\end{equation}
Here the following field is defined.
\begin{equation}
g_{(\mu\nu)\lambda} = \frac{1}{2} \, \mbox{tr} \ (e_{(\mu\nu)} \, e_{\lambda})
\label{g3}
\end{equation}
The parenthesis in the subscript of $g$ on the left-hand side is put 
to make the 
permutation symmetry manifest. This field is different from, but related to, 
the spin-3 gauge field $\phi_{\mu\nu\lambda}$ defined in (\ref{spin3}). 
\begin{equation}
\phi_{\mu\nu\lambda}= g_{(\mu\nu)\lambda}+\frac{1}{6} \, g_{\mu\nu} \,
 \mbox{tr}(\rho \,e_{\lambda})
\label{phig3}
\end{equation}
This new field satisfies $g^{\mu\nu} \, g_{(\mu\nu)\lambda}=0$.\footnote{
In what follows, when $ \phi_{\mu\nu\lambda}$ is contracted with a tensor 
$\psi^{\mu\nu \dots}$ which is traceless in $\mu\nu$, $\phi_{\mu\nu\lambda}$ will be 
replaced by $ g_{(\mu\nu)\lambda}$ without notice.} The matrix $\rho$ was
defined in (\ref{rho}). Note that the trace in the second term on the right 
can be represented in terms of the spin-three gauge field. 
\begin{equation}
\mbox{tr}(\rho \, e_{\lambda})= 2 \, \phi_{\mu\nu\lambda} \, g^{\mu\nu}
=2 \,{ \phi^{\mu}}_{\mu\lambda}
\label{rhoe}
\end{equation}
As shown in the above expression spacetime indices are raised and/or lowered by 
$g^{\mu\nu}$ and $g_{\mu\nu}$.\footnote{Exception: the indices of 
$e^a_{\mu}$, $e^a_{(\mu\nu)}$, $E_a^{\mu}$ and $E_a^{(\mu\nu)}$ will 
not be raised and/or lowered by the metric tensors.}

Now eq (\ref{delg}) can be solved to yield the following relation by using 
the usual method.
\begin{eqnarray}
\Gamma_{\mu\nu}^{\lambda}+\frac{1}{2} \, \Gamma_{\mu\nu}^{(\sigma\kappa)} \, 
{\phi_{\sigma\kappa}}^{\lambda} = \frac{1}{2} \, g^{\lambda\rho} \, 
(\partial_{\mu} \, 
g_{\nu\rho}+\partial_{\nu} \, g_{\mu\rho}-\partial_{\rho} \, g_{\mu\nu})
\equiv {\hat{\Gamma}}_{\mu\nu}^{\lambda}
\label{GammaGamma1}
\end{eqnarray}
${\hat{\Gamma}}_{\mu\nu}^{\lambda}$ is the ordinary Christoffel symbol 
(connection).

\subsection{Determination of $\Gamma^{\lambda\rho}_{\mu\nu}$}
\hspace{5mm}

Another relation is required to determine the two connections uniquely. 
Multiplying 
(\ref{delee}) by $e_{\rho}$ from the right and adding a similar eq 
obtained by multiplying (\ref{GammaGamma}) (with replacement 
$\nu \rightarrow \rho$) 
by $e_{\nu}e_{\lambda}$ from the left, the following eq is obtained.
\begin{eqnarray}
&&\partial_{\mu} \, (e_{\nu} \, e_{\lambda} \, e_{\rho})
+[\omega_{\mu}, e_{\nu} \, e_{\lambda} \, e_{\rho}] \nonumber \\
&=& \Gamma_{\mu\nu}^{\sigma} \ e_{\sigma} \, e_{\lambda} \, e_{\rho}+  
\Gamma_{\mu\lambda}^{\sigma} \ e_{\nu} \, e_{\sigma} \, e_{\rho}+  
\Gamma_{\mu\rho}^{\sigma} \ e_{\nu} \, e_{\lambda} \, e_{\sigma} \nonumber \\
&& +\frac{1}{2} \ \Gamma_{\mu\nu}^{(\sigma\kappa)} \, e_{(\sigma\kappa)} \, 
e_{\lambda} 
\, e_{\rho} 
 +\frac{1}{2} \ \Gamma_{\mu\lambda}^{(\sigma\kappa)} \, e_{\nu} \,
 e_{(\sigma\kappa)} 
\, e_{\rho}
 +\frac{1}{2} \ \Gamma_{\mu\rho}^{(\sigma\kappa)}  \, e_{\nu} 
\, e_{\lambda} \, e_{(\sigma\kappa)}
\end{eqnarray}
Now interchange $\lambda$ and $\rho$ in the above eq and add the 
result to the above. Taking the trace leads to the following eq.
\begin{eqnarray}
 \partial_{\mu} \ \mbox{tr}\ e_{\nu} \, \{e_{\lambda}, \, e_{\rho}\} 
&=& \Gamma_{\mu\nu}^{\sigma} \ \mbox{tr} \ e_{\sigma} \, \{e_{\lambda}, 
\, e_{\rho} \}
+\Gamma_{\mu\lambda}^{\sigma} \ \mbox{tr} \ e_{\nu} \, \{e_{\sigma}, \, 
e_{\rho} \}
+\Gamma_{\mu\rho}^{\sigma} \ \mbox{tr}\ e_{\nu} \, \{e_{\lambda}, \, 
e_{\sigma} \} 
\nonumber \\
&&+\frac{1}{2} \ \Gamma_{\mu\nu}^{(\sigma\kappa)} \, \mbox{tr}\ 
e_{(\sigma\kappa)} \, \{
e_{\lambda}, \, e_{\rho}\}
+\frac{1}{2} \ \Gamma_{\mu\lambda}^{(\sigma\kappa)} \, \mbox{tr}\
 e_{(\sigma\kappa)} \, \{
e_{\rho}, \, e_{\nu}\} \nonumber \\  &&
+\frac{1}{2} \ \Gamma_{\mu\rho}^{(\sigma\kappa)} \, \mbox{tr}\ 
e_{(\sigma\kappa)} \, \{
e_{\lambda}, \, e_{\nu}\}
\end{eqnarray}
In the above eq let us note that one can make the following 
substitutions. 
\begin{eqnarray}
&& \mbox{tr} \, e_{\sigma} \ \{e_{\lambda}, \, e_{\rho}\}
=4 \, \phi_{\sigma\lambda\rho}, \label{eeephi} \\
&& \mbox{tr} \, e_{(\sigma\kappa)} \, \{e_{\lambda}, \, e_{\rho}\}
= 4 \, g_{(\sigma\kappa)(\lambda\rho)}+
\frac{2}{3} \, g_{\lambda\rho} \ \mbox{tr} \ (e_{(\sigma\kappa)} \, \rho)
\label{eeeg}
\end{eqnarray}
Here $g_{(\sigma\kappa)(\lambda\rho)}$ is defined in appendix C.
Furthermore, the connection $\Gamma_{\mu\nu}^{\lambda}$ can be eliminated by use of 
(\ref{GammaGamma1}). This yields the eq.
\begin{eqnarray}
\hat{\nabla}_{\mu} \ \phi_{\nu\lambda\rho} &=& \frac{1}{2} \, 
\Gamma_{\mu\nu}^{(\sigma\kappa)} \left(g_{(\sigma\kappa)(\lambda\rho)}-
{g_{(\sigma\kappa)}}^{\tau} \, g_{\tau(\lambda\rho)}-\frac{1}{6} \, g_{\lambda\rho} \, 
{g_{(\sigma\kappa)}}^{\tau} \, \mbox{tr} \ (\rho \, e_{\tau})+\frac{1}{6} \, 
g_{\lambda\rho} \ \mbox{tr} \ (\rho e_{(\sigma\kappa)})\right) \nonumber \\
&& + (\mbox{cyclic permutations of } 
\nu, \lambda, \rho)
\label{delphi}
\end{eqnarray}
Here $\hat{\nabla}_{\mu}$ is the covariant derivative which uses the Christoffel 
connection. 

Let us introduce a 5 $\times$ 5 matrix $M_{(\sigma\kappa)(\lambda\rho)}$ by
\begin{equation}
M_{(\sigma\kappa)(\lambda\rho)}\equiv g_{(\sigma\kappa)(\lambda\rho)}-{g_{(\sigma\kappa)}}^{\tau} 
\, g_{\tau(\lambda\rho)}. \label{M}
\end{equation}
Its inverse matrix $J^{(\lambda\rho)(\sigma\kappa)}$ is assumed to 
exist\footnote{This is true for AdS$_3$ solution as shown in (\ref{AdSJ}). }
 and defined 
by the following eq.
\begin{equation}
\frac{1}{2} \  M_{(\mu\nu)(\lambda\rho)} \ J^{(\lambda\rho)(\sigma\kappa)} 
 = \delta_{\mu}^{\sigma}\delta_{\nu}^{\kappa}
+\delta_{\nu}^{\sigma}\delta_{\mu}^{\kappa}-\frac{2}{3} \ g_{\mu\nu} \ g^{\sigma\kappa}
\label{J}
\end{equation}
Then (\ref{delphi}) is rewritten as
\begin{eqnarray}
\hat{\nabla}_{\mu} \ \phi_{\nu\lambda\rho} &=& 
\frac{1}{2} \, \Gamma_{\mu\nu}^{(\sigma\kappa)} \ M_{(\sigma\kappa)(\lambda\rho)} 
+\frac{1}{2} \, \Gamma_{\mu\lambda}^{(\sigma\kappa)} \ M_{(\sigma\kappa)(\nu\rho)} 
+\frac{1}{2} \, \Gamma_{\mu\rho}^{(\sigma\kappa)} \ M_{(\sigma\kappa)(\nu\lambda)} 
\nonumber \\
&& +\frac{1}{12} \ \left( g_{\lambda\rho} \ \Gamma_{\mu\nu}^{(\sigma\kappa)}
+g_{\nu\rho} \ \Gamma_{\mu\lambda}^{(\sigma\kappa)}+g_{\lambda\nu} \ 
\Gamma_{\mu\rho}^{(\sigma\kappa)}   \right) \ W_{\sigma\kappa}
\label{delphi1}
\end{eqnarray}
Here $W_{\sigma\kappa}$ is defined by
\begin{equation}
W_{\sigma\kappa}= \mbox{tr} \ (e_{(\sigma\kappa)}-{g_{(\sigma\kappa)}}^{\tau} \ e_{\tau}) \, 
\rho. \label{W}
\end{equation}

We now introduce the following field.
\begin{equation}
\Phi_{\nu\lambda\rho} = \phi_{\nu\lambda\rho}-\frac{1}{5} \, (g_{\lambda\rho} \, 
{\phi_{\nu\sigma}}^{\sigma}
+g_{\rho\nu} \, {\phi_{\lambda\sigma}}^{\sigma}+g_{\nu\lambda} \, {\phi_{\rho\sigma}}^{\sigma})
\label{Phi}
\end{equation}
This tensor is traceless for each pair of indices, $g^{\lambda\rho} \, 
\Phi_{\nu\lambda\rho}=0$. In terms of this field, we can derive the following 
eq from (\ref{delphi1}).
\begin{eqnarray}
\hat{\nabla}_{\mu} \, \Phi_{\nu\lambda\rho} &=& 
\frac{1}{2} \, \Gamma_{\mu\nu}^{(\sigma\kappa)} \ M_{(\sigma\kappa)(\lambda\rho)}
-\frac{1}{5} \, g_{\lambda\rho} \, 
g^{\tau\eta} \, \Gamma^{(\sigma\kappa)}_{\mu\tau} \, M_{(\sigma\kappa)(\eta\nu)} \nonumber \\
&& + \  \mbox{cyclic permutations of } \ (\nu,\lambda,\rho)
\end{eqnarray}

This eq determines $\Gamma \, M$. 
\begin{eqnarray}
\Gamma_{\mu\nu}^{(\sigma\kappa)} \, M_{(\sigma\kappa)(\lambda\rho)}
&=& \frac{2}{3} \ \left(\hat{\nabla}_{\mu} \ \Phi_{\nu\lambda\rho}
+\hat{\nabla}_{\nu} \ \Phi_{\mu\lambda\rho}\right)-
\frac{1}{3} \ \left(\hat{\nabla}_{\lambda} \ \Phi_{\mu\nu\rho}
+\hat{\nabla}_{\rho} \ \Phi_{\mu\nu\lambda}\right) \nonumber \\
&&+\frac{2}{9} \, g_{\lambda\rho} \, \hat{\nabla}^{\kappa} \, 
\Phi_{\mu\nu\kappa}+S_{\mu\nu,  \lambda\rho}
\label{GM}
\end{eqnarray}
Here $S_{\mu\nu, (\lambda\rho)}$ is a function which satisfies 
\begin{eqnarray}
&& S_{\mu\nu, \lambda\rho}=S_{\nu\mu, \lambda\rho}=S_{\mu\nu, \rho\lambda}, 
\qquad g^{\lambda\rho} \, S_{\mu\nu, \lambda\rho}=0, \nonumber \\
&& S_{\mu\nu, \lambda\rho}-\frac{2}{5} \, g_{\lambda\rho} \, g^{\tau\eta} \, S_{\mu\tau, 
\eta\nu}+ \text{cyclic permutations of} \ (\nu,\lambda,\rho)=0. 
 \label{coefS}
\end{eqnarray}
By using the matrix $J$ (\ref{J}), (\ref{GM}) gives 
$\Gamma_{\mu\nu}^{(\lambda\rho)}$.
\begin{eqnarray}
\Gamma_{\mu\nu}^{(\lambda\rho)}&=&\frac{1}{6} \ (\hat{\nabla}_{\mu} \ \Phi_{\nu\kappa\sigma}
+\hat{\nabla}_{\nu} \ \Phi_{\mu\kappa\sigma}-\hat{\nabla}_{\kappa} \ \Phi_{\mu\nu\sigma} 
) \ J^{(\kappa\sigma)(\lambda\rho)}+\frac{1}{4} \, S_{\mu\nu, \kappa\sigma} \, 
J^{(\kappa\sigma)(\lambda\rho)} 
\label{Gamma2}
\end{eqnarray}

Finally, by contracting (\ref{delphi1}) with $g^{\lambda\rho}$ one obtains 
another algebraic eq which $S_{\mu\nu, (\lambda\rho)}$ must satisfy.
\begin{eqnarray}
\hat{\nabla}_{\mu} \, {\phi_{\nu\lambda}}^{\lambda} &=& 
\Gamma_{\mu\lambda}^{(\sigma\kappa)} \, M_{(\sigma\kappa)(\nu\rho)} \, 
g^{\lambda\rho}+\frac{5}{12} \, \Gamma_{\mu\nu}^{(\sigma\kappa)} \, 
W_{\sigma\kappa} \nonumber \\
&=&g^{\lambda\rho} \, S_{\mu\lambda, \nu\rho} +\frac{5}{48} \, S_{\mu\nu, \alpha\beta} \, 
J^{(\alpha\beta)(\sigma\kappa)} \, W_{\sigma\kappa}+\frac{5}{9} \, 
\hat{\nabla}^{\lambda} \, \Phi_{\mu\nu\lambda} \nonumber \\
&&+\frac{5}{72} \, (\hat{\nabla}_{\mu} \, \Phi_{\nu\alpha\beta}+\hat{\nabla}_{\nu} 
\, \Phi_{\mu\alpha\beta} - \hat{\nabla}_{\alpha} \, \Phi_{\mu\nu\beta}) 
\, J^{(\alpha\beta)(\sigma\kappa)} \, W_{\sigma\kappa}
\label{deltrace}
\end{eqnarray} 
The eqs (\ref{Gamma2}) and (\ref{GammaGamma1}) determine 
$\Gamma_{\mu\nu}^{\lambda}$.
\begin{eqnarray}
\Gamma_{\mu\nu}^{\lambda}= \hat{\Gamma}_{\mu\nu}^{\lambda}
-\frac{1}{12} \ \left(\hat{\nabla}_{\mu} \ \Phi_{\nu\kappa\sigma}
+\hat{\nabla}_{\nu} \ \Phi_{\mu\kappa\sigma}-\hat{\nabla}_{\kappa} \ 
\Phi_{\mu\nu\sigma} +\frac{3}{2} \, S_{\mu\nu, \kappa\sigma}
\right) \ J^{(\kappa\sigma)(\rho\tau)} 
\ {\phi_{\rho\tau}}^{\lambda}
\label{Gamma1}
\end{eqnarray}
The explicit (if formal) solution for $S_{\mu\nu,\lambda\rho}$ is much involved and is 
presented in appendix D.

\section{Spin connection as a solution to the torsion-free condition}
\hspace{5mm}
In this section the torsion-free condition (\ref{GammaGamma}) will be 
solved for the spin connection $\omega_{\mu}$. 
For this purpose covariant derivative of the vielbein will be introduced.  
This must be done in such a way that the derivative is compatible with 
the metric $g_{\mu\nu}$ and other gauge fields $g_{(\mu\nu)\lambda}$, 
$g_{(\mu\nu)(\lambda\rho)}$.  

\subsection{Covariant derivatives}
\hspace{5mm}
The torsion-free condition (\ref{GammaGamma}) takes the form of 
covariant constancy of $e^a_{\nu}$.
\begin{eqnarray}
D_{\mu} \ e^a_{\nu} &\equiv& \nabla_{\mu} \ e^a_{\nu} + {f^a}_{bc} \ 
\omega^b_{\mu} \
e^c_{\nu}=0, \label{covariantconst1} \\
\nabla_{\mu} \ e^a_{\nu} &\equiv& \partial_{\mu} \ e^a_{\nu}-
\Gamma_{\mu\nu}^{\lambda} \
e^a_{\lambda}-\frac{1}{2} \ \Gamma_{\mu\nu}^{(\lambda\rho)} \ 
e^a_{(\lambda\rho)}
\label{CovariantDerivative1}
\end{eqnarray}
The first eq is defining a full covariant derivative $D_{\mu}$ 
and the second eq is defining $\nabla_{\mu}$. 
Note that $\nabla_{\mu}$ is a new covariant derivative which differs from 
$\hat{\nabla}_{\mu}$ associated with the Christoffel symbol 
$\hat{\Gamma}^{\lambda}_{\mu\nu}$. The last term of 
(\ref{CovariantDerivative1}) can be rewritten as $(-1/4) \, 
\Gamma^{(\lambda\rho)}_{\mu\nu} \, {d^a}_{bc} \, e^b_{\lambda} \, e^c_{\rho}$. This 
definition keeps the covariance under the local frame rotations. 

By using (\ref{delg}) it can be shown that 
the effect of $\nabla_{\mu}$ on $g_{\nu\lambda}$ agrees with that of $
\hat{\nabla}_{\mu}$. 
\begin{eqnarray}
\nabla_{\mu} \, g_{\nu\lambda} &=& \nabla_{\mu} \, (e^a_{\nu} \, e_{a\lambda}) 
 =( \nabla_{\mu} \, e^a_{\nu}) \, e_{a\lambda}
+e^a_{\nu} \, (\nabla_{\mu} \, e_{a\lambda}) \nonumber \\
&=& \partial_{\mu} \, g_{\nu\lambda}
-\Gamma_{\mu\nu}^{\rho} \, g_{\rho\lambda}
-\Gamma_{\mu\lambda}^{\rho} \, g_{\rho\nu}
-\frac{1}{2} \, \Gamma^{(\rho\sigma)}_{\mu\nu} \, g_{(\rho\sigma)\lambda}
-\frac{1}{2} \, \Gamma^{(\rho\sigma)}_{\mu\lambda} \, g_{(\rho\sigma)\nu}=0.
\label{delgmn}
\end{eqnarray}
Note that this can be shown by using only (\ref{GammaGamma1}). The explicit 
expression for $\Gamma^{(\rho\sigma)}_{\mu\nu}$ is unnecessary. 

Let us next start with (\ref{delee}). By interchanging $\nu$ and $\lambda$ and 
adding the result to (\ref{delee}) one obtains an eq 
\begin{eqnarray}
\partial_{\mu} \ \{e_{\nu}, \, e_{\lambda}\} +[ \omega_{\mu}, \, \{ e_{\nu},
 \, e_{\lambda} \}]
&=&\Gamma_{\mu\nu}^{\sigma} \ \{e_{\sigma}, \, e_{\lambda}\} 
+\Gamma_{\mu\lambda}^{\sigma} \ \{e_{\nu}, \, e_{\sigma}\} 
+ \frac{1}{2} \, \Gamma_{\mu\nu}^{(\sigma\kappa)} \, \{e_{(\sigma\kappa)}, \, e_{\lambda}\}
\nonumber \\ &&
+\frac{1}{2} \, \Gamma_{\mu\lambda}^{(\sigma\kappa)}\,\{ e_{\nu}, \, e_{(\sigma\kappa)} \}. 
\label{delee2}
\end{eqnarray}

Let us subtract the terms proportional to the identity matrix from the above 
eq. The trace part was already studied just after (\ref{delee}). 
Owing to (\ref{emunu}) and (\ref{g3}) the result is
\begin{eqnarray}
&& \partial_{\mu} \, e_{(\nu\lambda)}+ [\omega_{\mu}, \, e_{(\nu\lambda)}] 
\nonumber \\
&& = -\frac{1}{3} \, \partial_{\mu} \ (g_{\nu\lambda} \ \rho)-\frac{1}{3} \, 
g_{\nu\lambda} \  
[\omega_{\mu}, \, \rho]+ \Gamma_{\mu\nu}^{\sigma} \, e_{(\sigma\lambda)}
+\frac{1}{3} \, 
\Gamma_{\mu\nu}^{\sigma} \ g_{\sigma\lambda} \ \rho+
\Gamma_{\mu\lambda}^{\sigma} \, e_{(\sigma\nu)}+\frac{1}{3} \, 
\Gamma_{\mu\lambda}^{\sigma} \ g_{\sigma\nu} \ \rho\nonumber \\
&&+
\frac{1}{4} \, \Gamma_{\mu\nu}^{(\sigma\kappa)} \ \{e_{(\sigma\kappa)}, \, 
e_{\lambda} \}
-\frac{1}{3}\, \Gamma_{\mu\nu}^{(\sigma\kappa)} \ g_{(\sigma\kappa)\lambda} 
\ \bm{I}
+\frac{1}{4} \, \Gamma_{\mu\lambda}^{(\sigma\kappa)} \ \{e_{(\sigma\kappa)}, 
\, e_{\nu} \}
-\frac{1}{3}\, \Gamma_{\mu\lambda}^{(\sigma\kappa)} \ g_{(\sigma\kappa)\nu} \ \bm{I}
\label{denl}
\end{eqnarray}
Expansion in terms of the basis $t_a$ yields the eq.
\begin{eqnarray}
&& {f^a}_{bc} \ {\omega^b}_{\mu} \ {e^c}_{(\nu\lambda)} \nonumber \\
&&= - \partial_{\mu} \ {e^a}_{(\nu\lambda)}+\Gamma^{\sigma}_{\mu\nu} \ e^a_{(\sigma\lambda)}+
\Gamma^{\sigma}_{\mu\lambda} \ e^a_{(\sigma\nu)} + \frac{1}{4} \, 
\Gamma_{\mu\nu}^{(\sigma\kappa)} \ {d^a}_{bc} \, e^b_{(\sigma\kappa)} \, e^c_{\lambda}
+\frac{1}{4} \, \Gamma_{\mu\lambda}^{(\sigma\kappa)} \ {d^a}_{bc} \, e^b_{(\sigma\kappa)} 
\, e^c_{\nu} \nonumber \\
&& -\frac{1}{3} \ \rho^a \ \partial_{\mu} \ g_{\nu\lambda}
-\frac{1}{3} \ g_{\nu\lambda} \ \partial_{\mu} \rho^a-\frac{1}{3} \ g_{\nu\lambda} \, 
{f^a}_{bc} \ \omega^b_{\mu} \ \rho^c+\frac{1}{3} \ g_{\lambda\sigma} \ 
\Gamma^{\sigma}_{\mu\nu} \ \rho^a+ \frac{1}{3} \ g_{\nu\sigma} \ 
\Gamma^{\sigma}_{\mu\lambda} \ \rho^a    
\label{fww}
\end{eqnarray}
Here $\rho^a$ is defined by $\rho=\rho^a \ t_a$. 
Contraction of the left-hand side with $g^{\nu\lambda}$ vanishes.  So the same 
must hold for the right-hand side. This leads to a differential eq 
for $\rho$. This eq can be interpreted as the covariant-constancy  
condition for $\rho^a$. 
\begin{eqnarray}
D_{\mu} \ \rho^a & \equiv & \nabla_{\mu} \, \rho^a +{f^a}_{bc} \ 
\omega^b_{\mu} \ \rho^c =0, \label{covariantconst3} \\
\nabla_{\mu} \ \rho^a & \equiv & \partial_{\mu} \, \rho^a
+g^{\nu\lambda} \  \Gamma^{(\kappa\sigma)}_{\mu\nu} \ \left({g_{(\kappa\sigma)}}^{\rho} 
\ e^a_{(\rho\lambda)}
-\frac{1}{2} \, {d^a}_{bc} \, 
e^b_{(\sigma\kappa)} \, e^c_{\lambda} \right) +\frac{1}{3} \, \rho^a 
\, \Gamma_{\mu\nu}^{(\sigma\kappa)} \,  {g_{(\sigma\kappa)}}^{\nu} 
\label{covariantDerivative3}
\end{eqnarray}
The first eq is defining a full covariant derivative $D_{\mu} \, \rho^a$ 
in terms of $\nabla_{\mu} \, \rho^a$ and the second eq is defining 
$\nabla_{\mu} \, \rho^a$. Here (\ref{delg}) is used to relate 
$\Gamma^{\lambda}_{\mu\nu}$ to $\Gamma^{(\sigma\kappa)}_{\mu\nu}$.  The expression 
(\ref{covariantDerivative3}) may look odd, because $\rho^a$ does not have any 
spacetime index. However, this eq can also be derived by starting with 
(\ref{rho}), {\em i.e.,} $\rho^a=\frac{1}{2} \, g^{\mu\nu} \, {d^a}_{bc} \,  
e^b_{\mu} \, e^c_{\nu}$, and using (\ref{CovariantDerivative1}), 
(\ref{delgmn}) and (\ref{emunu}).

The remaining traceless part of (\ref{fww}) can also be interpreted as 
 expressing the property of covariant-constancy of $e^a_{(\nu\lambda)}$.
\begin{eqnarray}
D_{\mu} \ e^a_{(\nu\lambda)} &\equiv& \nabla_{\mu} 
\ e^a_{(\nu\lambda)}+{f^a}_{bc} \ 
\omega^b_{\mu} \ e^c_{(\nu\lambda)}=0, \label{covariantconst2}\\
\nabla_{\mu} \ e^a_{(\nu\lambda)} &\equiv& \partial_{\mu} \  e^a_{(\nu\lambda)}
-\Gamma^{\sigma}_{\mu\nu} \  e^a_{(\sigma\lambda)}-\Gamma^{\sigma}_{\mu\lambda} 
\  
e^a_{(\nu\sigma)}-\frac{1}{4} \ \Gamma_{\mu\nu}^{(\sigma\kappa)} \ {d^a}_{bc} \ 
e^b_{(\sigma\kappa)} \ e^c_{\lambda} -\frac{1}{4} \ 
\Gamma_{\mu\lambda}^{(\sigma\kappa)} 
\ {d^a}_{bc} \ e^b_{(\sigma\kappa)} \ e^c_{\nu} \nonumber \\
&&+ \ g_{\nu\lambda} \ \left(\frac{1}{6} \ \Gamma_{\mu\rho}^{(\sigma\kappa)} 
\ {d^a}_{bc} \ e^b_{(\sigma\kappa)} \ e^c_{\tau} \ g^{\rho\tau}
-\frac{1}{3} \, \Gamma_{\mu\rho}^{(\tau\eta)} \, {g_{(\tau\eta)}}^{\sigma} \, 
g^{\kappa\rho} \,  
e^a_{(\sigma\kappa)}-\frac{1}{9} \ \rho^a \ \Gamma_{\mu\tau}^{(\sigma\kappa)}
 \ {g_{(\sigma\kappa)}}^{\tau}\right) \nonumber \\
&&  +\frac{1}{6} \, \, \rho^a \, \left(\Gamma_{\mu\nu}^{(\sigma\kappa)} \, 
g_{(\sigma\kappa)\lambda}+\Gamma_{\mu\lambda}^{(\sigma\kappa)} \, g_{(\sigma\kappa)\nu}\right)
\label{covariantDerivative2}
\end{eqnarray}

The right-hand side of (\ref{covariantDerivative2}) can be expanded in terms of  
$e^a_{\mu}$ and $e^a_{(\mu\nu)}$ as 
\begin{equation}
\nabla_{\mu} \, e^a_{(\nu\lambda)} = \partial_{\mu} \  e^a_{(\nu\lambda)}-
\Gamma_{\mu, (\nu\lambda)}^{\rho} \, e^a_{\rho} -\frac{1}{2} \, 
\Gamma_{\mu, (\nu\lambda)}^{(\rho\sigma)} 
\, e^a_{(\rho\sigma)}
\end{equation} 
and a new set of connections are defined.
\begin{eqnarray}
\Gamma_{\mu, (\nu\lambda)}^{\rho} &\equiv& 
E_a^{\rho} \, (\partial_{\mu} \, e^a_{(\nu\lambda)}
-\nabla_{\mu} \, e^a_{(\nu\lambda)})=
\frac{1}{4} \, \Gamma^{(\sigma\kappa)}_{\mu\nu} \, {d^a}_{bc} \, E^{\rho}_a \, 
e^b_{(\sigma\kappa)} \, 
e^c_{\lambda}-\frac{1}{12} \, g_{\nu\lambda} \, \Gamma_{\mu\eta}^{(\sigma\kappa)} \, 
{d^a}_{bc} \, E^{\rho}_a \, e^b_{(\sigma\kappa)} \, e^c_{\tau} \, g^{\eta\tau} \nonumber \\
&& +\frac{1}{18} \, g_{\nu\lambda} \, \rho^a \, E_a^{\rho} \, \Gamma^{(\sigma\kappa)}
_{\mu\tau} \, {g_{(\sigma\kappa)}}
^{\tau}-\frac{1}{6} \, \rho^a \, E_a^{\rho} \, \Gamma^{(\sigma\kappa)}_{\mu\nu} \, 
{g_{(\sigma\kappa)\lambda}}+(\nu \leftrightarrow \lambda),  \label{Gamma3} \\
\Gamma_{\mu, (\nu\lambda)}^{(\rho\sigma)} 
&\equiv& E_a^{(\rho\sigma)} \, (\partial_{\mu} \, e^a_{(\nu\lambda)}-
\nabla_{\mu} \, e^a_{(\nu\lambda)})=\Gamma^{\rho}_{\mu\nu} 
\, \delta^{\sigma}_{\lambda}+\Gamma^{\sigma}_{\mu\nu} 
\, \delta^{\rho}_{\lambda}-\frac{2}{3} \, \Gamma_{\mu\nu}^{\kappa} \, g_{\kappa\lambda} \, 
g^{\rho\sigma} \nonumber \\ &&+\frac{1}{4} \, \Gamma_{\mu\nu}^{(\kappa\eta)} \, {d^a}_{bc} \,
 E^{(\rho\sigma)}_a \, e^b_{(\kappa\eta)} \, e^c_{\lambda} -\frac{1}{12} \, 
g_{\nu\lambda} \, \Gamma_{\mu\alpha}^{(\beta\kappa)}\, {d^a}_{bc} \, E^{(\rho\sigma)}_a 
\, e^b_{(\beta\kappa)} \, e^c_{\tau} \, g^{\alpha\tau}  \nonumber \\ && 
+\frac{1}{6} \, g_{\nu\lambda} \, \Gamma_{\mu\alpha}^{(\tau\eta)} \, {g_{(\tau\eta)}}^{\rho}
g^{\sigma\alpha} 
+\frac{1}{6} \, g_{\nu\lambda} \, \Gamma^{(\tau\eta)}_{\mu\alpha} \, 
{g_{(\tau\eta)}}^{\sigma} \, g^{\rho\alpha}
-\frac{1}{9} \, g_{\nu\lambda} \, 
\Gamma^{(\tau\eta)}_{\mu\alpha} \, {g_{(\eta\tau)}}^{\alpha} \, g^{\rho\sigma} \nonumber \\ &&+
\frac{1}{18} \, g_{\nu\lambda} \, \rho^a \, E_a^{(\rho\sigma)} \, 
\Gamma_{\mu\tau}^{(\alpha\beta)} \, {g_{(\alpha\beta)}}^{\tau} -\frac{1}{6} 
\, \rho^a \, E_a^{(\rho\sigma)} \, \Gamma_{\mu\nu}^{(\alpha\beta)} \, g_{(\alpha\beta)\lambda}+
(\nu \leftrightarrow \lambda)  \label{Gamma4}
\end{eqnarray}
In the above eqs, $\rho \, E$'s are given by 
\begin{eqnarray}
\rho^a \, E_a^{\mu} &=& {\phi^{\mu\lambda}}_{\lambda}+\frac{1}{4} \, 
{g^{\mu}}_{(\tau\eta)} 
\, J^{(\tau\eta)(\sigma\kappa)} \, \left(
{\phi_{\sigma\kappa}}^{\nu} \, {\phi_{\nu\lambda}}^{\lambda}-
\phi_{(\sigma\kappa)(\lambda\rho)}\, g^{\lambda\rho} \right), 
\label{rhoE}  \\
\rho^a \, E_a^{(\mu\nu)} &=& \frac{1}{2} \, J^{(\mu\nu)(\lambda\rho)} \, 
(\phi_{(\lambda\rho)(\kappa\sigma)}-\phi_{\lambda\rho\tau} \, 
{\phi^{\tau}}_{\kappa\sigma}) \, g^{\kappa\sigma} \label{rhoEE}
\end{eqnarray}
To prove these eqs, $\rho^a=\frac{1}{2} \, g^{\mu\nu} \, {d^a}_{bc} \,  
e^b_{\mu} \, e^c_{\nu}$ and (\ref{ta}), (\ref{Gumun})-(\ref{G4}) and (\ref{GI}) 
must be used. Furthermore, ${d^a}_{bc}$ can be replaced by (\ref{dabc}) 
in appendix C.  In this way the right-hand sides of (\ref{Gamma3}), 
(\ref{Gamma4}) could be expressed solely in terms of the metric-like fields.  
This will, however, not be attempted in this paper.

Alternatively, the covariant derivative of $e^a_{(\nu\lambda)}$ 
(\ref{covariantDerivative2}) can be derived from that for 
$e^a_{\nu}$ (\ref{CovariantDerivative1}) by using $e^a_{(\nu\lambda)}=
\frac{1}{2} \, {d^a}_{bc} \, e^b_{\nu} \, e^c_{\lambda} -\frac{1}{3} 
\, g_{\nu\lambda} \, \rho^a$. Therefore, we can also write
\begin{equation}
\nabla_{\mu} \, e^a_{\nu\lambda} = \frac{1}{2} \, {d^a}_{bc} \, (e^b_{\nu} \, 
\nabla_{\mu} \, e^c_{\lambda}+e^b_{\lambda} \, \nabla_{\mu} \, e^c_{\nu})
-\frac{1}{3} \, g_{\nu\lambda} \, \nabla_{\mu} \, \rho^a.
\end{equation}
It can be shown that this eq agrees with (\ref{covariantDerivative2}). 
For $\rho^a$, 
we can write 
\begin{equation}
\nabla_{\mu} \, \rho^a = {d^a}_{bc} \, g^{\nu\lambda} \, e^b_{\nu} \, \nabla_{\mu} 
\, e^c_{\lambda}.
\end{equation}
 Because $D_{\mu} \, {d^a}_{bc}=0$\footnote{This can 
be proved 
by using Jacobi's identity containing $d_{abc}$ and $f_{abc}$.}  and 
$D_{\mu} \, g^{\lambda\rho}=-g^{\lambda\kappa} \, (\nabla_{\mu} \, 
g_{\kappa\sigma}) \, 
g^{\sigma\rho}=0$, the covariant constancy of $\rho^a$, $D_{\mu} \, \rho^a=0$, 
is a result of that of $e^a_{\mu}$.

\subsection{Covariant derivatives for general tensors}
\hspace{5mm}
Let $v^a$ be an vector in the local Lorentz frame.    
This vector can be expanded in terms of the vielbeins in either way 
\begin{equation}
v^a=v^{\mu} \, e^a_{\mu}+ \frac{1}{2} \, v^{(\mu\nu)} \, e^a_{(\mu\nu)},
\label{contravariantv}
\end{equation}
or
\begin{equation}
v^a=v_{\mu} \, E^{a\mu}+ \frac{1}{2} \, v_{(\mu\nu)} \, E^{a(\mu\nu)}.
\label{covariantv}
\end{equation}
(\ref{contravariantv}) defines the contravariant components and 
(\ref{covariantv}) the covariant ones. We choose $v^{(\mu\nu)}$ and 
$v_{(\mu\nu)}$ such that they satisfy traceless conditions,  
$v^{(\mu\nu)} \, g_{\mu\nu}=0$ and $v_{(\mu\nu)} \, g^{\mu\nu}=0$, respectively. 
Similar decomposition can be performed 
for tensors with an arbitrary number of local frame indices $a$'s. 
A general rule for decomposition is that for each local frame index $a$ 
there corresponds a pair of spacetime indices, $\mu$ and $(\mu\nu)$. The 
parenthesis in $(\mu\nu)$ implies the symmetry under interchange of $\mu$ 
and $\nu$. To avoid confusion, the indices of $v^{\mu}$, $v^{(\mu\nu)}$ will 
not be raised or lowered by $g_{\mu\nu}$.\footnote{Instead, this will be 
done in terms of $G_{MN}$ defined in appendix E.} 

In this subsection we will restrict the discussion to the case where the 
covariant 
derivative of $v^a$ is given by $\nabla_{\mu} \, v^a=\partial_{\mu} \, v^a$.\footnote{
 The rule of the covariant derivatives in this subsection cannot be 
used for $\rho^{\mu}$ and $\rho^{(\mu\nu)}$ defined by $\rho^a=\rho^{\mu} \, 
e^a_{\mu}+(1/2) \, \rho^{(\mu\nu)} \, e^a_{(\mu\nu)}$, because $\nabla_{\mu} \, \rho^a$ 
is not given by $\partial_{\mu} \, \rho^a$, but by (\ref{covariantDerivative3}).}
By use of (\ref{contravariantv}) this leads to the identity.
\begin{eqnarray}
& & (\nabla_{\mu} \, v^{\lambda}) \, e^a_{\lambda} + v^{\lambda} \, 
(\nabla_{\mu} \, 
e^a_{\lambda})+ \frac{1}{2} \, (\nabla_{\mu} \, v^{(\lambda\rho)}) \, 
e^a_{(\lambda\rho)}+
\frac{1}{2} \, v^{(\lambda\rho)} \, (\nabla_{\mu} \,  e^a_{(\lambda\rho)}) 
\nonumber \\
 &=& \partial_{\mu} \, v^{\lambda} \, e^a_{\lambda} + v^{\lambda} \, 
(\partial_{\mu} 
\, e^a_{\lambda})+ \frac{1}{2} \, (\partial_{\mu} \, v^{(\lambda\rho)}) \, 
e^a_{(\lambda\rho)}+\frac{1}{2} \, v^{(\lambda\rho)} \, (\partial_{\mu} \,  
e^a_{(\lambda\rho)})
\end{eqnarray}
By comparing the coefficients of $e^a_{\nu}$ and $e^a_{(\nu\lambda)}$ on both sides 
one obtains the definitions of the covariant derivatives of $v^{\mu}$ and 
$v^{(\mu\nu)}$. 
\begin{equation}
\nabla_{\mu} \, v^{\nu} = \partial_{\mu} \, v^{\nu}+\Gamma_{\mu\lambda}^{\nu} \, 
v^{\lambda}+\frac{1}{2} \, \Gamma_{\mu,(\lambda\rho)}^{\nu} \, v^{(\lambda\rho)}, 
\label{covrule1}
\end{equation}
\begin{eqnarray}
\nabla_{\mu} \, v^{(\nu\lambda)} & = & \partial_{\mu} \,  v^{(\nu\lambda)} 
+\Gamma_{\mu\rho}^{(\nu\lambda)} \, v^{\rho}
+\frac{1}{2} \, \Gamma_{\mu,(\rho\sigma)}^{(\nu\lambda)} \, v^{(\rho\sigma)}
\label{covrule2}
\end{eqnarray}

By using the expansion (\ref{covariantv}) and taking the similar steps, 
the covariant derivatives of the covariant components are also obtained.
\begin{equation}
\nabla_{\mu} \, v_{\nu} = \partial_{\mu} \, v_{\nu} -\Gamma^{\lambda}_{\mu\nu} 
\, v_{\lambda}
-\frac{1}{2} \, \Gamma_{\mu\nu}^{(\lambda\rho)} \, v_{(\lambda\rho)},
\label{delmvn}
\end{equation}
\begin{eqnarray}
\nabla_{\mu} \, v_{(\nu\lambda)} &=& \partial_{\mu} \, v_{(\nu\lambda)} 
-\Gamma_{\mu,(\nu\lambda)}^{\rho} \, v_{\rho} -\frac{1}{2} \, 
\Gamma_{\mu,(\nu\lambda)}^{(\rho\sigma)} \, v_{(\rho\sigma)} 
\label{covrule4}
\end{eqnarray}

An extension to the covariant derivatives for the tensors with more indices 
will be straightforward and clear. The rule is the same as in the spin-2  
gravity. For instance, by using (\ref{delmvn}) $\nabla_{\mu} \, 
g_{\nu\lambda}$ is calculated as 
\begin{eqnarray}
\nabla_{\mu} \, g_{\nu\lambda} =\partial_{\mu} \, g_{\nu\lambda}
-\Gamma_{\mu\nu}^{\rho} \, g_{\rho\lambda}-\frac{1}{2} \, 
\Gamma_{\mu\nu}^{(\rho\sigma)} \, g_{(\rho\sigma)\lambda}
-\Gamma_{\mu\lambda}^{\rho} \, g_{\nu\rho}-\frac{1}{2} \, 
\Gamma_{\mu\lambda}^{(\rho\sigma)} \, g_{\nu(\rho\sigma)},  \label{delgmn2}
\end{eqnarray}
because $g_{(\rho\sigma)\lambda}$ is paired with $g_{\rho\lambda}$. 
This vanishes as in (\ref{delgmn}). This is an important property of 
the metric tensor $G_{MN}$ introduced in appendix E, which can be used to 
raise and lower the indices, $\mu$ and $(\mu\nu)$. One can also explicitly 
check eq $\nabla_{\mu} \, g_{\nu(\lambda\rho)}=0$ by using (\ref{delphi}), 
(\ref{Gamma3}) and (\ref{Gamma4}). 

As we have seen, the indices of tensors  have the structure, $\mu, (\mu\nu)$:  
general tensors can be written as ${T_{MN...}}^{L ..}$, where $M$, $N$, $L$ 
take two types of indices, $\mu$, $(\mu\nu)$.

The covariant derivatives of the inverse vielbeins $E$ are special examples of 
the above ones. 
They can be obtained by using the definitions (\ref{covrule1})
-(\ref{covrule2}). 
\begin{eqnarray}
\nabla_{\mu} \, E_a^{\nu} &=& \partial_{\mu} \, E_a^{\nu}+\Gamma_{\mu\lambda}^{\nu} \, 
E_a^{\lambda}+\frac{1}{2} \, \Gamma_{\mu, (\lambda\rho)}^{\nu} \, E_a^{(\lambda\rho)}, 
\label{dE1} \\
\nabla_{\mu} \, E_a^{(\nu\lambda)} &= & \partial_{\mu} \, E_a^{(\nu\lambda)}+
\Gamma_{\mu\rho}^{(\nu\lambda)} \, E_a^{\rho}+\frac{1}{2} \, 
\Gamma_{\mu,(\rho\sigma)}^{(\nu\lambda)} \, E_a^{(\rho\sigma)} \label{dE2}
\end{eqnarray}
Although there exist $g_{\mu\nu}$ and $g^{\lambda\rho}$ on the right-hand sides of 
(\ref{Ee}), the above eqs are compatible with (\ref{Ee}), because 
one has $\nabla_{\mu} \, (e^a_N \, E_a^M)=(\nabla_{\mu} \, e^a_N) \, 
E_a^M+e^a_N \, (\nabla_{\mu} \, E_a^M)=\partial_{\mu} \, (e^a_N \, E^M_a)$ 
and $\nabla_{\mu} \, \delta_M^N= \partial_{\mu} \, \delta_M^N-\Gamma_{\mu M}^K \, 
\delta^N_K+\Gamma_{\mu K}^N \, \delta_M^K =\partial_{\mu} \, \delta_M^N$, where 
$\delta_M^N$ 
is defined around (\ref{deltaAC}) in appendix E.

\subsection{Spin connection}
\hspace{5mm}
It is now easy to solve for $\omega^a_{\mu}$. 
Multiplication of (\ref{covariantconst1}) and (\ref{covariantconst2}) by 
$E_d^{\nu}$ and $\frac{1}{2} \, E_d^{(\nu\lambda)}$, respectively, and adding the 
two we obtain the spin connection in the adjoint representation.
\begin{equation}
\omega^a_{\mu \ c} \equiv {f^a}_{bc} \, \omega^b_{\mu}=-E_c^{\nu} \, 
\nabla_{\mu} 
\, e^a_{\nu}-\frac{1}{2} \, E_c^{(\nu\lambda)} \, \nabla_{\mu} \, e^a_{(\nu\lambda)}
\label{omegaac}
\end{equation}
Then use of (\ref{hff}) yield 
\begin{equation}
\omega_{\mu}^a=\omega_{\mu}^a(e) \equiv \frac{1}{12} \ {f^{ab}}_{c} \ 
E_b^{\lambda} \ \nabla_{\mu} \ e^c_{\lambda}+\frac{1}{24} \ {f^{ab}}_{c} \ 
E_b^{(\lambda\rho)} \ \nabla_{\mu} \ e^c_{(\lambda\rho)}.
\label{spinconnection}
\end{equation}

Into the full covariant derivatives for $e^a_{\mu}$, $e^a_{(\mu\nu)}$ and 
$\rho^a$ introduced above the spin connection $\omega^a_{\mu}(e)$ 
(\ref{spinconnection}) is to be substituted. 

\subsection{The second-order action}
\hspace{5mm}
Now we substitute the solution (\ref{spinconnection}) into the action 
(\ref{CS}) and obtain the second-order action. 
\begin{eqnarray}
S_{\text{ 2nd order}} = \frac{k}{\pi} \ \int \mbox{tr} \ e \wedge 
\left(d\omega(e)+\omega(e) 
\wedge \omega(e)+\frac{1}{3} \  e \wedge e \right)
\label{CSsub}
\end{eqnarray} 
Here we do not consider the boundary terms.  
To derive classical eqs of motion from this action, 
$\omega^a_{\mu}(e)$ is to be varied as a functional of $e^a_{\mu}$. 

In sec. 7 we will derive the generalized Riemann curvature tensor 
${R^M}_{N\nu\lambda}$, (\ref{generalised Riemann}). The above action in the 
second-order formalism can be reexpressed in terms of this tensor.  
\begin{eqnarray}
S_{\text{ 2nd order}}& =& \frac{k}{12\pi} \ \int \, d^3x \, \left\{ -
\epsilon^{\mu\nu\lambda} \, 
{(f^a}_{bc} \,  e^c_{\mu} \, e^b_M \, E^N_a) \, {R^M}_{N\nu\lambda}+4 \, 
\epsilon^{\mu\nu\lambda} \, {f^a}_{bc} \, e^a_{\mu} \, e^b_{\nu} \, e^c_{\lambda} 
\right\}  \nonumber \\
&=&  \frac{k}{12\pi} \ \int \, d^3x \, \left\{ -\epsilon^{\mu\nu\lambda} \, 
{f^a}_{bc} \, e^c_{\mu} \, \left(e^b_{\rho} \, E^{\sigma}_a \, 
{R^{\rho}}_{\sigma\nu\lambda}+\frac{1}{2} \, e^b_{(\rho\zeta)} 
\, E^{\sigma}_a \, {R^{(\rho\zeta)}}_{\sigma\nu\lambda} \right. \right.  
\nonumber \\ &&
\left.\left. +\frac{1}{2} \, e^b_{\rho} \, E^{(\sigma\kappa)}_a \, 
{R^{\rho}}_{(\sigma\kappa)\nu\lambda}+\frac{1}{4} \, e^b_{(\rho\zeta)} \, 
E^{(\sigma\kappa)}_a \, {R^{(\rho\zeta)}}_{(\sigma\kappa)\nu\lambda}\right)+4 \, 
\epsilon^{\mu\nu\lambda} \, {f}_{abc} \, e^a_{\mu} \, e^b_{\nu} \, e^c_{\lambda} 
\right\} \nonumber \\ &&
 \label{CSsecond}
\end{eqnarray} 
See sec.7 for the derivation. 
Here, $ {f^a}_{bc} \,  e^c_{\mu} \, e^b_M \, E^N_a $ is a metric-like quantity 
which is insensitive to the local frame rotations, while 
$\epsilon^{\mu\nu\lambda} \, {f}_{abc} \, e^a_{\mu} \, e^b_{\nu} \, e^c_{\lambda}$ is 
the generalized cosmological term. So the action integral is expressed in 
terms of the connections $\Gamma_{\mu M}^N$ and the vielbeins. 
The remaining problem is to represent this action only in terms of the 
metric-like fields. This is not attempted in this paper. 

Under the local frame rotations the vielbein transforms as (\ref{elocal}), and  
it is easy to show that the spin connection $\omega^a_{\mu}(e)$ 
transforms as (\ref{local}). Next, under the generalized diffeomorphisms, 
the vielbein transforms as (\ref{e}). This can be rewritten as 
\begin{eqnarray}
\delta_1 \, e^a_{\mu} = D_{\mu} \, \Lambda_-^a = D_{\mu} \, (\tilde{\xi}^M \, e^a_M)
=D_{\mu} \, (\tilde{\xi}^{\nu} \, e^a_{\nu}+\frac{1}{2} \, \tilde{\xi}^{(\nu\lambda)} 
\, e^a_{(\nu\lambda)})
\end{eqnarray}
Here $\tilde{\xi}^M$ represents for $\tilde{\xi}^{\nu}$ and 
$\tilde{\xi}^{(\nu\lambda)}$, and they  are defined by 
\begin{equation}
\tilde{\xi}^M= \Lambda_-^a \, E_a^M =G^{MN} \, \xi_N . 
\end{equation}
These are the local parameters of the generalised diffeomorphisms. 
For the definition of $G^{MN}$ and the notation $M=\mu, (\mu\nu)$, 
see appendix E. The tilde in the notation $\tilde{\xi}^M$ means that the metric 
tensor $G^{MN}$ is used to raise the indices. Because $e^a_{\mu}$ and 
$e^a_{(\mu\nu)}$ are covariantly constant, we have
\begin{equation}
\delta_1 \, e^a_{\mu} = e^a_{\nu} \, \nabla_{\mu} \, \tilde{\xi}^{\nu}
+\frac{1}{2} 
\, e^a_{(\nu\lambda)} \, \nabla_{\mu} \, \tilde{\xi}^{(\nu\lambda)}. 
\label{transemu}
\end{equation}
This does not look like a transformation rule for a covariant vector. However, 
one can perform, additionally, a local frame rotation $\delta_2 \, e^a_{\mu}=
{f^a}_{bc} \, e^b_{\mu} \, \Lambda_+^c$ with $\Lambda_+^a=\omega^a_{\nu}(e)
\, \tilde{\xi}^{\nu}$. The combined transformation is 
\begin{equation}
\delta_{\text{diffeo}} \, e^a_{\mu}= \delta_1 \, e^a_{\mu}+ \delta_2 \, e^a_{\mu}=
e^a_{\nu} \, \nabla_{\mu} \, \tilde{\xi}^{\nu}+\tilde{\xi}^{\nu} \, \nabla_{\nu} \, 
e^a_{\mu}+\frac{1}{2} \, e^a_{(\nu\lambda)} \, \nabla_{\mu} \, 
\tilde{\xi}^{(\nu\lambda)}. 
\label{ediffeo}
\end{equation}
Here the torsion-free condition is used to replace ${f^a}_{bc} \, 
\omega^b_{\nu}(e) \, e^c_{\mu}$ by $-\nabla_{\nu} \, e^a_{\mu}$. This is the 
generalization of the diffeomorphism for the vielbein to the spin-3 gravity 
theory. If $\tilde{\xi}^{(\nu\lambda)}=0$, the above eq agrees with the 
transformation rule of a covariant vector. 

The new transformation rule of $\omega^a_{\mu}(e)$ can also be obtained by 
combining the generalized diffeomorphism $\delta_1 \, \omega^a_{\mu} (e)={f^a}_{bc} \, e^b_{\mu} \, \Lambda_-^c$ with the local frame rotation 
$\delta_2 \, \omega^a_{\mu}(e)=\partial_{\mu} \, \Lambda^a_++{f^a}_{bc} \, \omega^b_{\mu}(e) \, \Lambda^c_+$.
\begin{eqnarray}
\delta_{\text{diffeo}} \, \omega^a_{\mu}(e) &=& \delta_1 \, \omega^a_{\mu}(e)
+ \delta_2 \, \omega^a_{\mu}(e) \nonumber \\
&=&\omega^a_{\nu}(e) \, \nabla_{\mu} \, 
\tilde{\xi}^{\nu}+\tilde{\xi}^{\nu} \, \nabla_{\nu} \, \omega^a_{\mu}(e)
-\frac{1}{2} \, \tilde{\xi}^{(\lambda\rho)} \, \Gamma^{\nu}_{\mu, (\lambda\rho)} \, 
\omega^a_{\nu}(e)+ \frac{1}{2} \, \tilde{\xi}^{(\nu\lambda)} \, {f^a}_{bc} \, e^b_{\mu} \, 
e^c_{(\nu\lambda)}. \nonumber \\ &&
\label{delomega}
\end{eqnarray}
Here in computing $\nabla_{\nu} \, \omega^a_{\mu}$ we must set 
$\omega^a_{\mu\nu}=0$, since this extra component does not exist.\footnote{For 
the discussion of the  extra components $\omega^a_{(\mu\nu)}$ and 
$\nabla_{(\mu\nu)}$, see the comment at the end of sec.6.} On the right-hand 
side of (\ref{delomega}) a term $\tilde{\xi}^{\nu} \, ({R^a}_{\mu\nu} 
+{f^a}_{bc} \, e^b_{\mu} \, e^c_{\nu})$ is actually present, but this is 
dropped here, since this term vanishes when the eq of motion is used. 
In the spin-2 case the transformation rule obtained by dropping the eq of 
motion term coincides with the diffeomorphism of the spin connection in the 
second-order formalism.\cite{Carlip} We also expect that (\ref{delomega}) 
without the eq of motion term is similarly true in the second-order formalism, 
since the spin-3 gravity contains the spin-2 gravity.  
Anyway, the transformation rule of $\omega_{\mu}^a(e)$ must be checked 
explicitly by using the definition (\ref{spinconnection}) and the expressions 
for $\Gamma$'s. We will not attempt to directly prove (\ref{delomega}) in 
this paper. Whatever the transformation rule of $\omega^a_{\mu}(e)$ is, it 
is possible to show that the action integral in the second-order formalism, 
(\ref{CSsub}), is invariant under (\ref{ediffeo}) in the bulk. This is because when 
computing the variation of the action integral, $\delta \, \omega^a_{\mu}(e)$ 
is multiplied by the torsion (\ref{torsionless}), which vanishes.


\section{Generalized diffeomorphism for $g_{\mu\nu}$ and $\phi_{\mu\nu\lambda}$}
\hspace{5mm}
The CS theory (\ref{CS}) has generalized diffeomorphism invariance 
(\ref{diffeo}). In this section the transformation rules of the gauge fields 
$g_{\mu\nu}$ and $\phi_{\mu\nu\lambda}$ will be derived.

\subsection{Transformation of $g_{\mu\nu}$}
\hspace{5mm}
Let us first consider the metric field (\ref{metric}). This transforms as
\begin{eqnarray}
\delta g_{\mu\nu} &=& \frac{1}{2} \, \mbox{tr} \, \left(\partial_{\mu} \, 
\Lambda+[\omega_{\mu}, \, 
\Lambda]\right) \, e_{\nu} + (\mu \leftrightarrow \nu) \nonumber \\
& =& \frac{1}{2} \ \partial_{\mu} \, \mbox{tr} \, \Lambda \, e_{\nu}- 
\frac{1}{2} \, \mbox{tr} \, \Lambda \, (
\partial_{\mu} \, e_{\nu}+[\omega_{\mu}, \, e_{\nu}]) + (\mu \leftrightarrow 
\nu) \label{xixzeta}
\end{eqnarray}
For simplicity of notation, $\Lambda_-$ in (\ref{diffeo}) is here denoted as 
$\Lambda$. Now, two variation functions are introduced.
\begin{eqnarray}
&& \xi_{\mu} = \frac{1}{2} \,\mbox{tr} \, \Lambda \, e_{\mu}, \label{xi}\\
&& \zeta_{(\mu\nu)} + g_{(\mu\nu)\lambda} \, \xi^{\lambda} = \xi_{(\mu\nu)}= 
\frac{1}{2} \, \mbox{tr} \, \Lambda \, e_{(\mu\nu)} \label{zeta} 
\end{eqnarray}
To some extent one can regard $\xi_{\mu}$ and $\zeta_{(\mu\nu)}$ as the 
coordinate transformation parameter and the spin-3 gauge parameter, 
respectively. There are, however, some differences. Such differences can be 
observed at (\ref{dimension3}) and (\ref{del_phi}) below. This is because 
in a spin-3 gauge theory, diffeomorphisms and spin-3 gauge transformations 
are mixed and not easily disentangled. In eq (\ref{zeta}) an extra parameter 
$\xi_{(\mu\nu)}$ is also introduced.  With the help of (\ref{GammaGamma}), 
$\delta g_{\mu\nu}$ can be put into the form. 
\begin{eqnarray}
\delta g_{\mu\nu} &=& \partial_{\mu} \, \xi_{\nu} -\Gamma_{\mu\nu}^{\lambda}
 \, \xi_{\lambda} -\frac{1}{2} \, \Gamma_{\mu\nu}^{(\lambda\rho)} \, 
(\zeta_{(\lambda\rho)}+g_{(\lambda\rho)\sigma} \, \xi^{\sigma})
 + (\mu \leftrightarrow \nu) \nonumber \\
&=& \hat{\nabla}_{\mu} \, \xi_{\nu}+ \hat{\nabla}_{\nu} \, \xi_{\mu}
-\Gamma_{\mu\nu}^{(\lambda\rho)} \, \zeta_{(\lambda\rho)}
\label{del_g}
\end{eqnarray}
Here (\ref{GammaGamma1}) is used and $\hat{\nabla}_{\mu}$ is the ordinary 
covariant derivative that uses Christoffel symbol  
$\hat{\Gamma}_{\mu\nu}^{\lambda}$. Therefore those parts which depend on 
$\xi_{\mu}$ are the ordinary diffeomorphism. The remaining term, which depends 
on $\zeta_{(\mu\nu)}$, is the new spin-3 gauge transformation. This term  
depends non-linearly on gauge fields such as $g_{(\mu\nu)(\lambda\rho)}$, via 
$J^{(\mu\nu)(\kappa\sigma)}$, since $\Gamma_{\mu\nu}^{(\lambda\rho)}$ does also.

Interestingly, this infinitesimal transformation can also be written as
\begin{equation}
\delta \, g_{\mu\nu} = \nabla_{\mu} \, \xi_{\nu}+\nabla_{\nu} \, \xi_{\mu}
\label{gdiffeo}
\end{equation}
by adopting the covariant derivative (\ref{delmvn}) introduced in sec.4. 
Here $\xi_{(\mu\nu)}$ is used as the partner of $\xi_{\mu}$ in computing 
the derivative. The above result can also be derived by using $g_{\mu\nu}=
e^a_{\mu} \, e_{a\nu}$ and the transformation rule of $e^a_{\mu}$. 

\subsection{Transformation of $\phi_{\mu\nu\lambda}$}
\hspace{5mm}
Let us next turn to the spin-3 gauge field $\phi_{\mu\nu\lambda}$. 
In this case the variation can be rewritten as follows. 
\begin{eqnarray}
\delta \, \phi_{\mu\nu\lambda} &=& \frac{1}{4} \, \mbox{tr} \, 
\left(\partial_{\mu} \, \Lambda+[\omega_{\mu}, \, 
\Lambda]\right) \, \{e_{\nu}, \, e_{\lambda} \} \qquad +    
(\mbox{2 cyclic permutations of} \ \  
\mu, \nu, \lambda) \nonumber \\
&=& \frac{1}{2} \, \partial_{\mu} \, \mbox{tr} \, \Lambda \, 
e_{(\nu\lambda)}+\frac{1}{6} \, \partial_{\mu} \, (
g_{\nu\lambda} \, \mbox{tr} \, \rho \Lambda) 
-\frac{1}{2} \, \Gamma_{\mu\nu}^{\rho} \, \mbox{tr} \, \Lambda \,  
e_{\rho\lambda}
-\frac{1}{6} \, \Gamma_{\mu\nu}^{\rho} \, g_{\rho\lambda} \, \mbox{tr} 
\, \Lambda \,  \rho \nonumber \\
&& -\frac{1}{2} \, \Gamma_{\mu\lambda}^{\rho} \, \mbox{tr} \, \Lambda \, 
e_{(\rho\nu)}
-\frac{1}{6} \, \Gamma_{\mu\lambda}^{\rho} \, g_{\rho\nu} \, \mbox{tr} \, 
\Lambda \,  \rho 
-\frac{1}{8} \, \Gamma_{\mu\nu}^{(\rho\sigma)} \, \mbox{tr} \, \Lambda 
\, \{ e_{(\rho\sigma)}, e_{\lambda} \} \nonumber \\ &&
-\frac{1}{8} \, \Gamma_{\mu\lambda}^{(\rho\sigma)} \, \mbox{tr} \, \Lambda 
\, \{ e_{(\rho\sigma)}, e_{\nu} \}
+ (\mbox{permutations})
\label{variationphi}
\end{eqnarray}
Here (\ref{emunu}) and (\ref{delee2}) are used. 

In this expression $\mbox{tr} \, \Lambda \, e_{(\nu\lambda)}$ is rewritten 
by means of (\ref{zeta}). To compute other terms involving $\Lambda$, 
(\ref{xi}) and (\ref{zeta}) must be solved for $\Lambda=t_a \, \Lambda^a$.
Multiplying  $\xi_{\mu}=\Lambda_a \, e^a_{\mu}$ and $\zeta_{(\mu\nu)}
+g_{(\mu\nu)\lambda} \, \xi^{\lambda}
= \Lambda_a \, e^a_{(\mu\nu)}$ by $E_b^{\mu}$ and $\frac{1}{2} \, 
E_b^{(\mu\nu)}$, respectively, 
and adding the two, the following formula is derived.
\begin{equation}
\Lambda_a= \xi_{\mu} \, E_a^{\mu}+ \frac{1}{2} \, \left(\zeta_{(\mu\nu)}
+g_{(\mu\nu)\lambda} \, \xi^{\lambda}\right) \, E_a^{(\mu\nu)}
=\xi_{\mu} \, E_a^{\mu}+ \frac{1}{2} \, \xi_{(\mu\nu)}  \, E_a^{(\mu\nu)}
\label{Lambda}
\end{equation} 
The trace $\mbox{tr} \, \Lambda \rho$ is then reexpressed as follows.
\begin{eqnarray}
&&\mbox{tr} \, \Lambda \rho = 2\, \Lambda_a \, \rho^a=
\zeta_{(\rho\sigma)} \, E_a^{(\rho\sigma)} \, \rho^a+ 2 \, \xi^{\sigma} \, 
(g_{\sigma\lambda} \,  E_a^{\lambda}+ \frac{1}{2} \, 
g_{(\rho\kappa)\sigma}\, E_a^{(\rho\kappa)}) \, \rho^a \nonumber \\
&&= \zeta_{(\rho\sigma)} \, E_a^{(\rho\sigma)} \, \rho^a+
 \xi^{\sigma} \, \rho^a \, \mbox{tr} \, e_{\sigma} \, t_a   
=\zeta_{(\rho\sigma)} \, E_a^{(\rho\sigma)} \, \rho^a+  \xi^{\sigma} \, 
\mbox{tr} \ e_{\sigma} \, \rho
\end{eqnarray}
Here formula (\ref{ta}) for $t_a$ given in appendix C is used. 
The term $\mbox{tr} \, \Lambda \, \{e_{(\rho\sigma)}, \, 
e_{\lambda}\}$ in (\ref{variationphi}) is similarly computed as follows.
\begin{eqnarray}
\mbox{tr} \, \Lambda \, \{e_{(\rho\sigma)}, \, e_{\lambda}\} &=& 
2 \, \xi_{\mu} \, {d^a}_{bc}
\, E_a^{\mu} \, e^b_{(\rho\sigma)} \, e^c_{\lambda} \, + \, \xi^{\alpha} 
\, g_{(\mu\nu)\alpha}
\, {d^a}_{bc} \, e^b_{(\rho\sigma)} \, e^c_{\lambda} \, E_a^{(\mu\nu)} 
\nonumber \\
&&+ \zeta_{(\mu\nu)} \, 
{d^a}_{bc} \, E_a^{(\mu\nu)} \, e^b_{(\rho\sigma)} \, e^c_{\lambda}
\end{eqnarray}

The variation $\delta \phi_{\mu\nu\lambda}$ will be decomposed into 
$\delta_{\xi} \, \phi_{\mu\nu\lambda}+ \delta_{\zeta} \, 
\phi_{\mu\nu\lambda}$.  Let us first concentrate on those terms which depend 
on $\xi$. After some calculation one obtains 
\begin{eqnarray}
\delta_{\xi} \, \phi_{\mu\nu\lambda} &=& \hat{\nabla}_{\mu} \, 
(\xi^{\sigma} \, \phi_{
\nu\lambda\sigma} ) 
-\frac{1}{2} \, \xi^{\alpha} \, \Gamma_{\mu\nu}^{(\sigma\kappa)} \, 
M_{(\sigma\kappa)(\lambda\alpha)} -\frac{1}{2} \, \xi^{\alpha} \, 
\Gamma_{\mu\lambda}^{(\sigma\kappa)} \, M_{(\sigma\kappa)(\nu\alpha)} 
\nonumber \\
 &&
-\frac{1}{12} \, \Gamma_{\mu\nu}^{(\sigma\kappa)} \, \xi_{\lambda} \, 
W_{\sigma\kappa}
-\frac{1}{12} \, \Gamma_{\mu\lambda}^{(\sigma\kappa)} \, \xi_{\nu} \, 
W_{\sigma\kappa}
+ (\mbox{cyclic permutations})
\end{eqnarray}
The last two terms which contain $W_{\sigma\kappa}$ can be rewritten using 
(\ref{GM}) as
\begin{eqnarray}
-\frac{1}{12} \, \Gamma_{\mu\nu}^{(\sigma\kappa)} \, \xi_{\lambda} \, W_{\sigma\kappa}
=-\frac{1}{5} \, \xi_{\lambda} \, \hat{\nabla}_{\mu} \, {\phi_{\nu\alpha}}^{\alpha}
+\frac{1}{9} \, \xi_{\lambda} \,  \hat{\nabla}^{\alpha} \, 
\Phi_{\mu\nu\alpha}+g^{\alpha\beta} \, S_{\mu\alpha,\nu\beta}.
\end{eqnarray}
Here $\Phi_{\mu\nu\alpha}$ (\ref{Phi}) is the traceless part of $\phi_{\mu\nu\alpha}$. 
Those terms which contain $M_{(\sigma\kappa)(\lambda\alpha)}$ can be rewritten 
by using (\ref{GM}), as 
\begin{eqnarray}
-\frac{1}{2} \, \xi^{\alpha} \, \Gamma_{\mu\nu}^{(\sigma\kappa)} \,  
M_{(\sigma\kappa)(\lambda\alpha)}&=& -\frac{1}{3} \, \xi^{\alpha} \, 
\left(\hat{\nabla}_{\mu} \, \Phi_{\nu\lambda\alpha}
+\hat{\nabla}_{\nu} \, \Phi_{\mu\lambda\alpha}\right)
+\frac{1}{6} \, \xi^{\alpha} \, \left(\hat{\nabla}_{\lambda} \, 
\Phi_{\mu\nu\alpha}+\hat{\nabla}_{\alpha} \, \Phi_{\mu\nu\lambda}\right) \nonumber \\
&&-\frac{1}{9} \, 
\xi_{\lambda} \, \hat{\nabla}^{\kappa} \, \Phi_{\mu\nu\kappa}-\frac{1}{2} \, 
\xi^{\alpha} \, S_{\mu\nu,\lambda\alpha}.
\end{eqnarray}
Finally, the variation $\delta_{\xi} \, \phi_{\mu\nu\lambda}$ is given by 
\begin{eqnarray}
\delta_{\xi} \, \phi_{\mu\nu\lambda} &=& 
\xi^{\sigma} \, \hat{\nabla}_{\sigma} \, \phi_{\mu\nu\lambda}
+ \hat{\nabla}_{\mu} \, \xi^{\sigma} \, \phi_{\sigma\nu\lambda}+
\hat{\nabla}_{\lambda} \,  \xi^{\sigma} \, \phi_{\sigma\nu\mu}+
\hat{\nabla}_{\nu} \, \xi^{\sigma} \, \phi_{\sigma\mu\lambda} \nonumber \\
&&+\frac{1}{5}  \, g_{\mu\nu} \, \xi^{\sigma} \, \left(
 \hat{\nabla}_{\lambda} \,  {\phi_{\sigma\kappa}}^{\kappa}-
\hat{\nabla}_{\sigma} \,  {\phi_{\lambda\kappa}}^{\kappa}
\right) 
+\frac{1}{5}  \, g_{\nu\lambda} \, \xi^{\sigma} \, \left(
 \hat{\nabla}_{\mu} \,  {\phi_{\sigma\kappa}}^{\kappa}-
\hat{\nabla}_{\sigma} \,  {\phi_{\mu\kappa}}^{\kappa}
\right) \nonumber \\ && 
+\frac{1}{5}  \, g_{\mu\lambda} \, \xi^{\sigma} \, \left(
 \hat{\nabla}_{\nu} \,  {\phi_{\sigma\kappa}}^{\kappa}-
\hat{\nabla}_{\sigma} \,  {\phi_{\nu\kappa}}^{\kappa}
\right) \nonumber \\
&& +\left\{ g^{\alpha\beta} \,( S_{\mu\alpha,\nu\beta} \, \xi_{\nu}+
 S_{\mu\alpha,\lambda\beta} \, \xi_{\nu})-\xi^{\alpha} \, S_{\mu\nu,\lambda\alpha}+ \text{cyclic permutations of} \  \mu,\nu, \lambda \right\}. \nonumber \\ &&
\label{dimension3}
\end{eqnarray}
Therefore except for the trace parts and the terms containing $S_{\mu\nu,\lambda\rho}$, the 
spin-3 gauge field $\phi_{\mu\nu\lambda}$ 
transforms as a spin-3 tensor under ordinary diffeomorphisms ($\xi_{\mu}$).

Those terms which depend on $\zeta_{(\mu\nu)}$ can also be worked out. 
After certain amount of calculation the $\zeta$ transformation of 
the spin-3 gauge field is obtained. 
\begin{eqnarray}
\delta_{\zeta} \phi_{\mu\nu\lambda} &=& 
\hat{\nabla}_{\mu} \, \zeta_{(\nu\lambda)}+ \frac{1}{6} \, g_{\nu\lambda} \, 
\hat{\nabla}_{\mu} 
\, (\zeta_{(\rho\sigma)} \, E_a^{(\rho\sigma)} \, \rho^a) +\frac{1}{12} \, 
(\zeta_{(\alpha\beta)} 
\, E_a^{(\alpha\beta)} \, \rho^a) \, (\Gamma_{\mu\nu}^{(\sigma\kappa)}\, 
\phi_{\sigma\kappa\lambda}
+\Gamma_{\mu\lambda}^{(\sigma\kappa)}\, \phi_{\sigma\kappa\nu}) \nonumber \\
&& +\frac{1}{2} \, {\phi_{\sigma\kappa}}^{\rho} \, (\Gamma_{\mu\nu}^{(\sigma\kappa)} \, 
\zeta_{(\rho\lambda)}+\Gamma_{\mu\lambda}^{(\sigma\kappa)} \, \zeta_{(\rho\nu)}
) \nonumber \\
&&+\frac{1}{8} \, (\Gamma_{\mu\nu}^{(\kappa\sigma)} \, g_{(\kappa\sigma)
(\lambda\rho)}+\Gamma_{\mu\lambda}^{(\kappa\sigma)} \, g_{(\kappa\sigma)(\nu\rho)}) 
\, {\phi^{\rho}}_{(\tau\eta)} \, J^{(\tau\eta)(\alpha\beta)} \, \zeta_{(\alpha\beta)} 
\nonumber \\
&&-\frac{1}{16} \, (\Gamma_{\mu\nu}^{(\kappa\sigma)} \, g_{(\kappa\sigma)(\rho\tau)\lambda}
+\Gamma_{\mu\lambda}^{(\kappa\sigma)} \, g_{(\kappa\sigma)(\rho\tau)\nu}) \, 
J^{(\rho\tau)(\alpha\beta)} \, \zeta_{(\alpha\beta)} \nonumber \\
&&+\frac{1}{24} \, (\Gamma_{\mu\nu}^{(\kappa\sigma)} \, \phi_{\lambda\tau\eta}
+ \Gamma_{\mu\lambda}^{(\kappa\sigma)} \, \phi_{\nu\tau\eta}  )\, 
\phi_{\kappa\sigma,\rho\gamma} \, g^{\rho\gamma} \, J^{(\tau\eta)
(\alpha\beta)} \, 
\zeta_{(\alpha\beta)}  \nonumber \\ 
&& + (\mbox{2 cyclic permutations})
\label{del_phi}
\end{eqnarray}
Here those terms which include $\rho$ can be simplified further by using 
(\ref{rhoE})-(\ref{rhoEE}). The gauge fields 
$g_{(\kappa\sigma)(\rho\tau)\lambda}$, $g_{(\kappa\sigma)(\nu\rho)}$ and 
$\phi_{\kappa\sigma,\rho\gamma}$ are defined in appendix C. 
Under a new spin-3 gauge transformation ($\zeta_{(\mu\nu)})$,  
$\phi_{\mu\nu\lambda}$ transforms in a complicated way which depends on 
higher-indexed gauge fields. Transformations of these gauge 
fields must also be studied. However, in this paper this will not be attempted.

At the beginning of this section it was shown that by using the new covariant 
derivative $\nabla_{\mu}$, the transformation $\delta g_{\mu\nu}$ can be 
compactly expressed as (\ref{gdiffeo}) just like in Einstein gravity. Then 
one may expect that due to relations among gauge fields, the transformation 
$\delta \phi_{\mu\nu\lambda} =\delta_{\xi} \phi_{\mu\nu\lambda} +\delta_{\zeta} 
\phi_{\mu\nu\lambda}$ might also be succinctly written. 

Actually,  using $\delta \, e^a_{\mu} = D_{\mu} \, \Lambda^a$ and 
$ \delta \, e^a_{(\mu\nu)} = \frac{1}{2} \, {d^a}_{bc} \, 
(e^b_{\mu} \, D_{\nu} \,\Lambda^c+e^b_{\nu} \, D_{\mu} \,\Lambda^c)
-\frac{1}{3} \, \rho^a \, (\nabla_{\mu} \, \xi_{\nu}
+\nabla_{\nu} \, \xi_{\mu}) 
-\frac{1}{3} \, g_{\mu\nu} \, \delta \rho^a$, 
one can show that
\begin{eqnarray}
\delta \, \phi_{\mu\nu\lambda}  
&=& \nabla_{\mu} \, \left(\xi_{(\nu\lambda)}+ \frac{1}{3} \, \rho^a \, 
\Lambda_a 
\, g_{\nu\lambda}\right)    
+\nabla_{\nu} \, \left(\xi_{(\lambda\mu)}+ \frac{1}{3} \, \rho^a \, 
\Lambda_a \, g_{\lambda\mu}\right) \nonumber \\ &&
+\nabla_{\lambda} \, \left(\xi_{(\mu\nu)} + \frac{1}{3} 
\, \rho^a \, \Lambda_a \, g_{\mu\nu}\right).  
 \label{spin3gauge}
\end{eqnarray}
Except for the trace terms this agrees with the expected transformation rule 
of the spin-3 gauge field. 

\section{Parallel transport and  Curvature tensor}
\hspace{5mm}
To investigate the spin-3 geometry, it is useful to introduce a parallel 
transport matrix, holonomy matrix and curvature tensor. This will be done 
in this section. 

Let $v^a(x)$ be a vector field in the local Lorentz frame. For an arbitrary 
curve $x^{\mu}=x^{\mu}(s)$, this vector is said to be parallel transported along 
the curve,\cite{Carlip} if it satisfies the equation
\begin{eqnarray}
\frac{d v^a}{ds} +{{\omega_{\mu}}^a}_b(e) \, \frac{d x^{\mu}}{ds} \, v^b=0.
\label{ParaTr}
\end{eqnarray}
This equation can be solved in terms of the ordered exponential
\begin{eqnarray}
v^a(x(s)) &=& {U^a}_b(s,0) \, v^b(x(0)), \\
{U^a}_b(s,0) &=& {\left( P \, \exp \left\{- \int_0^s \omega_{\mu} \, 
\frac{d x^{\mu}}{ds'} \, ds' \right\} \right)^a}_b. \label{PO}
\end{eqnarray}
Here, as usual,  the symbol $P$ denotes path ordering. 
\begin{eqnarray}
P (A(s_1) \, B(s_2)) = \left\{   \begin{array}{cc}
A(s_1) \, B(s_2) \qquad (\text{if} \ \ s_1 > s_2), \\
 B(s_2) \, A(s_1) \qquad (\text{if} \ \  s_2 > s_1).
\end{array} \right.
\end{eqnarray}

These relations can be converted into that for spacetime quantities 
by means of the vielbeins. Firstly, we perform the following GL(8,R) gauge 
transformation on the spin connection matrix $\omega_{\mu}(e)$.\footnote{In 
the case of Einstein gravity a similar transformation is used.\cite{Jackiw}} 
\begin{eqnarray}
{{\omega_{\mu}}^a}_b(e) \rightarrow {{\Upsilon_{\mu}}^M}_N = E^M_a \, 
{{\omega_{\mu}}^a}_b(e) \, e^b_N+ E^M_a \, \partial_{\mu} \, e^a_N
\label{gaugetrGL8}
\end{eqnarray}
Here $M, N=\mu, (\mu\nu)$ are the indices explained in appendix E. By 
(\ref{omegaac}) the new spin connection can be written as 
\begin{eqnarray}
{{\Upsilon_{\mu}}^M}_N = -E^M_a \, \nabla_{\mu} \, e^a_N + E^M_a \, 
\partial_{\mu} \, e^a_N.
\end{eqnarray}
This agrees with the connections defined in sec.4, 
\begin{eqnarray}
{{\Upsilon_{\mu}}^{\lambda}}_{\nu}&=&\Gamma^{\lambda}_{\mu\nu}, \qquad 
{{\Upsilon_{\mu}}^{(\lambda\rho)}}_{\nu}=\Gamma^{\lambda\rho}_{(\mu\nu)}, \nonumber \\ 
{{\Upsilon_{\mu}}^{\rho}}_{(\nu\lambda)} & =& \Gamma_{\mu, (\nu\lambda)}^{\rho} , \qquad 
{{\Upsilon_{\mu}}^{(\rho\sigma)}}_{(\nu\lambda)} = \Gamma_{\mu, (\nu\lambda)}^{(\rho\sigma)}. 
\end{eqnarray}

Under the gauge transformation (\ref{gaugetrGL8}) the path-ordered 
exponential (\ref{PO}) transforms into a spacetime quantity, 
\begin{equation}
{U^M}_N(s,0) = {\left( P \, \exp \left\{- \int_0^s \Upsilon_{\mu} \, 
\frac{d x^{\mu}}{ds'} \, ds' \right\} \right)^M}_N. 
\end{equation}
The parallel transport eq (\ref{ParaTr}) is also rewritten as 
\begin{eqnarray}
\frac{d v^M}{ds} +{{\Upsilon_{\mu}}^M}_N \, \frac{d x^{\mu}}{ds} \, v^N=0.
\end{eqnarray}

If the curve $x^{\mu}(s), (0 \leq s \leq 1)$ is closed, the matrix $U^M_N(1,0)$ 
defines a holonomy matrix. For an infinitesimal closed curve $\gamma$ which 
encloses a small surface $S$, this holonomy can be evaluated by expansion 
of the exponential. 
By using Stokes's theorem this yields a generalization of the Riemann 
curvature tensor at the lowest order of expansion. 
\begin{eqnarray}
{U^M}_N(1,0) = {\delta^M}_N+\int_S \, {R^M}_{N\mu\nu} \, d\Sigma^{\mu\nu} +\dots
\end{eqnarray}
Here 
\begin{eqnarray}
{R^M}_{N\mu\nu} \equiv \partial_{\mu} \, {{\Gamma_{\nu}}^M}_N
- \partial_{\nu} \, 
{{\Gamma_{\mu}}^M}_N- {{\Gamma_{\nu}}^M}_K  \, {{\Gamma_{\mu}}^K}_N
+{{\Gamma_{\mu}}^M}_K  \, {{\Gamma_{\nu}}^K}_N. \label{generalised Riemann}
\end{eqnarray}

The action integral in the second-order formalism (\ref{CSsub}) can be 
expressed in terms of this curvature tensor.  To do this, we  perform the 
$GL(8,R)$ gauge transformation $\omega_{\mu b}^a=e^a_M \, 
\Gamma^M_{\mu N} \, E_b^N-
E_b^N \, \partial_{\mu} \, e^a_N$ on the curvature tensor ${R^a}_{b\mu \nu}=
\partial_{\mu} \, \omega^a_{\nu b}-\partial_{\nu} \, \omega^a_{\mu b}
+\omega^a_{\mu c} \, \omega^c_{\nu b}- \omega^a_{\nu c} \, \omega^c_{\mu b}$. 
Since the curvature 2-form is gauge covariant, we obtain
\begin{eqnarray}
{R^a}_{b\mu \nu}= e^a_M \, E^N_b \, {R^M}_{N\mu\nu}.  
\end{eqnarray}
The identity ${R^a}_{b\mu\nu}={f^a}_{cb} \, R^c_{\mu\nu}$, where $R^c_{\mu\nu}=
\partial_{\mu} \, \omega^c_{\nu}-\partial_{\nu} \, \omega^c_{\mu}+{f^c}_{de} 
\, \omega^d_{\mu} \, \omega^e_{\nu}$, leads to (\ref{CSsecond}). 

In the spin-2 gravity theory, the Riemann curvature tensor also defines the 
commutator of the covariant derivatives, $[\nabla_{\nu}, \, \nabla_{\mu}] 
\, v^{\lambda} ={R^{\lambda}}_{\rho\nu\mu} \, v^{\rho}$. In this theory, 
this eq can be derived 
by starting with the curvature 2-form ${R^a}_{b\nu\mu} \, v^b=
[D_{\nu}, D_{\mu}] 
\, v^a$ and by projecting onto the base space using the vielbein as 
$v^{\mu}=v^a \, E_a^{\mu}$. 
In the spin-3 case we also expect similar formulae such as 
\begin{eqnarray}
[\nabla_{\nu}, \, \nabla_{\mu}] \, v^{\lambda} &=&
{R^{\lambda}}_{\rho\mu\nu} \, v^{\rho}
+\frac{1}{2} \, {R^{\lambda}}_{(\rho\sigma)\mu\nu} \, v^{(\rho\sigma)}, 
\label{ccR1} \\
 \ [\nabla_{\nu}, \, \nabla_{\mu}] \, v^{(\lambda\rho)} & = &
{R^{(\lambda\rho)}}_{\sigma\mu\nu} 
\, v^{\sigma}+\frac{1}{2} \, {R^{(\lambda\rho)}}_{(\sigma\kappa)\mu\nu} \, 
v^{(\sigma\kappa)}. \label{ccR}
\end{eqnarray} 
However, there is an obstacle in deriving such formulae, because one does not 
know how to compute $\nabla_{\nu} \, \nabla_{\mu} \, v^{\lambda}$, and the 
covariant derivative does not have the component \lq $\nabla_{(\mu\nu)}$' 
in the new direction $(\mu\nu)$. We will speculate on these formulae in the 
remaining part of this subsection. 

If this component exists, it is possible to compute the commutators of 
the covariant derivatives. Actually, we have $\nabla_{\nu} \, \nabla_{\mu} 
\, v^{\lambda}=\partial_{\nu} \, \partial_{\mu} \, v^{\lambda}+(\partial_{\nu} \, 
\Gamma_{\mu M}^{\lambda}) \, v^M+\Gamma_{\mu M}^{\lambda} \, \partial_{\nu} \, v^M
-\Gamma_{\nu\mu}^M \, \nabla_M \, v^{\lambda}+\Gamma_{\nu M}^{\lambda} \, \partial_{\mu} 
\, v^M+\Gamma^{\lambda}_{\nu M} \, \Gamma_{\mu N}^M \, v^N $ and then 
$[\nabla_{\nu}, \nabla_{\mu}] \, v^{\lambda}
= (\partial_{\nu} \, \Gamma_{\mu M}^{\lambda}-\partial_{\mu} \, \Gamma_{\nu M}^{\lambda}
+\Gamma_{\nu N}^{\lambda} \, \Gamma_{\mu M}^N-\Gamma_{\mu N}^{\lambda} \, \Gamma_{\nu M}^N)
\, v^M={R^{\lambda}}_{M\nu\mu} \, v^M$. The term $\Gamma_{\nu\mu}^M \, \nabla_M \, 
v^{\lambda} $ cancels out in the commutator. The actual value of 
$\nabla_{(\mu\nu)} \, v^{\lambda} $ does not matter. It is important to notice 
that {\em it can even be zero}: $\nabla_{(\mu\nu)} \, v^{\lambda} =0 $. 

In order to define $\nabla_{(\mu\nu)}$, then, we would need to introduce new 
coordinates $x^{\mu\nu}$ and set $\nabla_{(\mu\nu)} \, v^{\lambda} =\partial_{(\mu\nu)} 
\, v^{\lambda}+\Gamma^{\lambda}_{(\mu\nu), M} \, v^M$. This, however, would  
make the spacetime have dimension 8, and one would need to cope with a 
problem of integrating over the new coordinates. 
So one of possible prescriptions would be to avoid introducing $x^{\mu\nu}$ 
and set $\nabla_{(\mu\nu)} \, v^{\lambda}=\Gamma^{\lambda}_{(\mu\nu), M} \, 
v^M$. We would then also need to introduce a new component of the spin 
connection, $\omega^a_{(\mu\nu)}$, and impose a torsion-free 
condition, $\nabla_{(\mu\nu)} \, e^a_M+{f^a}_{bc} \, \omega^b_{(\mu\nu)} \, 
e^c_M=0$. However, compatibility of this covariant derivative 
$\nabla_{(\mu\nu)}$ with $g_{\lambda\rho}$ and $g_{(\lambda\rho)\sigma}$ 
would inevitably lead to the conclusion $\Gamma^N_{(\mu\nu),M}=0$ and 
$\omega^a_{(\mu\nu)}=0$. To define $\nabla_{(\mu\nu)}$, introduction of 
extra coordinates seems unavoidable. Therefore, we will set $\nabla_{(\mu\nu)} 
=0$ in this paper. Even in this case the rules (\ref{ccR1})-(\ref{ccR}) of the commutators 
of the covariant derivatives still apply.

\section{Gravitational CS term}
\hspace{5mm}
In 3D there also exists a gravitational Chern-Simons term.\cite{DJT} 
It is given by
\begin{eqnarray}
S_{\text{ GCS}}(\omega) &=& \frac{k}{8 \pi \mu} \, \int \, d^3x \, 
\epsilon^{\mu\nu\lambda} \, 
\left(\omega(e)^a_{\mu b} \, \partial_{\nu} \, \omega (e)^b_{\lambda a}
+\frac{2}{3} \, \omega(e)^a_{\mu b}\, \omega(e)^b_{\nu c}\, 
\omega(e)^c_{\lambda a} \right).  
\label{GCS}
\end{eqnarray}
Here $\mu$ is a constant. In this action, the spin connection 
$\omega^a_{\mu b}(e)$ is a functional of the vielbein $e^a_{\mu}$, 
$e^a_{(\mu\nu)}$ 
as defined in (\ref{spinconnection}). This action is invariant in the bulk 
under both the local frame transformation and the generalized diffeomorphism. 
The invariance is broken at the boundary.  
If this term is added to the CS action in the second-order formalism 
(\ref{CSsecond}), the action of a topological massive spin-3 gravity 
(a generalization of the topological massive gravity\cite{DJT}) is obtained. 
In the gravity/CFT correspondence the gravitational action with the 
gravitational CS term corresponds to a left-right asymmetric (chiral) CFT. 
This action has derivatives of cubic 
order and hence the eqs of motion will contain terms with cubic-order 
derivatives. Let us note 
that if the solution for the spin connection is not substituted into the 
action integral, and the vielbein and the spin connection were treated 
independently, the torsion-free eq would be modified. In order to avoid this, 
the torsion-free condition may be imposed by means of a Lagrange multiplier 
field.\cite{DX}\cite{C2}  However, the generalized diffeomorphism invariance 
(\ref{diffeo}) will be broken by the multiplier term.\footnote{A linearized 
action in the topological massive higher-spin gravity is studied in 
\cite{TPHSM1}. Topological massive higher-spin gravity
 with a multiplier field is studied in 
\cite{TPHSM2}. } 
 
It is known that in the case of ordinary 3D spin-2 gravity, the gravitational 
Chern-Simons term can also be expressed in terms of the Christoffel 
connection up to a winding number term; $S_{\text{GCS}}(\omega)=
S_{\mbox{\tiny GCS}}^{\mbox{\tiny spin-2}}(\hat{\Gamma})$+ 
(winding number term).\cite{DJT}\cite{Jackiw} 
\begin{eqnarray}
S_{\mbox{\tiny GCS}}^{\mbox{\tiny spin-2}}(\hat{\Gamma}) = \frac{k}{8\pi \mu} \, \int 
\, d^3x \, \epsilon^{\mu\nu\lambda} \, 
(\hat{\Gamma}^{\rho}_{\mu\sigma} \, \partial_{\nu} \, 
\hat{\Gamma}^{\sigma}_{\lambda\rho} 
+\frac{2}{3} \, \hat{\Gamma}^{\sigma}_{\mu\kappa} \, 
\hat{\Gamma}^{\kappa}_{\nu\rho} \,
\hat{\Gamma}^{\rho}_{\lambda\sigma} ) \label{EGCS}
\end{eqnarray}
Actually, this last form of the gravitational CS term must be used 
in the second-order formalism. 
In the case of spin-3 gravity, a similar expression for the action can be 
derived by using the gauge transformation (\ref{gaugetrGL8}). After 
substitution we have, up to winding number terms, 
\begin{eqnarray}
S_{\text{GCS}}^{\text{spin-3}}(\Gamma) &=& \frac{k}{8\pi \mu} \, \int \, d^3x \, 
\epsilon^{\mu\nu\lambda} \, 
\left({{\Upsilon_{\mu}}^{M}}_{N} \, \partial_{\nu} \, {{\Upsilon_{\lambda}}^{N}}_{M}
+\frac{2}{3} \, {{\Upsilon_{\mu}}^{M}}_{N} \, {{\Upsilon_{\nu}}^{N}}_{K}   \, 
{{\Upsilon_{\lambda}}^{K}}_{M} \right) \nonumber \\
&=&   \frac{k}{8\pi \mu} \, \int \, d^3x \, \epsilon^{\mu\nu\lambda} \, 
\left(\Gamma^{\rho}_{\mu\sigma} \partial_{\nu} \, \Gamma^{\sigma}_{\lambda \rho}
+\frac{1}{2} \, \Gamma^{(\rho\sigma)}_{\mu\kappa} \, \partial_{\nu} \, 
\Gamma^{\kappa}_{\lambda, (\rho\sigma)}+\frac{1}{2} \, 
\Gamma^{\kappa}_{(\rho\sigma)} \, \partial_{\nu} \, 
\Gamma_{\lambda\kappa}^{(\rho\sigma)}\right. \nonumber \\
&&+\frac{1}{4} \, \Gamma^{(\rho\sigma)}_{\mu, (\kappa\tau)} \, 
\partial_{\nu} \, \Gamma^{(\kappa\tau)}_{\lambda, (\rho\sigma)}
+\frac{2}{3} \, \Gamma_{\mu\sigma}^{\rho} \, 
\Gamma_{\nu \kappa}^{\sigma} \, 
\Gamma_{\lambda\rho}^{\kappa} + \, \Gamma_{\mu, \sigma}^{(\rho\tau)} 
\, \Gamma_{\nu \kappa}^{\sigma} \, 
\Gamma_{\lambda, (\rho\tau)}^{\kappa} \nonumber \\ && \left. +\frac{1}{2} \, 
\Gamma_{\mu, (\sigma\eta)}^{(\rho\tau)} \, \Gamma_{\nu \kappa}^{(\sigma\eta)} \, 
\Gamma_{\lambda, (\rho\tau)}^{\kappa}+\frac{1}{12} \, \Gamma_{\mu,
 (\sigma\eta)}^{(\rho\tau)} \, \Gamma_{\nu (\kappa\alpha)}^{(\sigma\eta)} \, 
\Gamma_{\lambda, (\rho\tau)}^{(\kappa\alpha)} \right).
\end{eqnarray}
In the spin-2 gravity theory, solutions such as BTZ black hole\cite{BTZ} 
in the theory without the gravitational CS term are known to be also 
solutions of the eqs of motion of the topologically massive gravity theory. 
Therefore the natural questions to ask are: do the solutions in the 
spin-3 gravity without the gravitational CS term, such as the spin-3 black 
hole\cite{GK}, also solve the eqs of motion in the spin-3 topologically 
massive gravity? If it is the case, how the central charges of the W$_3$ 
algebras in the boundary CFT and the value of the entropy will be modified 
in the presence of the gravitational CS term? 

The black hole solution with spin-3 charge is asymptotically 
AdS$_3$ with AdS radius $1/2$.\cite{GK} Therefore it may be interesting to 
study the existence of propagating gravitons with this asymptotic boundary 
condition.  
These problems are left for the future studies.

\section{Summary and discussion}
\hspace{5mm}
In this paper a second-order formalism of the 3D spin-3 gravity is addressed 
and it is shown that many of the notions and geometrical quantities of 
Einstein gravity theory can be introduced into this theory. 
Extra vielbeins $e^a_{(\mu\nu)}$ (\ref{emunu}) are introduced in order to 
eliminate the spin connection from the CS formulation of the 3D spin-3 
gravity in a way covariant under the local frame rotations.  It is shown that 
new connections $\Gamma^{N}_{\mu M}$ can be expressed in terms of the metric-like 
fileds and that a covariant derivative $\nabla_{\mu}$ 
(\ref{CovariantDerivative1}), (\ref{covariantDerivative2}) can be defined. 
The torsion-free condition is solved for the spin connection $\omega^a_{\mu}$ 
as (\ref{spinconnection}) in terms of the generalized vielbein and its 
inverse.  In terms of this solution, the action integral in the second-order 
formalism (\ref{CSsub}) is presented, although in a somewhat implicit form.  
Many metric-like fields other than $g_{\mu\nu}$ and $\phi_{\mu\nu\lambda}$ 
are shown to exist. Although they are expected to be expressed in terms of 
$g_{\mu\nu}$ and $\phi_{\mu\nu\lambda}$ at least in the case of fluctuations around 
AdS$_3$ vacuum, a precise relation among these fields needs to be worked out 
in the future study. 
Then a generalised Riemann curvature tensor for the spin-3 gravity is also 
defined. 
The explicit form of the generalized diffeomorphism of the metric 
$g_{\mu\nu}$ and the spin-3 gauge field $\phi_{\mu\nu\lambda}$ is presented. 
Finally, the action integral for topologically massive spin-3 
gravity is presented explicitly.

In the present paper, the transformation rules of the connections 
$\Gamma_{\mu M}^N$ under the generalized diffeomorphisms are not studied 
explicitly. This is because the expression for $S_{\mu\nu,\lambda\rho}$ 
in $\Gamma^{(\lambda\rho)}_{\mu\nu}$ is complicated. This problem must be 
studied in the future. However, by assuming the transformation rule of 
$\omega^a_{\mu}(e)$ as (\ref{delomega}) and using the relation between 
$\Gamma$'s and $\omega^a_{\mu}$ it is possible to derive the transformation 
rule of $\Gamma$'s. 

For other future work we would like to consider the coupling of matter 
fields to the spin-3 gravity. For this purpose it is necessary to define 
density and tensors which transform appropriately under the generalized 
diffeomorphisms. Then it must be shown that the covariant derivatives of 
the general tensors also transform as tensors. At present, this remains an
unsolved problem. 

Finally, there will be several directions for future investigations. To   
enumerate a few, the geometry of the 3D spin-3 gravity is still not 
well-understood. This must be studied further and the spin-3 gravity 
must be formulated from scratch without relying on CS theory. In the 
case of supergravity, where gravity theory is likewise extended by 
supersymmetry transformations, one can understand the theory 
geometrically by introducing supercoordinates, a superspace and superfields. 
Likewise, it might be possible to better understand the spin-3 gravity 
analogously by introducing a \lq spin-3 space'. 

A generalization of the work in this paper to spin-N($ \geq 4$) gravity theories 
will be straightforward. For example, in the spin-4 
gravity theory, extended vielbeins $e_{\mu}$, $e_{(\mu\nu)}$ and 
$e_{(\mu\nu\lambda)}$, which are completely symmetric in the indices and satisfy traceless 
conditions, will provide 3+5+7=15 basis vectors. This number agrees with the dimension of 
$sl(4,R)$. The case of spin-N gravity works similarly.


\newpage
\setcounter{section}{0}
\renewcommand{\thesection}{\Alph{section}}
\section{sl(3,R) algebra}
\hspace{5mm}
Let the generators $L_i \ (i=-1,0,1)$, $W_n \ (n=-2, \ldots 2)$ satisfy 
an $sl(3, R)$ algebra.
\begin{eqnarray}
\ [L_i,L_j] &=& (i-j) \, L_{i+j}, \qquad 
\ [L_i,W_n] = (2i-n) \, W_{i+n}, \nonumber \\
\ [W_m,W_n] &=& -\frac{1}{3} (m-n) \, \{2m^2+2n^2-mn-8\} \, L_{m+n}
\end{eqnarray}
We use the same three-dimensional representation as in \cite{Campoleoni} with 
the parameter $\sigma=-1$.
\begin{eqnarray}
L_1 &=& \left(\begin{array}{ccc}
        0  & 0& 0 \\
        1 & 0 & 0 \\
        0 & 1 & 0  \end{array}\right), \qquad 
L_0  = \left(\begin{array}{ccc}
        1  & 0& 0 \\
        0 & 0 & 0 \\
        0 & 0 & -1 \end{array} \right), \qquad 
L_{-1}  = \left(\begin{array}{ccc}
        0  & -2& 0 \\
        0 & 0 & -2 \\
        0 & 0 & 0  \end{array}\right), \nonumber \\
W_2 &=& \left(\begin{array}{ccc}
        0  & 0& 0 \\
        0 & 0 & 0 \\
        2 & 0 & 0  \end{array}\right), \qquad 
W_1  = \left(\begin{array}{ccc}
        0  & 0& 0 \\
        1 & 0 & 0 \\
        0 & -1 & 0 \end{array} \right), \qquad 
W_0  = \frac{2}{3} \, \left(\begin{array}{ccc}
        1  & 0& 0 \\
        0 & -2 & 0 \\
        0 & 0 & 1  \end{array}\right), \nonumber \\
W_{-1} &=& \left(\begin{array}{ccc}
        0  & -2 & 0 \\
        0 & 0 & 2 \\
        0 & 0 & 0 \end{array} \right), \qquad 
W_{-2}  = \left(\begin{array}{ccc}
        0  & 0& 8 \\
        0 & 0 & 0 \\
        0 & 0 & 0  \end{array}\right)
\label{generators}
\end{eqnarray}
Nonvanishing norms of these matrices are given by 
\begin{eqnarray}
\mbox{tr} \ (L_0)^2=2, \quad \mbox{tr} \ (L_{-1}L_1)=-4, \quad \mbox{tr} \ 
 (W_0)^2=\frac{8}{3}, \quad \mbox{tr} \ (W_1W_{-1})=-4, \quad \mbox{tr} \
 (W_2W_{-2})=16.
\end{eqnarray}
These generators will also be collectively denoted as $t_a, (a=1, \ldots,8)$,
\begin{eqnarray}
t_1&=&L_1, \quad t_2=L_0, \quad t_3= L_{-1}, \nonumber \\
t_4 &=&W_2, \quad t_5=W_1, \quad t_6= W_0, \quad t_7= W_{-1}, \quad  
t_8= W_{-2}. 
\end{eqnarray}
The structure constants ${f_{ab}}^c$ are defined by 
\begin{eqnarray}
 [ t_a, t_b]= {f_{ab}}^c \, t_c.
\end{eqnarray}
The Killing metric $h_{ab}$ for the local frame is defined by 
\begin{eqnarray}
h_{ab}=\frac{1}{2} \, \mbox{tr} \, (t_a t_b)
\end{eqnarray}
Its nonzero components are given by $h_{22}=1, \ h_{13}=h_{31}=-2, 
h_{48}=h_{84}=8, \ h_{57}=h_{75}=-2, \ h_{66}=4/3$. 
This metric tensor has a signature $(3,5)$. 
Indices of the local frame are raised and lowered by $h_{ab}$ and its inverse 
$h^{ab}$. Then $f_{abc} \equiv {f_{ab}}^d \, h_{dc}$ 
is completely anti-symmetric in the three indices. It can be shown that 
$f_{abc}$ and $h_{ab}$ are related by
\begin{equation}
h_{ab}= -\frac{1}{12} \, {f_a}^{cd} \, f_{bcd}. \label{hff}
\end{equation}
The structure constants are given by
\begin{eqnarray}
&& f_{123} = -2, \ f_{158}=8, \ f_{167}=-4, \ f_{248}=-16, \nonumber \\
&& f_{257} = 2, \ f_{347}=8, \ f_{356}=-4
\end{eqnarray}

The invariant tensor ${d_{ab}}^c$ is defined by 
\begin{eqnarray}
 \{t_a, t_b \} = {d_{ab}}^c \, t_c+ {d_{ab}}^0 \, t_0,
\end{eqnarray}
where $t_0=\bm{I}$ is an identity matrix. The constant with the lowered 
index $d_{abc}= {d_{ab}}^d \, h_{dc}$ is completely symmetric in all the 
indices. 
These constants are given by
\begin{eqnarray}
&& d_{127}=d_{235}=-2, \quad  d_{136}= d_{226}=d_{567}=\frac{4}{3},  \quad 
d_{118}= d_{334}=8, \nonumber \\
&& d_{468}=\frac{32}{3}, \quad d_{477}= d_{558}=-8, \quad  d_{666}=-\frac{16}{9}, 
\qquad {d_{ab}}^0=\frac{4}{3} \, h_{ab} 
\end{eqnarray}

\section{AdS$_3$}
\hspace{5mm}
The flat connections which yield $AdS_3$ spacetime are given by
\begin{eqnarray}
A &=& e^r \ L_1 \,dx^++L_0 \, dr, \nonumber \\
\bar{A} &=& -e^r \ L_{-1} \, dx^--L_0 \, dr.
\end{eqnarray}
In this appendix, to avoid confusion of $\mu(=0,1,2)$ with $a(=1,2,...)$ 
a different notation $\mu=t,\phi,r$ will be used, and $dx^{\pm} \equiv dt 
\pm d 
\phi$. The corresponding vielbein and spin connection 
are given by
\begin{eqnarray}
e_t = \omega_t &=& \frac{1}{2} \, e^r \, (L_1+L_{-1}), \nonumber \\
e_{\phi} = \omega_{\phi} & = & \frac{1}{2} \, e^r \, (L_1-L_{-1}), \nonumber \\
e_r &=& L_0, \qquad \omega_r=0.
\label{AdS3}
\end{eqnarray}
The metric is $ds^2=g_{\mu\nu} \, dx^{\mu} \, dx^{\nu}=dr^2+e^{2r}(-dt^2+
d\phi^2)$. 

By (\ref{AdS3}) one obtains 
\begin{eqnarray}
(e_r)^2 &=&\frac{1}{2} \ W_0+\frac{2}{3}\, \bm{I} \qquad \qquad  \rightarrow 
\qquad \qquad \qquad  \hat{e}_{rr}=\frac{1}{2} \ W_0, \nonumber \\
(e_t)^2 &=& \frac{1}{8} \, e^{2r} (W_2+W_{-2}+2 \, W_0-\frac{16}{3}\, \bm{I})
 \quad 
\rightarrow \quad \hat{e}_{tt}=\frac{1}{8} \, e^{2r} (W_2+W_{-2}+2 \, W_0), 
\nonumber \\
(e_{\phi})^2 &=& \frac{1}{8} \ e^{2r} (W_2+W_{-2}-2 \, W_0+\frac{16}{3}\, 
\bm{I}) \quad \rightarrow \quad \hat{e}_{\phi\phi}=\frac{1}{8} \, e^{2r} 
(W_2+W_{-2}-2 \, W_0). \nonumber \\ &&
\end{eqnarray}
Since $\hat{e}_{\mu\nu}$ satisfies 
\begin{equation}
\rho=g^{\mu\nu} \, \hat{e}_{\mu\nu}=0, 
\end{equation}
one has $e_{(\mu\nu)}=\hat{e}_{\mu\nu}$ for this special geometry. The other 
components are given by 
\begin{eqnarray}
e_{(rt)} &=& \hat{e}_{rt}= \frac{1}{4} \, e^r \, (W_1+W_{-1}), \nonumber \\
e_{(r\phi)} &=& \hat{e}_{r\phi}=\frac{1}{4} \, e^r \, (W_1-W_{-1}), \nonumber \\
e_{(t\phi)} &=& \hat{e}_{t\phi}=\frac{1}{8} \, e^{2r} \, (W_2-W_{-2}).
\end{eqnarray}

The non-vanishing components of the vielbein in terms of the basis $t_a$ are 
\begin{eqnarray}
&& e^2_r=1, \ e^1_t=\frac{1}{2} \, e^r, \ e^3_t =\frac{1}{2} \,e^r, \ 
e^1_{\phi}=\frac{1}{2} \, e^r, \ e^3_{\phi}=-\frac{1}{2} \, e^r, \nonumber \\
&& e^5_{(rt)}=\frac{1}{4} \, e^r, \ e^7_{(rt)}=\frac{1}{4} \, e^r, \ e^5_{(r\phi)}
=\frac{1}{4} \, e^r, \ e^7_{(r\phi)}=-\frac{1}{4} \, e^r, \ e^4_{(t\phi)}
=\frac{1}{8} \, e^{2r}, \nonumber \\  &&
\ e^8_{(t\phi)}=-\frac{1}{8} \, e^{2r},  e^6_{(rr)} = \frac{1}{2}, \ e^4_{(tt)}= \frac{1}{8} \, e^{2r}, \ e^8_{(tt)}
=\frac{1}{8} \, e^{2r}, \ e^6_{(tt)}=\frac{1}{4} \, e^{2r}, \nonumber \\
&& e^4_{(\phi\phi)}= \frac{1}{8} \, e^{2r},\ e^8_{(\phi\phi)}= \frac{1}{8} \, 
e^{2r},\ e^6_{(\phi\phi)}=- \frac{1}{4} \, e^{2r}.
\end{eqnarray}
Then the inverse vielbein exists. An explicit calculation shows that
\begin{eqnarray}
&& E^r_2=1, \ E^t_1=e^{-r}, \ E^t_3=e^{-r}, \ E^{\phi}_1=e^{-r}, \ E^{\phi}_3
=-e^{-r}, 
\nonumber \\
&& E^{(tt)}_4=4 \, e^{-2r}, \ E^{(\phi\phi)}_4=4 \, e^{-2r}, \ E^{(t\phi)}_4=
4 \, e^{-2r}, \ 
E^{(rt)}_5=2 \, e^{-r}, \ E^{(r\phi)}_5=2 \, e^{-r}, \,\nonumber \\ &&
E^{(rr)}_6=\frac{8}{3}, 
 E^{(tt)}_6=\frac{4}{3} \, e^{-2r},  
 E_6^{(\phi\phi)}=-\frac{4}{3} \, e^{-2r}, E^{(rt)}_7=2 \, e^{-r}, \ 
E^{(r\phi)}_7=-2 \, e^{-r}, \nonumber \\ &&
 E^{(tt)}_8=4 \, e^{-2r}, 
 E^{(\phi\phi)}_8=4 \, e^{-2r}, \ E^{(t\phi)}_8=-4 \, e^{-2r}. 
\end{eqnarray}
Therefore, the 8D local frame spanned by $e^a_{\mu}$ and 
$e^a_{(\mu\nu)}$ is actually non-degenerate. 

The spin-3 gauge field vanishes.
\begin{equation}
\phi_{\mu\nu\lambda}= g_{(\mu\nu)\lambda}=0
\end{equation}
$M_{(\mu\nu)(\lambda\rho)}=g_{(\mu\nu)(\lambda\rho)}$ (\ref{M}) have the 
following non-vanishing components. 
\begin{eqnarray}
g_{(rr)(rr)} &=& \frac{1}{3}, \ \  g_{(rt)(rt)}= \frac{-1}{4} \, e^{2r}, 
g_{(r\phi)(r\phi)} = \frac{1}{4} \, e^{2r}, \ \  g_{(tt)(tt)}=
\frac{1}{3} \, e^{4r}, \ \ 
g_{(t\phi)(t\phi)} =\frac{-1}{4} \, e^{4r}, \nonumber \\
g_{(\phi\phi)(\phi\phi)}&=&\frac{1}{3} \, e^{4r}, \ \ 
g_{(rr)(tt)}= \frac{1}{6} \, e^{2r}, \ \ g_{(rr)(\phi\phi)}=
\frac{-1}{6} \, e^{2r}, \ \ g_{(tt)(\phi\phi)}=\frac{1}{6} \, e^{4r} \label{Mg4}
\end{eqnarray}
So the tensor $J^{(\mu\nu)(\lambda\rho)}$ (\ref{J}) is given by
\begin{eqnarray}
J^{(rr)(rr)}&=& \frac{16}{3}, \ \ J^{(rt)(rt)}= -4 \, e^{-2r}, \ \ 
J^{(r\phi)(r\phi)}=4 \, e^{-2r}, \ J^{(t\phi)(t\phi)}= -4 \, e^{-4r}, \nonumber \\
J^{(tt)(tt)}&=& \frac{16}{3} \, e^{-4r}, 
J^{(\phi\phi)(\phi\phi)}=\frac{16}{3}\, e^{-4r}, \ \ 
J^{(rr)(tt)}= \frac{8}{3} \, e^{-2r}, \ \ J^{(rr)(\phi\phi)}=
\frac{8}{3} \, e^{-2r}, \nonumber \\
J^{(tt)(\phi\phi)}&=& \frac{8}{3} \, e^{-4r}. 
\label{AdSJ}
\end{eqnarray}

The killing vectors determine the generalized diffeomorphisms which do not 
change the metric-like quantities, $g_{\mu\nu}$, $g_{\mu(\nu\lambda)}$ and 
$g_{(\mu\nu)(\lambda\rho)}$. In the spin-3 geometry there exist two types: 
Killing vectors $\xi_{\mu}$ and Killing tensors $\xi_{(\mu\nu)}$. They are 
determined by eqs $\delta g_{\mu\nu}=0$ and $\delta \phi_{\mu\nu\lambda}=0$. By the 
results (\ref{gdiffeo}) and (\ref{spin3gauge}), they are determined by the 
following set of eqs.
\begin{eqnarray}
&&\nabla_{\mu} \, \xi_{\nu}+\nabla_{\nu} \, \xi_{\mu} = 0, \\
&& \nabla_{\mu} \, \left(\xi_{(\nu\lambda)}+ \frac{1}{3} \, \rho^a \, 
\Lambda_a 
\, g_{\nu\lambda}\right)    
+\nabla_{\nu} \, \left(\xi_{(\lambda\mu)}+ \frac{1}{3} \, \rho^a \, 
\Lambda_a \, g_{\lambda\mu}\right) \nonumber 
+\nabla_{\lambda} \, \left(\xi_{(\mu\nu)}\right.  \nonumber \\ &&  \left.+ 
\frac{1}{3} 
\, \rho^a \, \Lambda_a \, g_{\mu\nu}\right)=0
\end{eqnarray}
Generally, these are coupled eqs for $\delta g_{\mu\nu}=0$ and 
$\delta \phi_{\mu\nu\lambda}=0$. However, if the background geometry is 
AdS$_3$, $\delta \phi_{\mu\nu\lambda}$ and $\rho^a$ vanish, and the eqs are 
decoupled.  Then  $\xi_{\mu}$ and $\xi_{(\mu\nu)}$ are determined by 
\begin{eqnarray}
\hat{\nabla}_{\mu} \, \xi_{\nu}+ \hat{\nabla}_{\nu} \, \xi_{\mu} &=&0, \\ 
\hat{\nabla}_{\mu} \, \xi_{(\nu\lambda)}+ \hat{\nabla}_{\nu} \, \xi_{(\lambda\mu)}+ 
\hat{\nabla}_{\lambda} \, \xi_{(\mu\nu)} &=&0
\end{eqnarray}
Here $\hat{\nabla}_{\mu}$ is the Christoffel symbol for AdS$_3$ background. 
The Killing tensors {\em cannot} be expressed in terms of the Killing 
vectors as 
$\xi_{(\mu\nu)} = \xi_{\mu} \, \xi_{\nu}-\frac{1}{3} \, g_{\mu\nu} \, g^{\lambda\rho} \, 
\xi_{\lambda} \, \xi_{\rho}$. 

In this case of AdS$_3$ geometry, there exist 6 Killing vectors $\xi^{(i)}_{\mu}$ 
$(i=1, \dots, 6)$ and 10 Killing tensors $\xi^{(\alpha)}_{(\mu\nu)}$ 
$(\alpha=1, \dots, 10)$. 
The Killing vectors correspond to the isometry $SO(2,2)$, and are 
the same as those in the spin-2 gravity: 
$\xi^{(i)} \equiv \xi^{(i)}_{\mu} \, g^{\mu\nu} \, \partial_{\nu}$
\begin{eqnarray}
&&\xi^{(1)}= \partial_t, \qquad 
\xi^{(2)}=\partial_{\phi}, \qquad \xi^{(3)}= \phi \, \partial_t-t \, \partial_{\phi} 
\nonumber \\
&& \xi^{(4)}=\partial_r- t \, \partial_t-\phi \, \partial_{\phi}, \qquad 
\xi^{(5)}=t \, \partial_r- \frac{1}{2} \, (t^2+\phi^2) \, \partial_t
-\frac{1}{2} \, e^{-2r} \, \partial_t  -t \, \phi \, \partial_{\phi}, 
\nonumber  \\
&& \xi^{(6)}=\phi \, \partial_r-t \, \phi \, \partial_t-\frac{1}{2} \, 
(t^2+\phi^2) \, 
\partial_{\phi}-\frac{1}{2} \,
e^{-2r} \, \partial_{\phi}
\end{eqnarray}
The Killing tensors are given by 
\begin{eqnarray}
&& \xi^{(1)}_{(rr)}=1, \quad \xi^{(1)}_{(tt)}=\frac{3}{2} \, (t^2+\phi^2) \, e^{4r}
+\frac{1}{2} \, e^{2r}, \quad \xi^{(1)}_{(\phi\phi)}=\frac{3}{2} \, (t^2+\phi^2) \, 
e^{4r}-\frac{1}{2} \, e^{2r}, \nonumber \\
&&\xi^{(1)}_{(rt)}=\frac{3}{2} \, t \, e^{2r}, \quad \xi^{(1)}_{(r\phi)}=-\frac{3}{2} 
\, \phi \, e^{2r}, \quad \xi^{(1)}_{(t\phi)}=-3 \, t\, \phi \, e^{4r} \nonumber 
\end{eqnarray}
\begin{eqnarray}
&& \xi^{(2)}_{(rr)}=t, \ \xi^{(2)}_{(tt)}=\frac{1}{2} \,t \, (t^2+3 \, 
\phi^2) \, e^{4r}
+\frac{1}{2} \,t\,  e^{2r}, \ \xi^{(2)}_{(\phi\phi)}=\frac{1}{2} \, t \, 
(t^2+3 \, \phi^2) \, 
e^{4r}-\frac{1}{2} \, t \, e^{2r}, \nonumber \\
&&\xi^{(2)}_{(rt)}=\frac{3}{4} \, t\, (t^2+\phi^2) \, e^{2r}+\frac{1}{4}, 
\ \xi^{(2)}_{(r\phi)}=-\frac{3}{2} \, t
\, \phi \, e^{2r}, \ \xi^{(2)}_{(t\phi)}=-\frac{1}{2} \, \phi \, (3 \, t^2+\phi^2) 
\, e^{4r}
-\frac{3}{4} \, \phi \, e^{2r} \nonumber 
\end{eqnarray}
\begin{eqnarray}
&& \xi^{(3)}_{(rr)}=\phi, \ \xi^{(3)}_{(tt)}=\frac{1}{2} \, \phi \, 
(3 \, t^2+ \phi^2) \, e^{4r}
+\frac{1}{2} \, \phi \,  e^{2r}, \ \xi^{(3)}_{(\phi\phi)}=\frac{1}{2} \, \phi \, 
(3 \, t^2+\phi^2) \, 
e^{4r}-\frac{1}{2} \, \phi \, e^{2r}, \nonumber \\
&&\xi^{(3)}_{(rt)}=\frac{3}{2} \, t \phi \, e^{2r}, 
\ \xi^{(3)}_{(r\phi)}=-\frac{3}{4} \, (t^2+\phi^2)
\, e^{2r}+\frac{1}{4}, \ \xi^{(3)}_{(t\phi)}=-\frac{1}{2} \, t \, 
( t^2+3 \, \phi^2) \, e^{4r} 
\nonumber 
\end{eqnarray}
\begin{eqnarray}
&& \xi^{(4)}_{(rr)}=t \, \phi, \ \xi^{(4)}_{(tt)}=\frac{1}{2} \, t\phi \, 
(t^2+ \phi^2) \, e^{4r}
+\frac{1}{2} \, t \, \phi \,  e^{2r}, \ \xi^{(4)}_{(\phi\phi)}=\frac{1}{2} \, 
t\phi \, 
(t^2+\phi^2) \, 
e^{4r}-\frac{1}{2} \,t \phi \, e^{2r}, \nonumber \\
&&\xi^{(4)}_{(rt)}=\frac{1}{4} \, \phi \, (\phi^2+3 \, t^2) \, e^{2r}
+\frac{1}{4} \, \phi, 
\ \xi^{(4)}_{(r\phi)}=-\frac{1}{4} \, t(t^2+3\, \phi^2)
\, e^{2r}+\frac{1}{4} \, t, \nonumber \\ &&
 \xi^{(4)}_{(t\phi)}=-\frac{1}{8} \,  ( t^4+\phi^4+6\, t^2\phi^2) \, e^{4r} 
+\frac{1}{8} \nonumber 
\end{eqnarray}
\begin{eqnarray}
&& \xi^{(5)}_{(rr)}=t^2 +\phi^2, \quad \xi^{(5)}_{(tt)}=\frac{1}{4} \, (t^4+6\, 
t^2 \, \phi^2+\phi^4) \, e^{4r}
+\frac{1}{2} \, (t^2+\phi^2) \, e^{2r}+\frac{1}{4}, 
\nonumber \\ && \xi^{(5)}_{(\phi\phi)}=\frac{1}{4} \,  (t^4+6\, t^2\, \phi^2
+\phi^4) \, 
e^{4r}-\frac{1}{2} \, (t^2+ \phi^2) \, e^{2r}+\frac{1}{4}, \nonumber \\
&&\xi^{(5)}_{(rt)}=\frac{1}{2} \, t \, (3\, \phi^2+t^2) \, e^{2r}+\frac{1}{2} \, t, 
\quad  \xi^{(5)}_{(r\phi)}=-\frac{1}{2} \, \phi \, (3\, t^2+\phi^2)
\, e^{2r}+\frac{1}{2}\, \phi, \quad \xi^{(5)}_{(t\phi)}=-t\, \phi ( t^2+\phi^2) 
\, e^{4r} \nonumber 
\end{eqnarray}
\begin{eqnarray}
&& \xi^{(6)}_{(tt)}=e^{4r}, \qquad  \xi^{(6)}_{(\phi\phi)}=e^{4r} \nonumber 
\end{eqnarray}
\begin{eqnarray}
&& \xi^{(7)}_{(t\phi)}=e^{4r},  \nonumber 
\end{eqnarray}
\begin{eqnarray}
&& \xi^{(8)}_{(tt)}=\xi^{(8)}_{(\phi\phi)}=2t\phi \, e^{4r}, 
\qquad  \xi^{(8)}_{(rt)}= \phi \, e^{2r}, 
\qquad \xi^{(8)}_{(r\phi)}=-t \, e^{2r}, \qquad \xi^{(8)}_{(t\phi)}
=-(t^2+\phi^2) \, e^{4r} \nonumber 
\end{eqnarray}
\begin{eqnarray}
&& \xi^{(9)}_{(tt)}=2t\phi \, e^{4r}, \qquad  \xi^{(9)}_{(\phi\phi)}
= 2t \, e^{4r}, \qquad \xi^{(9)}_{(rt)}= e^{2r}, \qquad \xi^{(9)}_{(t\phi)}
=-2\, \phi \, e^{4r} \nonumber 
\end{eqnarray}
\begin{eqnarray}
&& \xi^{(10)}_{(tt)}=-2 \, \phi \, e^{4r}, \qquad  \xi^{(10)}_{(r\phi)}=  e^{2r}, 
\qquad \xi^{(10)}_{(t\phi)}=2t \, e^{4r}, \qquad \xi^{(10)}_{(\phi\phi)}=
-2 \, \phi \, e^{4r} \nonumber \\ &&
\end{eqnarray}
Those components which are not presented vanish.

The Killing vectors (tensors) 
are related to the $SL(3,R)$ matrices $\Lambda_a \, t^a$, which generate 
the generalized diffeomorphisms, 
by $\xi_{\mu}= \Lambda_a \, e^a_{\mu}$ 
and  $\xi_{(\mu\nu)}= \Lambda_a \, e^a_{(\mu\nu)}$. 
The Killing vectors can also be obtained by solving eqs 
\begin{eqnarray}
\delta \, e^a_{\mu} = \partial_{\mu} \, \Lambda^a+{f^a}_{bc} \, 
\omega^b_{\mu} \, \Lambda^c={f^a}_{bc} \, \Sigma^b \, e^c_{\mu}.
\end{eqnarray}
Here  $\Sigma^b$ are some functions to be determined by $\Lambda^a$.

\section{ Metric-like fields}
\hspace{5mm}
In this appendix various metric-like fields are defined.

Let us recall that a product of two generators of $sl(3,R)$, $t_a$ and $t_b$, 
can be reduced to terms which are linear in $t_c$ or proportional to an 
identity matrix by using the structure constants.
\begin{equation}
t_a \, t_b = \frac{1}{2} \, [t_a, \, t_b] +\frac{1}{2} \, \{t_a, \, t_b\}
= \frac{1}{2} \, {f_{ab}}^c \, t_c+ \frac{1}{2} \, ({d_{ab}}^c \, 
t_c+{d_{ab}}^0 \, \bm{I})
\end{equation}
Therefore all invariants of the local frame transformations can be 
constructed by contracting $f_{abc}$, $d_{abc}$ and $h_{ab}$ with $e^a_{\mu}$ 
and $e^a_{(\mu\nu)}$. 

Now it is easy to expand $t_a$ in terms of $e_{\mu}$ and $e_{(\mu\nu)}$. 
\begin{equation}
t_a = t_b \, \delta_a^b= t_b \, (e^b_{\mu} \, E_a^{\mu}+\frac{1}{2} \, 
e^b_{(\mu\nu)} 
\, E_a^{(\mu\nu)})= e_{\mu} \,  E_a^{\mu}+\frac{1}{2} \, e_{(\mu\nu)} \, 
E_a^{(\mu\nu)}
\label{ta}
\end{equation}
By using this eq, then the vielbein $e_{\mu}^a$, $e^a_{(\mu\nu)}$ 
can be expressed 
in terms of $E$'s. 
\begin{eqnarray}
e^a_{\mu} &=& \frac{1}{2} \, \mbox{tr} \, t^a \, e_{\mu} = \frac{1}{2} \, 
\mbox{tr} (e_{\nu} \, 
E^{a\nu}+
\frac{1}{2} \, e_{(\nu\lambda)} \, E^{a(\nu\lambda)}) \, e_{\mu}
= g_{\mu\nu} \, E^{a\nu}+\frac{1}{2} \, g_{\mu(\nu\lambda)} \, 
E^{a(\nu\lambda)}, \\
e^a_{(\mu\nu)} &=& \frac{1}{2} \, \mbox{tr} \, t^a \, e_{(\mu\nu)} =  
\frac{1}{2} \, \mbox{tr} 
(e_{\lambda} \, E^{a\lambda}+
\frac{1}{2} \, e_{(\lambda\rho)} \, E^{a(\lambda\rho)}) \, e_{(\mu\nu)} 
\nonumber \\
&=&  g_{(\mu\nu)\lambda} \, E^{a\lambda}+\frac{1}{2} \, 
g_{(\mu\nu)(\lambda\rho)} \, 
E^{a(\lambda\rho)}
\end{eqnarray}

Now, by using this formula (\ref{ta})  $h_{ab}$, $d_{abc}$ and $f_{abc}$ 
are expressed 
in terms of $E$, $e$ and gauge fields. For $h_{ab}$ one obtains
\begin{eqnarray}
h_{ab} 
&=&\frac{1}{2} \, \mbox{tr} \, t_a \, t_b 
= \frac{1}{2} \, \mbox{tr} \left(e_{\mu} \,  E_a^{\mu}+\frac{1}{2} \, 
e_{(\mu\nu)} \, E_a^{(\mu\nu)}\right) \, \left(e_{\lambda} \,  
E_b^{\lambda}+\frac{1}{2} 
\, e_{(\lambda\rho)} \, E_b^{(\lambda\rho)
}\right) \nonumber \\
&=& g_{\mu\nu} \, E_a^{\mu} \, E_b^{\nu}+ \frac{1}{2} \, g_{(\mu\nu)\lambda} \, 
(E_a^{(\mu\nu)} \, E_b^{\lambda}+E_b^{(\mu\nu)} \, E_a^{\lambda})+\frac{1}{4} \, 
g_{(\mu\nu)(\lambda\rho)} E_a^{(\mu\nu)} \, E_b^{(\lambda\rho)}
\end{eqnarray}
For $d_{abc}$ one obtains
\begin{eqnarray}
d_{abc} &=& \frac{1}{2} \, \mbox{tr} \, \{t_a, \, t_b\} \, t_c \nonumber \\
&=& 2 \, E_a^{\mu} \, E_b^{\nu} \, E_c^{\lambda} \, \phi_{\mu\nu\lambda} 
+E_a^{\mu} \, E_b^{\nu} \, E_c^{(\lambda\rho)} \,
 \left(g_{(\mu\nu)(\lambda\rho)} \, 
+\frac{1}{6} \, g_{\mu\nu} \mbox{tr} \, \rho \, e_{(\lambda\rho)}\right)
\nonumber \\
&& +E_a^{\mu} \, E_b^{(\nu\rho)} \, E_c^{\lambda} \, 
\left(g_{(\mu\lambda)(\nu\rho)} \,
 +\frac{1}{6} \, g_{\mu\lambda} \mbox{tr} \, \rho \, e_{(\nu\rho)}\right)
\nonumber \\ &&
+E_a^{(\mu\rho)} \, E_b^{\nu} \, E_c^{\lambda} \, 
\left(g_{(\mu\rho)(\nu\lambda)} \, 
+\frac{1}{6} \, g_{\nu\lambda} \mbox{tr} \, \rho \, e_{(\mu\rho)}\right) 
\nonumber \\
&& +\frac{1}{4} \, E_a^{\mu} \, E_b^{(\nu\sigma)} \, E_c^{(\lambda\rho)} \, 
g_{\mu(\nu\sigma)(\lambda\rho)}
+\frac{1}{4} \, E_a^{(\mu\sigma)} \, E_b^{\nu} \, 
E_c^{(\lambda\rho)} \, g_{\nu(\mu\sigma)(\lambda\rho)} \nonumber \\&&
+\frac{1}{4} \, 
E_a^{(\mu\rho)} \, 
E_b^{(\nu\sigma)} \, E_c^{\lambda} \, g_{\lambda(\nu\sigma)(\mu\rho)} 
+\frac{1}{8} \, E_a^{(\mu\rho)} \, E_b^{(\nu\sigma)} \, E_c^{(\lambda\kappa)} \, 
g_{(\mu\rho)(\nu\sigma)(\lambda\kappa)}
\label{dabc}
\end{eqnarray}
Here the following manipulation is used. 
\begin{eqnarray}
\mbox{tr} \, \{e_{\mu}, \, e_{\nu} \} \, e_{(\lambda\rho)}= \mbox{tr} \, (2 \, 
e_{(\mu\nu)}+\frac{4}{3} \, g_{\mu\nu} \, \bm{I}+\frac{2}{3} \, g_{\mu\nu} \, 
\rho) \,  e_{(\lambda\rho)} =4 \, g_{(\mu\nu)(\lambda\rho)}+\frac{2}{3} \, 
g_{\mu\nu} \, \mbox{tr} \, \rho \, e_{(\lambda\rho)}
\end{eqnarray}
Extra gauge fields $g_{(\mu\nu)(\lambda\rho)}$ .... are defined as follows.
\begin{eqnarray}
g_{(\mu\nu)(\lambda\rho)} &=& \frac{1}{2} \, \mbox{tr} \, e_{(\mu\nu)} \, 
e_{(\lambda\rho)}, \\
g_{(\mu\nu)(\lambda\rho)\sigma} &=& \frac{1}{2} \, \mbox{tr} \, \{e_{(\mu\nu)},
 \, 
e_{(\lambda\rho)}\} \, e_{\sigma}, \\
g_{(\mu\nu)(\lambda\rho)(\sigma\kappa)} &=& \frac{1}{2} \, \mbox{tr} \, 
\{e_{(\mu\nu)}, \, 
e_{(\lambda\rho)}\} \, e_{(\sigma\kappa)}
\end{eqnarray}
So for the spin-3 gravity gauge fields with up to 6 indices must be 
introduced. 
Note that one can also define a gauge field such as 
\begin{eqnarray}
\phi_{\mu\nu, \lambda\rho} &=& \frac{1}{2} \, \mbox{tr} \, 
\hat{e}_{\mu\nu} \, \hat{e}_{\lambda\rho} 
\nonumber \\
&=& g_{(\mu\nu)(\lambda\rho)}+\frac{1}{6} \, g_{\mu\nu} \, \mbox{tr} \, 
\rho \, e_{(\lambda\rho)} 
 +\frac{1}{6} \, g_{\lambda\rho} \, \mbox{tr} \, \rho \, 
e_{(\mu\nu)}+\frac{1}{18} \, g_{\mu\nu} \, 
g_{\lambda\rho} \,  \mbox{tr} \, \rho^2. \label{phi4}
\end{eqnarray}

Similarly, for $f_{abc}$ one has 
\begin{eqnarray}
f_{abc} &=& \frac{1}{2} \, \mbox{tr} \, [t_a, \, t_b] \, t_c \nonumber \\
&=& \frac{1}{2} \, E_a^{\mu} \, E_b^{\lambda} \, E_c^{\sigma} \, \mbox{tr} 
\, [e_{\mu}, 
\, e_{\lambda}] \, e_{\sigma} 
\nonumber \\
&& +\frac{1}{4} \, E_a^{\mu} \, E_b^{\lambda} \, E_c^{(\sigma\kappa)} \, 
\mbox{tr} \, [e_{\mu}, 
\, e_{\lambda}] \, e_{(\sigma\kappa)}+ \ \mbox{permutations}
\nonumber \\
&& +\frac{1}{8} \, E_a^{\mu} \, E_b^{(\lambda\rho)} \, E_c^{(\sigma\kappa)} 
\, \mbox{tr} \, [e_{\mu}, 
\, e_{(\lambda\rho)}] \, e_{(\sigma\kappa)}+ \ \mbox{permutations} \nonumber \\ 
&& +\frac{1}{16} \, E_a^{(\mu\nu)} \, E_b^{(\lambda\rho)} \, E_c^{(\sigma\kappa)} 
\, \mbox{tr} 
\, [e_{(\mu\nu)}, \, e_{(\lambda\rho)}] \, e_{(\sigma\kappa)}
\label{fabc}
\end{eqnarray}
Here some terms which can be obtained by permutation of indices are not 
written explicitly. 

Therefore for spin-3 gravity, partly anti-symmetric gauge fields with 
up to 6 indices such as
\begin{equation}
F_{\mu\lambda\sigma} \equiv \frac{1}{2} \, \mbox{tr} \, [e_{\mu}, \, 
e_{\lambda}] \, 
e_{\sigma}, \qquad 
F_{\mu\lambda(\sigma\kappa)} \equiv \frac{1}{2} \, \mbox{tr} \, 
[e_{\mu}, \, e_{\lambda}] 
\, e_{(\sigma\kappa)}, 
\quad \ldots 
\label{FF}
\end{equation}
must also be introduced. To remove the local frame indices $a,b, ..$ from 
the action integral (\ref{CSsecond}) and the various relations obtained in 
this paper it is necessary to use the metrix-like fields defined in this 
appendix. However, not all these fields will be independent. They will be 
expressed in terms of fewer fields. At present, explicit relations among 
these fields are not known and we cannot carry out this program. 


In the remaining part of this appendix, it will be argued that 
these metric-like fields can be expressed in terms of $g_{\mu\nu}$ and 
$\phi_{\mu\nu\lambda}$, when the spin-3 geometry is in the neighborhood of 
AdS$_3$ vacuum. 

This is performed by using perturbation expansion around the AdS$_3$ vacuum.
Let us expand the vielbein as $e^a_{\mu}=\bar{e}^a_{\mu}+\psi^a_{\mu}$, where 
$\bar{e}^a_{\mu}$ is the AdS$_3$ vacuum (\ref{AdS3}) and $\psi^a_{\mu}$ is a small 
fluctuation around it. By computing the metric $g_{\mu\nu}=
\bar{g}_{\mu\nu}+ \tilde{g}_{\mu\nu}$ and the spin-3 gauge field 
$\phi_{\mu\nu\lambda}=0+\tilde{\phi}_{\mu\nu\lambda}$ up to first order in 
$\psi^a_{\mu}$, and gauge fixing the local Lorentz rotaion by imposing 8 
conditions on $\psi^a_{\mu}$, one can express $\psi^a_{\mu}$ in terms 
of the fluctuations $\tilde{g}_{\mu\nu}$ and $\tilde{\phi}_{\mu\nu\lambda}$. 
\begin{eqnarray}
\psi^2_r &=& \frac{1}{2} \, \tilde{g}_{rr}, \qquad \psi^6_r= \frac{1}{2} \, 
\tilde{\phi}_{rrr}, \nonumber \\
\psi^2_i &=& \tilde{g}_{ri}, \qquad \psi^6_i = \frac{3}{2} \,\tilde{\phi}_{rri}, 
\nonumber \\
\psi^1_i &=& \epsilon_i \,  e^{-r} \, (-\frac{1}{8} \, \tilde{g}_{tt}+
\frac{1}{4} \, \tilde{g}_{t\phi}-\frac{1}{8} \, \tilde{g}_{\phi\phi}), \nonumber \\
\psi^3_i &=&   e^{-r} \, (-\frac{1}{2} \, \tilde{g}_{ii}-
\frac{1}{4} \, \tilde{g}_{t\phi}+\frac{1}{8} \, (\tilde{g}_{tt}
+\tilde{g}_{\phi\phi})), \nonumber \\
\psi^4_i &=&  -\frac{1}{2} \, \epsilon_i \, \tilde{\phi}_{rri}
+\frac{1}{4} \, \epsilon_i  \, (\tilde{\phi}_{rrt}+\tilde{\phi}_{rr\phi})
+\frac{1}{3} \, e^{-2r} \, \tilde{\phi}_{iii}-\frac{1}{4} \, e^{-2r} \, (
\tilde{\phi}_{iit}+\tilde{\phi}_{ii\phi}) \nonumber \\ 
&&+\frac{1}{12} \, e^{-2r} \, 
(\tilde{\phi}_{ttt}+\tilde{\phi}_{\phi\phi\phi}), \nonumber \\
\psi^5_i &=& \epsilon_i \,  e^{-r} \, (-\frac{1}{4} \, (\tilde{\phi}_{rtt}
+\tilde{\phi}_{r\phi\phi})+\frac{1}{2} \, \tilde{\phi}_{rt\phi}), \nonumber \\
\psi^7_i &=& \frac{1}{6} \, \epsilon_i \,  e^{r} \, \tilde{\phi}_{rrr}
-e^{-r} \, \tilde{\phi}_{rii}-\frac{1}{2} \, e^{-r} \, \tilde{\phi}_{rt\phi}
+\frac{1}{4} \, e^{-r} \, (\tilde{\phi}_{rtt}+\tilde{\phi}_{r\phi\phi}), \nonumber 
\\ \psi^8_i &=& -\frac{1}{4} \, \epsilon_i \, (\tilde{\phi}_{rrt}
+\tilde{\phi}_{rr\phi})+\frac{1}{4} \, e^{-2r} \, (\tilde{\phi}_{iit}
+\tilde{\phi}_{ii\phi})-\frac{1}{12} \, e^{-2r} \, (\tilde{\phi}_{ttt}
+\tilde{\phi}_{\phi\phi\phi})
\end{eqnarray}
Here the index $i$ takes two values, $i=t,\phi$, and $\epsilon_t=+1$, 
$\epsilon_{\phi}=-1$. The remaining components vanish; $\psi_r^{1,3,4,5,7,8}=0$. 
Then by substituting the result into the other metric-like fields one obtains 
them in terms of $\tilde{g}_{\mu\nu}$ and $\tilde{\phi}_{\mu\nu\lambda}$. 
For example, the case of $g_{(\mu\nu)(\lambda\rho)}
=\bar{g}_{(\mu\nu)(\lambda\rho)}+\tilde{g}_{(\mu\nu)(\lambda\rho)}$ is presented below.
$\bar{g}_{(\mu\nu)(\lambda\rho)}$ is the background (\ref{Mg4}) and 
$\tilde{g}_{(\mu\nu)(\lambda\rho)}$ is the fluctuation. 
\begin{eqnarray}
\tilde{g}_{(rr)(rr)} &=& \frac{2}{3} \, \tilde{g}_{rr}, \qquad 
\tilde{g}_{(rr)(ri)} = \frac{1}{3} \, \tilde{g}_{ri}, \qquad 
\tilde{g}_{(rr)(t\phi)} =- \frac{1}{6} \, \tilde{g}_{t\phi}, \nonumber \\
\tilde{g}_{(rr)(ii)} &=& \frac{1}{6} \, \epsilon_i \, e^{2r} \, \tilde{g}_{rr}
-\frac{1}{6} \, \tilde{g}_{ii}, \qquad 
\tilde{g}_{(ri)(ri)} = -\frac{1}{4} \, \epsilon_i \, e^{2r} \, \tilde{g}_{rr}+\frac{1}{4} \, \tilde{g}_{ii}, \nonumber \\
\tilde{g}_{(ri)(ii)} &=& -\frac{1}{3} \, \epsilon_i \, e^{2r} \, \tilde{g}_{ri}, 
\qquad \tilde{g}_{(rt)(r\phi)} = \frac{1}{4} \, \tilde{g}_{t\phi}, \qquad 
\tilde{g}_{(rt)(t\phi)} = -\frac{1}{4} \, e^{2r} \, \tilde{g}_{r\phi}, \nonumber \\
\tilde{g}_{(r\phi)(t\phi)}& =& \frac{1}{4} \, e^{2r} \, \tilde{g}_{rt}, \qquad 
\tilde{g}_{(rt)(\phi\phi)} = -\frac{1}{6}  \, e^{2r} \, \tilde{g}_{rt},
\qquad \tilde{g}_{(ii)(ii)} = -\frac{2}{3} \, \epsilon_i \, e^{2r} \, \tilde{g}_{ii}, \nonumber \\
\tilde{g}_{(t\phi)(ii)} &=& -\frac{1}{3} \, \epsilon_i \, e^{2r} \, \tilde{g}_{t\phi},
\qquad \tilde{g}_{(tt)(\phi\phi)} = -\frac{1}{6} \, e^{2r} \, \tilde{g}_{tt}+
\frac{1}{6} \, e^{2r} \, \tilde{g}_{\phi\phi}, \nonumber \\
\tilde{g}_{(t\phi)(t\phi)} &=& \frac{1}{4} \, e^{2r} \, \tilde{g}_{tt}-
\frac{1}{4} \, e^{2r} \, \tilde{g}_{\phi\phi}, \qquad 
\tilde{g}_{(r\phi)(tt)} = \frac{1}{6}  \, e^{2r} \, \tilde{g}_{r\phi}
\end{eqnarray}
Other extra metric-like fields can also be worked out similarly. 

Although this argument is far from the all-order proof, this at least supports 
the conjecture that all metric-like fields can be expressed in terms of 
$g_{\mu\nu}$ and $\phi_{\mu\nu\lambda}$ in the neighborhood of AdS$_3$ vacuum.


\section{Solution for $S_{\mu\nu,\lambda\rho}$}
\hspace{5mm}
In  this appendix a solution to the eqs for $S_{\mu\nu,\lambda\rho}$, (\ref{coefS}), 
(\ref{deltrace}) are presented. Let us define matrices ${\cal A}_{\mu\nu}$ 
and ${\cal B}^{\mu\nu}$ by 
\begin{eqnarray}
{\cal B}^{\mu\nu} &=& \frac{5}{48} \, J^{(\mu\nu)(\lambda\rho)} \, W_{\lambda\rho}, \\
{\cal A}_{\mu\nu} &=& \hat{\nabla}_{\mu} \, {\phi_{\nu\lambda}}^{\lambda}-\frac{5}{9} 
\, \hat{\nabla}^{\lambda} \, \Phi_{\mu\nu\lambda}-\frac{5}{72} \, (\hat{\nabla}_{\mu} 
\, \Phi_{\nu\alpha\beta}+\hat{\nabla}_{\nu} \,\Phi_{\mu\alpha\beta}-\hat{\nabla}_{\alpha} 
\,\Phi_{\mu\nu\beta}) \, J^{(\alpha\beta)(\sigma\kappa)} \, W_{\sigma\kappa} \nonumber \\ &&
\end{eqnarray}
The matrix ${\cal B}^{\mu\nu}$ is symmetric traceless but ${\cal A}_{\mu\nu}$ is 
not symmetric. In terms of these matrices, eq (\ref{deltrace}) is written as 
\begin{eqnarray}
g^{\lambda\rho} \, S_{\mu\lambda,\rho\nu}+ S_{\mu\nu,\lambda\rho} \, {\cal B}^{\lambda\rho}= 
{\cal A}_{\mu\nu}. \label{SAB}
\end{eqnarray}
If $\phi_{\mu\nu\lambda}=0$, then the connections reduce to the Christoffel symbol
in Einstein gravity. In this case ${\cal A}_{\mu\nu}=0$ and the solution will be 
given by $S_{\mu\nu,\lambda\rho}=0$.  In what follows the above matrices 
are considered to be small and $S_{\mu\nu,\lambda\rho}$ will be obtained as 
a power series in these matrices. Then the second term on the left-hand side 
of (\ref{SAB}) is second order in ${\cal A}$, ${\cal B}$. 

Let us first consider the equation.
\begin{equation}
g^{\lambda\rho} \, S^{(0)}_{\mu\lambda,\rho\nu}={\cal A}_{\mu\nu} \label{A0}
\end{equation}
The solution to this eq and (\ref{coefS}) is obtained as
\begin{eqnarray}
S^{(0)}_{\mu\nu,\lambda\rho} &=& -\frac{3}{5} \, g_{\mu\nu} \, ({\cal A}_{\lambda\rho}+{\cal 
A}_{\rho\lambda})
-\frac{2}{5} \, g_{\lambda\rho} \,  ({\cal A}_{\mu\nu}+{\cal A}_{\nu\mu})  
\nonumber \\ &&
  -\frac{3}{10} \, (g_{\mu\lambda} \, g_{\nu\rho}+ g_{\mu\rho} \, 
g_{\nu\lambda}-2 \, g_{\mu\nu} \, g_{\lambda\rho} ) \, {{\cal A}_{\alpha}}^{\alpha} 
\nonumber \\
&& +\frac{1}{5} \, g_{\mu\lambda} \, (2 \, {\cal A}_{\nu\rho}+ {\cal A}_{\rho\nu})
+\frac{1}{5} \, g_{\nu\lambda} \, (2 \, {\cal A}_{\mu\rho}+ 
{\cal A}_{\rho\mu}) \nonumber \\
&&+\frac{1}{5} \, g_{\mu\rho} \, (2 \, {\cal A}_{\nu\lambda}+ 
{\cal A}_{\lambda\nu})
+\frac{1}{5} \, g_{\nu\rho} \, (2 \, {\cal A}_{\mu\lambda}+
{\cal A}_{\lambda\mu})
\label{SA0}
\end{eqnarray}

Now we split $S_{\mu\nu,\lambda\rho}$ as $S_{\mu\nu,\lambda\rho}=S^{(0)}_{\mu\nu,\lambda\rho}+
 \tilde{S}_{\mu\nu,\lambda\rho}$. 
Owing to (\ref{A0}) this remaining part $\tilde{S}_{\mu\nu,\lambda\rho}$ satisfies 
\begin{equation}
g^{\lambda\rho} \, \tilde{S}_{\mu\lambda,\rho\nu}+ \tilde{S}_{\mu\nu,\lambda\rho} \, 
{\cal B}^{\lambda\rho}=
{\cal A}^{(1)}_{\mu\nu},
\end{equation}
where ${\cal A}^{(1)}_{\mu\nu}=-S^{(0)}_{\mu\nu,\lambda\rho} \, {\cal B}^{\lambda\rho}$. 
This equation has the same structure as (\ref{SAB}). Because the term 
$\tilde{S}_{\mu\nu,\lambda\rho} \, {\cal B}^{\lambda\rho}$ is subleading, 
we can drop this term to leading order, replace $g^{\lambda\rho} \, 
\tilde{S}_{\mu\lambda,\rho\nu}$ 
by $g^{\lambda\rho} \, S^{(1)}_{\mu\lambda,\rho\nu}$, and solve eq $g^{\lambda\rho} \, 
S^{(1)}_{\mu\lambda,\rho\nu}={\cal A}^{(1)}_{\mu\nu}$, which has the same form as 
(\ref{A0}). 

We will repeat this procedure, obtain $S^{(n)}_{\mu\nu,\lambda\rho}$ 
at each step and by assuming convergence finally sum up $S^{(n)}_{\mu\nu,
\lambda\rho}$ to have $S_{\mu\nu,\lambda\rho}=\sum_{n=0}^{\infty} \, 
S^{(n)}_{\mu\nu,\lambda\rho}$.  
${\cal A}^{(n)}_{\mu\nu}$ is defined by 
\begin{equation}
{\cal A}^{(n)}_{\mu\nu}=-S^{(n-1)}_{\mu\nu,\lambda\rho} \, {\cal B}^{\lambda\rho} \qquad 
({\cal A}^{(0)}_{\mu\nu}={\cal A}_{\mu\nu}). \label{An+1} 
\end{equation}
The equation for $S^{(n)}_{\mu\nu,\lambda\rho}$ is given by 
\begin{equation}
g^{\lambda\rho} \, S^{(n)}_{\mu\lambda,\rho\nu}={\cal A}^{(n)}_{\mu\nu}.
\end{equation}
and the solution is given by (\ref{SA0}) with $S^{(0)}_{\mu\nu,\lambda\rho}$ and 
${\cal A}_{\mu\nu}$ replaced by $S^{(n)}_{\mu\nu,\lambda\rho}$  and 
${\cal A}^{(n)}_{\mu\nu}$, respectively. 

The above procedure yields recursion relations between ${\cal A}^{(n)}_{\mu\nu}$ 
and ${\cal A}^{(n+1)}_{\mu\nu}$, and their solution for ${\cal A}^{(n)}_{\mu\nu}$ 
takes the following form.
\begin{eqnarray}
{\cal A}^{(n)}_{\mu\nu} &=& g_{\mu\nu} \, F^{(n)}+\sum_{m=1}^{n-1} \,
 ({\cal B}^m)_{\mu\nu} \, X^{(n)}_m \nonumber \\
&& + \sum_{m=0}^n \, \{ ({\cal B}^m \, {\cal A} \, {\cal B}^{n-m})_{\mu\nu}+ 
({\cal B}^m \, 
{\cal A} \, {\cal B}^{n-m})_{\nu\mu} \} \, Y^{(n)}_m \label{Aansatz}
\end{eqnarray}
Here $g_{\mu\nu}$ is inserted to construct the powers of ${\cal B}^{\lambda\rho}$. 
By computing $S^{(n)}_{\mu\nu,\lambda\rho}$ 
by using (\ref{SA0}) with suitable replacements and obtaining ${\cal 
A}^{(n+1)}_{\mu\nu}$ by using (\ref{An+1}), the 
recursion relations for $F^{(n)}$, $X^{(n)}_m$ and $Y^{(n)}_m$ are obtained. 
\begin{eqnarray}
F^{(n+1)}&=& \frac{6}{5} \, \sum_{m=1}^{n-1} \, \text{tr} ({\cal B}^{m+1})  \, X^{(n)}_m+\frac{12}{5} 
\, \sum_{m=0}^n \, Y^{(n)}_m \, 
\text{tr} ({\cal A} \, {\cal B}^{n+1}), \\ 
X^{(n+1)}_1 &=& -\frac{3}{5} \, F^{(n)}+\frac{6}{5}\, \text{tr}({\cal A}{\cal B}^n) \, \sum_{m=0}^n \, Y_m^{(n)}+\frac{3}{5}\sum_{m=1}^{n-1} \, (\text{tr} \, {\cal B}^m) \, X^{(n)}_m, \\ 
X^{(n+1)}_{m+1} &=&-\frac{12}{5} \, X^{(n)}_m, \qquad 
Y^{(n+1)}_m = -\frac{6}{5} \, Y^{(n)}_m-\frac{6}{5} \, Y^{(n)}_{m-1}
\end{eqnarray}
The initial condition at $n=1$ is 
\begin{equation}
Y^{(1)}_0= - \frac{4}{5}, \qquad Y^{(1)}_1=-\frac{2}{5}, \qquad F^{(1)}= \frac{6}{5} \, 
\text{tr} ({\cal B}{\cal A}), \qquad X^{(1)}_1=\frac{3}{5} \, \text{tr}{\cal A}
\end{equation}
The solution for $Y^{(n)}_m$ is given by 
\begin{equation}
Y^{(n)}_m = \frac{1}{3} \, \left(-\frac{6}{5}\right)^n \, \frac{(n-1)!}{m! \, (n-m)!} 
\, (2n-m).
\end{equation}
$X^{(n)}_m$ is determined in terms of $F^{(m)}$ by $X^{(n)}_m= (3/4) \, (-12/5)^{m} \, 
F^{(n-m)}$, and $F^{(n)}$ is 
the solution to the recursion relation
\begin{eqnarray}
F^{(n+1)}=\frac{6}{5} \,  \, \sum_{m=1}^{n-1} \, (-12/5)^m \, \text{tr} \, 
({\cal B}^{m+1}) \, 
F^{(n-m)}-\frac{1}{2} \, (-12/5)^{n+1}  \, \text{tr}({\cal A} {\cal B}^{n+1}).
\end{eqnarray}
This last eq can be solved by iterations.
\begin{eqnarray}
F^{(n)}=-\frac{1}{2} \, (-12/5)^n \, \text{tr} ({\cal A}{\cal B}^n)+\frac{1}{4} \, 
\sum_{m=1}^{n-2} \, (-12/5)^{n} \, \text{tr}({\cal B}^{n-m}) \, \text{tr}({\cal A}{\cal B}^m) 
\nonumber \\
+(25/576) \, \sum_{m=1}^{n-2} \, \sum_{k=1}^{m-2} \, (-12/5)^{n-k} \, \text{tr}
({\cal B}^{n-m}) \, \text{tr}({\cal B}^{m-k}) \, F^{(k)}= \dots
\end{eqnarray}
For example,  $F^{(2)}=-(72/25) \, \text{tr} ({\cal A}{\cal B}^2)$, 
$F^{(3)}=-(432/125) \, 
\text{tr} ({\cal B}^2) \, \text{tr} ({\cal A}{\cal B})+(864/125) \, \text{tr} 
({\cal A}{\cal B}^3)$, $F^{(4)}=(5184/625) \, \text{tr}({\cal B}^2) \, \text{tr}
({\cal A}{\cal B}^2) +(5184/625) \, \text{tr} ({\cal B}^3) \, 
\text{tr} ({\cal A}{\cal B})-(10368/625) \, \text{tr} ({\cal A}{\cal B}^4)$ and  then 
we obtain  
$X^{(2)}_1=-(54/25) \, \text{tr} ({\cal B}{\cal A})$, $X^{(3)}_1=(648/125) \, \text{tr} 
({\cal B}^2{\cal A})$, $X^{(3)}_2=(648/125) \, \text{tr} ({\cal B}{\cal A})$.  
Then, ${\cal A}^{(n)}_{\mu\nu}$ can be computed using (\ref{Aansatz}) to any 
desired larger value of $n$. 

Finally, $S_{\mu\nu,\lambda\rho}$ is given by 
\begin{eqnarray}
S_{\mu\nu,\lambda\rho} &=& \sum_{n=0}^{\infty} \, S^{(n)}_{\mu\nu,\lambda\rho} \nonumber \\
&=& \sum_{n=0}^{\infty} \,  \left\{-\frac{3}{5} \, g_{\mu\nu} \, ({\cal A}^{(n)}_{\lambda\rho}
+{\cal A}^{(n)}_{\rho\lambda})
-\frac{2}{5} \, g_{\lambda\rho} \,  ({\cal A}^{(n)}_{\mu\nu}+{\cal A}^{(n)}_{\nu\mu}) \right.   \nonumber \\
&&
+\frac{3}{5} \, g_{\mu\nu} \, g_{\lambda\rho} \, {{\cal A}^{(n)}_{\alpha}}^{\alpha} -\frac{3}{10} \, (g_{\mu\lambda} \, g_{\nu\rho}+g_{\mu\rho} \, g_{\nu\lambda}) \, {\cal A}^{(n)\alpha}_{\alpha} \nonumber \\
&& +\frac{1}{5} \, g_{\mu\lambda} \, (2 \, {\cal A}^{(n)}_{\nu\rho}+
{\cal A}^{(n)}_{\rho\nu}) 
+\frac{1}{5} \, g_{\nu\lambda} \, (2 \, {\cal A}^{(n)}_{\mu\rho}+ 
{\cal A}^{(n)}_{\rho\mu}) \nonumber \\
&& \left. +\frac{1}{5} \, g_{\mu\rho} \, (2 \, {\cal A}^{(n)}_{\nu\lambda}
+ {\cal A}^{(n)}_{\lambda\nu})
+\frac{1}{5} \, g_{\nu\rho} \, (2\, {\cal A}^{(n)}_{\mu\lambda}+ 
{\cal A}^{(n)}_{\lambda\mu}) \right\}
\end{eqnarray}

\section{Metric $G_{MN}$ for \lq 8D space'}
\hspace{5mm}
The metric tensor $g_{\mu\nu}$ is constructed in terms of $e^a_{\mu}$. By 
a similar construction one can extend the metric tensor to that for an 
extended 8D space by combining $e^a_{\mu}$ and $e^a_{(\mu\nu)}$. Let us define 
a new metric $$ \left( \begin{array}{cc} 
G_{\mu\lambda} & G_{\mu(\rho\kappa)} \\
G_{(\nu\sigma)\lambda} & G_{(\nu\sigma)(\rho\kappa)}
\end{array} \right) $$
by the eqs.
\begin{eqnarray}
G_{\mu\nu} &=& g_{\mu\nu}=e^a_{\mu} \, e_{a\nu}, \\
G_{\mu(\nu\lambda)}=G_{(\nu\lambda)\mu} &=&  g_{\mu(\nu\lambda)}= e^a_{\mu} 
\, e_{a(\nu\lambda)}, \\
G_{(\mu\nu)(\lambda\rho)} &=&  g_{(\mu\nu)(\lambda\rho)}= e^a_{(\mu\nu)} \, 
e_{a(\lambda\rho)}
\end{eqnarray}
This tensor is a metric tensor in a fictitious 8D space which contains the 
ordinary spacetime. The inverse metric is easily obtained.
\begin{eqnarray}
G^{\mu\nu} &=& g^{\mu\nu}+\frac{1}{4} \, {g^{\mu}}_{(\lambda\rho)} \, 
J^{(\lambda\rho)(\sigma\kappa)} \, {g_{(\sigma\kappa)}}^{\nu}, \label{Gumun}\\
G^{\mu(\nu\lambda)} &=& G^{(\nu\lambda)\mu} = -\frac{1}{2} \, 
{g^{\mu}}_{(\rho\sigma)} \, J^{(\rho\sigma)(\nu\lambda)}, \\
G^{(\mu\nu)(\lambda\rho)} &=&  J^{(\mu\nu)(\lambda\rho)} \label{G4}
\end{eqnarray}
$J^{(\lambda\rho)(\sigma\kappa)}$ is defined in (\ref{J}). 
They satisfy the following relations.
\begin{eqnarray}
&& G^{\mu\nu} \, G_{\nu\lambda}+\frac{1}{2} \, G^{\mu(\nu\rho)} \, 
G_{(\nu\rho)\lambda}= \delta_{\lambda}^{\mu}, \label{GG1}\\
&& G^{\mu\nu} \, G_{\nu(\lambda\rho)}+\frac{1}{2} \, G^{\mu(\nu\sigma)} \, 
G_{(\nu\sigma)(\lambda\rho)}=0, \\
&& G^{(\mu\nu)\lambda} \, G_{\lambda\rho}+\frac{1}{2} \, 
G^{(\mu\nu)(\lambda\sigma)} \, G_{(\lambda\sigma)\rho}=0, \\
&& G^{(\mu\nu)\sigma} \, G_{\sigma(\lambda\rho)}+ \frac{1}{2} \, 
G^{(\mu\nu)(\kappa\sigma)} \, G_{(\kappa\sigma)(\lambda\rho)}
= \delta^{\mu}_{\lambda} \, \delta^{\nu}_{\rho}+\delta^{\nu}_{\lambda} \, 
\delta^{\mu}_{\rho}-\frac{2}{3} \, g^{\mu\nu} \, g_{\lambda\rho}
\label{GG4}
\end{eqnarray}
This inverse metric can be expressed in terms of the inverse vielbein. 
One can show that
\begin{eqnarray}
G^{\mu\nu} &=& E_a^{\mu} \, E^{a\nu}, \nonumber \\
G^{\mu(\nu\lambda)} &=&  E_a^{\mu} \, E^{a(\nu\lambda)}, \nonumber \\
G^{(\mu\nu)(\lambda\rho)} &=&  E_a^{(\mu\nu)} \, E^{a(\lambda\rho)}. \label{GI}
\end{eqnarray}

By using the above formulae one can show that $E$'s can be expanded in terms 
of $e$'s as follows.
\begin{eqnarray}
E^{a\mu} &=& G^{\mu\nu} \, e^a_{\nu}+\frac{1}{2} \, G^{\mu(\nu\lambda)} \, 
e^a_{(\nu\lambda)}, \\
E^{a(\mu\nu)} &=&  G^{(\mu\nu) \lambda} \, e^a_{\lambda}+\frac{1}{2} \, 
G^{(\mu\nu)(\lambda\rho)} \, e^a_{(\lambda\rho)}
\end{eqnarray} 

Let us denote the above metric as $G_{MN}$, where $M$ and $N$ take two 
types of indices $\mu$ and $(\nu\lambda)$. Then, the relation 
(\ref{GG1})-(\ref{GG4}) can be succinctly written as 
\begin{equation}
G_{MN} \, G^{NL} = {\delta_M}^L,
\end{equation}
where ${\delta_M}^L$ is the ordinary Kronecker's $\delta$ symbol for 
$M=\mu$ and $L=\nu$. Otherwise, 
\begin{equation}
{\delta_{\mu}}^{(\nu\lambda)}={\delta_{(\mu\nu)}}^{\lambda}=0, \qquad 
{\delta_{(\mu\nu)}}^{(\lambda\rho)}= \delta_{\mu}^{\lambda} \, 
\delta_{\nu}^{\rho}+\delta_{\nu}^{\lambda} \, \delta_{\mu}^{\rho}-\frac{2}{3}
 \, g_{\mu\nu} \, g^{\lambda\rho}. \label{deltaAC}
\end{equation}

Finally, the covariant derivative $\nabla_{\mu}$ is compatible with this 
metric tensor. As was shown in (\ref{delgmn2}), $G_{\mu\nu}=g_{\mu\nu}$ satisfies 
$\nabla_{\lambda} \, g_{\mu\nu}=0$. This property is true for all components. 
\begin{equation}
\nabla_{\mu} \, G_{MN} = \partial_{\mu} \, G_{MN}-\Gamma_{\mu M}^K \, G_{KN}
-\Gamma_{\mu N}^K \, G_{KM}= 0.
\label{metricity}
\end{equation}
This is because $G_{MN}$ is given by $e^a_M \, e_{aN}$ and the vielbeins are 
covariantly constant, $D_{\mu} \, e^a_N=0$. The above relation is not 
sufficient to determine $\Gamma_{\mu M}^N$ completely, since this is not 
symmetric under interchange of the lower indices. 

When $\nabla_{\mu} \, G_{(\nu\lambda)(\rho\sigma)}$ is computed explicitly, 
 this does not seemingly vanish. The result contains  
$g_{(\mu\nu)(\lambda\rho)\sigma}$. As was discussed at the end of sec.2, 
however, not all the gauge fields are independent. Those relations among these 
fields will be such that these covariant derivatives of $G_{MN}$ actually 
vanish. Thus by using the condition of metric-compatibility  
(\ref{metricity}), some of these relations may be obtained.


\newpage

\begin{thebibliography}{99}
\bibitem{HR} M. Henneaux and S-J Rey, {\it Nonlinear $W_{\infty}$ 
as asymptotic symmetry of three-dimendional higher spin AdS gravity}, 
[ArXiv:1008.4579 [hep-th]].
\bibitem{Campoleoni} A. Campoleoni, S. Fredenhagen, S. Pfenninger and S. 
Theisen, {\it Asymptotic symmetries 
of three-dimensional gravity coupled to higher-spin fields}, [arXiv:1008.4744 
[hep-th]].
\bibitem{GGS} M. R. Gaberdiel, R. Gopakumar and A. Saha, {\it Quantum 
W-symmetry in AdS$_3$ }, 
JHEP 1102 (2011) 004 [arXiv:1009.6087 [hep-th]].
\bibitem{GG} M. R. Gaberdiel and R. Gopakumar, {\it An AdS$_3$ dual for minimal
model CFTs}, [arXiv:1011.2986 [hep-th]].
\bibitem{GaHa} M. R. Gaberdiel and T. Hartman, {\it Symmetries of 
holographic minimal models}, [arXiv: 1101.2910 [hep-th]]
\bibitem{GK} M. Gutperle and P. Kraus, {\it Higher spin black holes}, 
[arXiv:1103.4304 [hep-th]].
\bibitem{Ahn} C. Ahn, {\it The large N 't Hooft limit of coset minimal
 models}, [arXiv: 1106.0351 [hep-th]].
\bibitem{GGHR} M. R. Gaberdiel, R. Gopakumar, T. Hartman and S. Raju, {\it 
Partition functions of holographic minimal models}, [arXiv: 1106.1897 [hep-th]].
\bibitem{CY} C-M. Chang and X. Yin, {\it Higher spin gravity with matter 
in AdS$_3$ and its CFT dual}, [arXiv:1106.2580
 [hep-th]].
\bibitem{AGKP} M. Ammon, M. Gutperle, P. Kraus and E. Perlmutter, 
{\it Spacetime geometry and in higher 
spin gravity}, [arXiv:1106.4788 [hep-th]].
\bibitem{KP} P. Kraus and E. Perlmutter, {\it Partition functions of higher 
spin black holes and their 
CFT duals}, [arXiv:1108.2567 [hep-th]].
\bibitem{C2} A. Campoleoni, {\it Higher spins in D=2+1}, [arXiv:1110.5841 
[hep-th]].
\bibitem{GGR} M. Gary, D. Grumiller and R. Rashkov, {\it Towards non-AdS 
holography in 3-dimensional 
higher spin gravity}, [arXiv:1201.0013 [hep-th]].
\bibitem{GHJ} M. R. Gaberdiel, T. Hartman and K. Jin, {\it Higher spin black 
holes from CFT}, [arXiv:1203.0015 [hep-th]].
\bibitem{HGJR} M. Henneaux, G. L. G\'omez, J. Park and S-J Rey, {\it 
Super-W$_{\infty}$ asymptotic symmetry of higher-spin AdS$_3$ Super gravity}, 
[arXiv: 1203.5152 [hep-th]].
\bibitem{BCT} M. Ba\~nados, R. Canto and S. Theisen, {\it The action for 
higher spin black holes in three dimensions}, [arXiv: 1204.5105 [hep-th]].
\bibitem{CHLM} A. Castro, E. Hijano, A. Lepage-Jutier and A. Maloney, 
{\it Black holes and singularity resolution in higher spin gravity}, 
[arXiv:1110.4117 [hep-th]].
\bibitem{PTT} A. P\'{e}rez, D. Tempo and R. Troncoso, {\it Higher spin 
gravity in 3D black holes, global charges and thermodynamics}, [arXiv:
1207.2844 [hep-th]].
\bibitem{review} M. Ammon, M. Gutperle, P. Kraus and E. Perlmutter, 
{\it Black holes in three dimensional higher spin gravity: a review}, 
[arXiv:1208.5182 [hepth]].
\bibitem{Gubser} I.R. Klebanov and A.M. Polyakov, {\it AdS dual of the 
critical O(N) vector model}, Phys. Lett B 350, 213 (2002), 
[arXiv:hep-th/0210114].
\bibitem{Fronsdal} C. Fronsdal, {\it Massless fields with integer spin}, Phys. 
Rev. D 18 (1978) 3624.
\bibitem{FV1} E. S. Fradkin and M. A. Vasiliev, {\it On the gravitational 
interaction of massless higher 
spin fields}, Phys. Lett. B 189 (1987) 89.
\bibitem{FV2} E. S. Fradkin and M. A. Vasiliev, {\it  Candidate for the 
role of higher-spin gravity}, Ann. 
Phys. 177 (1987) 63.
\bibitem{Vasiliev} M. A. Vasiliev, {\it Progress in higher spin gauge 
theories}, [arXiv:hep-th/0104246]. 
\bibitem{Blencowe} M. P. Blencowe, {\it A consistent interacting massless 
higher-spin field theory in D=2+1}, 
Class. Quantum Grav. 6 (1989) 443-452.
\bibitem{AT} A. Ach\'ucarro and P. K. Townsend, {\it A Chern-Simons action 
for three-dimensional anti-de Sitter supergravity theories}, Phys. Lett. B180 
(1986) 89.
\bibitem{WittenCS} E. Witten, {\it 2+1 dimensional gravity as an exactly 
soluble system}, Nucl. Phys. B311 (1988) 46.
\bibitem{DJT} S. Deser, R. Jackiw and S. Templeton, {\it Three-dimensional 
Massive gauge theories}, Phys. Rev. Lett. 48
 (1982) 975; {\it Topologically massive gauge theories}, Ann. Phys. 140 
(1982) 372. 
\bibitem{DX} S. Deser and X. Xiang, {\it Canonical formulations of full 
non-linear topologically massive gravity}, Phys. 
Lett. B263 (1991) 39; S. Carlip, Nucl. Phys. B362 (1991) 111. 
\bibitem{Jackiw} R. Jackiw, {\it Fifty years of Yang-Mills theory and my 
contributions to it}, [arXiv:0403109 [physics]].
\bibitem{Carlip} S. Carlip, {\it Quantum gravity in 2+1 dimensions}, Cambridge
Univ. Press 1998.
\bibitem{TPHSM1} A. Bagchi, S. Lai, A. Saha and B. Sahoo, {\it Topologically 
massive higher spin gravity}, [arXiv:1107.0915 [hep-th]]; 
{\it One loop partition function for topologically massive higher spin 
gravity}, [arXiv:1107.2063 [hep-th]]. 
\bibitem{TPHSM2} B. Chen, J. Long and J-b Wu, {\it Spin-3 topologically 
massive gravity}, [arXiv:1106.5141 [hep-th]]; B. Chen and J. Long, 
{\it High spin topologically massive gravity}, 
[arXiv:1110.5113 [hep-th]]. 
\bibitem{BTZ} M. Ba\~{n}ados, C. Teitelboim and J, Zanelli, Phys. Rev. 
Lett. 69, 1849 (1992), hep-th/9204099.
\bibitem{CFPT} A. Campoleoni, S. Fredenhagen, S. Pfenninger and S. Theisen, 
{\it Towards metric-like higher-spin gauge theories in three dimensions}, 
[arXiv:1208.1851 [hep-th]].












\end{thebibliography}
\end{document}